\documentclass[moor]{informs1}      

\usepackage{natbib}
 \NatBibNumeric
 \def\bibfont{\small}%
 \bibpunct[, ]{[}{]}{,}{n}{}{,}%

\usepackage[colorlinks=true,breaklinks=true,bookmarks=true,urlcolor=blue,
     citecolor=blue,linkcolor=blue,bookmarksopen=false,draft=false]{hyperref}

\def\EMAIL#1{\href{mailto:#1}{#1}}

\TheoremsNumberedThrough     

\EquationsNumberedThrough    

\usepackage[utf8]{inputenc}
\usepackage{pgfplots}
\pgfplotsset{compat=1.15}
\usepackage{mathrsfs}
\usetikzlibrary{arrows}
\usepackage{amsmath}
\usepackage{dsfont}
\usepackage{amssymb}
\title{Online Matching in Geometric Random Graphs}

\author{%
   Flore Sentenac \\
  CREST, ENSAE \\
  Palaiseau, France\\
  \texttt{flore.sentenac@ensae.fr} \\
 \And
   Nathan Noiry \\
 S2A\\
  Telecom Paris, France \\
 \texttt{nathan.noiry@telecom-paris.fr} \\
   \AND
  Matthieu Lerasle \\
  CREST, ENSAE \\
  Palaiseau, France\\
   \texttt{matthieu.lerasle@ensae.fr} \\
  \And
  Laurent Ménard \\
  Modal’X, UMR CNRS 9023, UPL, Univ. Paris Nanterre\\ 
  F92000 Nanterre, France\\
 \texttt{ laurent.menard@normalesup.org} \\
   \And
   Vianney Perchet \\
  CREST, ENSAE Paris\\
  Palaiseau, France\\
  CRITEO AI Lab, Paris, France\\
  \texttt{vianney.perchet@normalesup.org} \\
}

\date{\today}

\usepackage{algorithm}
\usepackage{algpseudocode}
\usepackage{hyperref}
\usepackage{natbib}
\usepackage{physics}

\usepackage{tikz}
\usepackage{standalone}

\newcommand{\cU}{\mathcal{U}}

\newcommand{\ranking}{\textsc{ranking}}

\newcommand{\closest}{\textsc{closest}}

\newcommand{\leftfirst}{\textsc{small-first}}

\newcommand{\leftfirstgenerative}{\textsc{small-first-generative}}

\newcommand{\graphrounding}{\textsc{graph-rounding}}

\usepackage{wrapfig}

\usepackage{cleveref}

\begin{document}




\RUNTITLE{ Matching in Geometric  Graphs}

\TITLE{Online Matching in Geometric Random Graphs}

\ARTICLEAUTHORS{%
\AUTHOR{Flore Sentenac}
\AFF{CREST, ENSAE, Paris, France \EMAIL{flore.sentenac@gmail.com}}
\AUTHOR{Nathan Noiry }
\AFF{S2A, Telecom Paris, France, \EMAIL{nathan.noiry@telecom-paris.fr}}
\AUTHOR{Matthieu Lerasle  }
\AFF{CREST, ENSAE , Paris, France \EMAIL{matthieu.lerasle@ensae.fr}}
\AUTHOR{Laurent Ménard }
\AFF{ Modal’X, UMR CNRS 9023, UPL, Univ. Paris Nanterre , France \EMAIL{laurent.menard@normalesup.org}}
\AUTHOR{   Vianney Perchet  }
\AFF{ CREST, ENSAE, CRITEO AI Lab Paris , France \EMAIL{vianney.perchet@normalesup.org}}
} 

\ABSTRACT{We investigate online maximum cardinality matching, a central problem in ad allocation. In this problem, users are revealed sequentially, and each new user can be paired with any previously unmatched campaign that it is compatible with. Despite the limited theoretical guarantees, the greedy algorithm, which matches incoming users with any available campaign, exhibits outstanding performance in practice. Some theoretical support for this practical success has been established in specific classes of graphs, where the connections between different vertices lack strong correlations – an assumption not always valid in real-world situations. To bridge this gap, we focus on the following model: both users and campaigns are represented as points uniformly distributed in the interval $[0,1]$, and a user is eligible to be paired with a campaign if they are "similar enough," meaning the distance between their respective points is less than $c/n$, where $c>0$ is a model parameter. As a benchmark, we determine the size of the optimal offline matching in these bipartite one-dimensional random geometric graphs. We achieve this by introducing an algorithm that constructs a maximum matching and analyzing it. We then turn to the online setting and investigate the number of matches made by the online algorithm \closest, which pairs incoming points with their nearest available neighbors in a greedy manner. We demonstrate that the algorithm's performance can be compared to its fluid limit, which is completely characterized as the solution of a specific partial differential equation (PDE). From this PDE solution, we can compute the competitive ratio of \closest, and our computations reveal that it remains significantly better than its worst-case guarantee. This model turns out to be closely related to the online minimum cost matching problem, and we can extend the results obtained here to refine certain findings in that area of research. Specifically, we determine the exact asymptotic cost of \closest\ in the small excess regime, providing a more accurate estimate than the previously known loose upper bound.
  }


\maketitle

%


%
%
%
%






\section{Introduction}

\hfill \break

Online maximum cardinality matching\footnote{This setting is usually referred to as ``online matching" in the literature. We will use online maximum cardinality matching to avoid any confusion with online metric matching, the second setting considered in the paper.} is motivated, among others, by its application to ad allocation on the internet. 
Advertising platforms handle multiple companies paying for relevant ad space,  and their goal is to maximize the number of valid ad-user allocations while adhering to budget constraints.

In this dynamic setting, users generate web pages sequentially, and ads must be allocated immediately to avoid forfeiting available ad slots. This dynamic allocation process can be represented as an online maximum cardinality matching problem.

Consider a bipartite graph with two sets of vertices: one set representing companies and the other representing users. In this graph, edges represent compatibility between companies and users. In the language of online matching, the companies are the ``offline'' vertices, meaning they are present from the start. The users, on the other hand, are revealed one by one over time. Each time a new user appears, the connections (edges) between this user and the companies are revealed. Based on those edges, the vertex can then be matched irrevocably to a previously unmatched neighbor. 
The excellent survey \cite{Mehta} details in length applications, results, and techniques of online maximum cardinality matching.

A first line of work studied online maximum cardinality matching in the adversarial framework, where the algorithm is evaluated on the worst possible instance and vertex arrival order. 
It is folklore that greedy algorithms, which match incoming vertices to any available neighbor have a competitive ratio of $1/2$ in the worst case. 
However, they achieve $1- 1/e$ as soon as the incoming vertices arrive in Random Order \citep{GoelMehta}. 
The \ranking\ algorithm is the worst-case optimal, it achieves at least $1- 1/e$ on any instance \citep{ranking,rankingprimaldualanalysis,rankingmadesimple}, and also has a higher competitive ratio in the Random Order setting \citep{rankingrandomorder}. Beyond this worse-case setting, the known i.i.d.\ model assumes there exists a probability distribution over types of vertices, from which the incoming vertex is drawn i.i.d. at every iteration. 
With the knowledge of that distribution, algorithms with much better competitive ratios than \ranking\ were designed \citep{Manshadi,JailletLu,brubach2019online,10.1145/3519935.3520046}, the best one to date achieving a competitive ratio of $0.711$.

This known i.i.d.\ model fits some situations and is certainly interesting, but still very general. The algorithms designed are tailored to the worst known i.i.d. model and fail to handle additional graph knowledge. Moreover, as the guarantee is given for the worst possible input distribution, it does not always reflect the average performance of those algorithms. It has been highlighted in \cite{borodin2018experimental} that on many average-case and practical input families, simple greedy strategies outperform or perform comparably to state-of-the-art algorithms designed for that known i.i.d. setting. They thus call for the formulation of new stochastic input models that better match practical inputs for certain
application domains, e.g., online advertising.

As a consequence, another line of work considers standard online algorithms on some classes of random graphs, representing situations where some properties of the underlying graph are known. The seminal example would be online matching in Erdos-Renyi graphs \citep{MastinJaillet}, or more generally in the configuration model that specifies a law on the degrees of the vertices  \citep{Noiry2021OnlineMI,aamand2022optimal}. The idea behind the latter is that one can ``estimate'' the typical popularity of a campaign (say, some of them target a large number of users while others are more selective). In those instances, greedy strategies can be precisely analyzed. For instance, in \citep{MastinJaillet}, they show that the competitive ratio of greedy is larger than $0.837$, which is much higher than the worst-case guarantee.

Unfortunately, these approaches fail to model correlations between edges. Campaigns that are ``similar'' tend to target the same users, and vice-versa (for instance, luxury products will target users with high incomes, while baby products target families). A possible approach to model these correlations is through space embedding techniques:   ads and users are represented by points in an Euclidian space, typically feature vectors, and an edge is present between two vertices if the points are close enough. 

We shall in the following introduce and analyze the online maximum cardinality matching problem for space-embedded graphs, called geometric graphs. As it is already challenging and interesting,  we shall focus on the one-dimensional geometric graph.\\

Although it is not the primary focus of this paper, this model is strongly related to a prevalent stochastic model in minimum cost metric matching \citep{https://doi.org/10.48550/arxiv.1904.09284,https://doi.org/10.48550/arxiv.2104.03219,balkanski2022power}.

In minimum cost metric matching, a set of $n$ online servers are embedded in a metric space. At every iteration, a request arrives and has to be matched to one of the free servers. The cost of the match is the distance between the request and the server it is matched to. The goal is to minimize the total cost of the matching, with the constraint that no request can be ignored if a server is available upon arrival (otherwise the algorithm ignoring all requests would trivially get a cost of zero). One notable application is ride-hailing, where the servers are drivers, the requests riders, and the cost is the distance the rider has to drive before getting to the rider.

Due to the spatial nature of the problem, random geometric graphs are regularly chosen as a stochastic framework for that problem. The \closest\ algorithm, which consists in matching greedily the incoming vertex to its closest available neighbor, has received some attention. A first work, \citep{https://doi.org/10.48550/arxiv.2104.03219}, focuses on metric matching for a single set of $n$ points drawn i.i.d. on the line. The authors show that the greedy algorithm that successively matches the two closest points together, has a competitive ratio of $O(\log(n))$. Later, back in the bipartite setting, some works focused on the so-called excess supply setting, where $n$ offline points are drawn i.i.d. on the line and $n\tau$, with $\tau<1$, vertices arrive sequentially. In that setting, the cost of \closest\ was first shown to be $O\left(\log(n)^3\right)$  \citep{https://doi.org/10.48550/arxiv.2104.03219}. That result was later refined to $O(\frac{1}{1-\tau})$ in \citep{balkanski2022power}. In that work, they also demonstrate a constant competitive ratio for \closest\ when there is an equal number of vertices on both sides, as well as under a semi-random model, where the incoming vertices are picked adversarially.

Extending the techniques developed for the maximum cardinality matching case, we can precise the result in the case where there is some excess on the offline side and give the exact asymptotic cost of \closest\ under that setting.\\

Note that, although related, the two settings we are about to study are quite different: in metric matching, the incoming vertex is necessarily matched at every iteration, it is not the case in maximum cardinality matching. The cost function is also completely different: in the online maximum cardinality matching setting, any valid match generates the same reward, which is not the case in the online metric matching setting. Thus no algorithm tailored for one of those problems generalizes easily to the other, which implies that the works on both topics are relatively independent in the literature. 

\subsection{Model and Contributions} 

\hfill \break

 A one-dimensional bipartite geometric random graph $\mathbb{G}[n,c]$ has two sets of $n$ vertices, $\mathcal{X} \text{ and }\mathcal{Y}$, drawn independently and uniformly in $[0,1]$:
\[
\mathcal{X}= (x_i)_{i\in [n]}\stackrel{\text{i.i.d.}}{\sim} \cU[0,1] \quad \text{ and } \quad \mathcal{Y}= (y_i)_{i\in [n]}\stackrel{\text{i.i.d.}}{\sim} \cU[0,1].
\]
Moreover, there is an edge between $x_i\in \mathcal{X}$ and $y_j \in \mathcal{Y}$ iff they are close enough, i.e.,
\[
\left|x_i-y_j\right|<\frac{c}{n}.
\]
If there is an edge between $x_i\in \mathcal{X}$ and $y_j \in \mathcal{Y}$, we say $x_i$ and $y_j$  are neighbors.

Note that under this parametrization, the expected degree of a vertex is of order $c$ (neglecting the boundary effects). In particular, the graph remains sparse and the online matching problem is not trivial.

We denote $\textsc{G}[U,V,n,c]$
the one-dimensional bipartite geometric graph with pre-specified sets of vertices $U=\left(u_i\right)_{i \in [|U|]}$ and $V=\left(v_j\right)_{i \in [|V|]}$ and edges generated by the same process as in $\mathbb{G}[n,c]$. Note that $\mathbb{G}[n,c]$ is random while $\textsc{G}[U,V,n,c]$ is not.\\

In online \textit{maximum cardinality} matching, the set of vertices $U$ is known from the beginning, while the vertices in $V$ are revealed sequentially. Upon arrival, a vertex $v_j \in V$ can be matched irrevocably to a previously unmatched neighbor in $\mathcal{U}$. The goal is to maximize the total number of matched vertices. 

For an algorithm $\textsc{ALG}$  and a graph $\textsc{G}[U,V,n,c]$, for any $t\leq |V|$ we denote $\mu_\textsc{ALG}\left(\textsc{G}[U,V,n,c]\right)(t)$ the cardinality of the matching generated by  algorithm $\textsc{ALG}$ in graph $\textsc{G}[U,V,n,c]$ after the arrival of the  $t$ first online vertices $v_1, \ldots, v_t$. We use the shorthand  $\mu_\textsc{ALG}\left(\textsc{G}[U,V,n,c]\right)(|V|)=\mu_\textsc{ALG}\left(\textsc{G}[U,V,n,c]\right)$. We also denote $\mu_*\left(\textsc{G}[U,V,n,c]\right)$ the size of a maximum matching in graph $\textsc{G}[U,V,n,c]$. The competitive ratio of an algorithm $\textsc{ALG}$ in $\textsc{G}[U,V,n,c]$ is defined as:
\[
\frac{\mu_\textsc{ALG}\left(\textsc{G}[U,V,n,c]\right)}{\mu_*\left(\textsc{G}[U,V,n,c]\right)}.
\]

 A \textit{weighted} one-dimensional bipartite geometric random graph $\mathbb{G}_\rho[n]$ also has two sets of $n$ vertices, $\mathcal{X} \text{ and }\mathcal{Y}$, drawn independently and uniformly in $[0,1]$. Unlike in the previous model, there is an edge between any two vertices $x_i \in \mathcal{X}$ and $y_j \in \mathcal{Y}$. This edge is weighted, the weight of the edge is denoted $w_{(x_i,y_j)}$ and is equal to the distance between the two endpoints:
 \[
 w_{(x_i,y_j)}=|x_i-y_j|.
 \]
As previously, we denote $\textsc{G}_\rho[U,V]$
the weighted geometric graph with pre-specified sets of vertices $U$ and $V$ and edges generated by the same process as in $\mathbb{G}_\rho[n]$.\\

In online metric matching, as in online maximum cardinality matching, the set of vertices $U$ is known from the beginning, while the vertices in $V$ are revealed sequentially. Unlike before, upon arrival, a vertex $v_j\in V$ \textit{has to} (instead of can) be matched to a previously unmatched vertex in $U$. This match has a cost of $w_{(u_i,v_j)}$, and the goal is to minimize the total cost generated by the matching process. Note that the total cost equals the sum of the weights of the edges selected in the matching.

For an algorithm $\textsc{ALG}$  and a graph $\textsc{G}_\rho[U,V]$, for any $t\leq |V|$ we denote $\kappa_\textsc{ALG}\left(\textsc{G}_\rho[U,V]\right)(t)$ the cost of the matching generated by algorithm $\textsc{ALG}$ in graph $\textsc{G}_\rho[U,V]$ after the arrival of $t$ online vertices.\\

Our contribution is threefold. First, we design an algorithm constructing a maximum matching in any one-dimensional geometric graph $G[U,V,n,c]$. We also show that the fraction of matched vertices in a maximum matching of $\mathbb{G}[n,c]$  converges w.h.p. to $c/(c+1/2)$. This bound is obtained through the study of the algorithm, via the construction of a potential function which is then treated as a random walk. More precisely, we shall prove the following theorem.\\

\textsc{Theorem \ref{thm:sizemaxmath1D}} \textbf{(informal)}.
\textit{ With probability at least $1-O\left(\frac{1}{n}\right)$,
\[
\mu_*\left(\mathbb{G}\left[n,c\right]\right)=\frac{c}{c+\frac{1}{2}}n + O\left(\sqrt{n \ln(n)}\right).
\]}

\begin{figure}[htb]
\begin{minipage}[]{0.1\textwidth}
\hfill \break
  \end{minipage}
  \begin{minipage}[]{0.4\textwidth}
    \includegraphics[width=6cm]{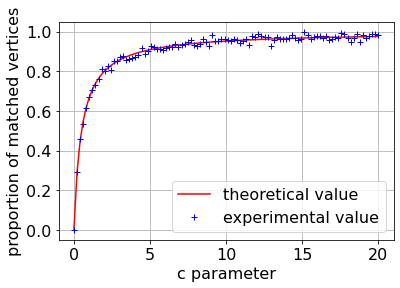}
  \end{minipage}
  \begin{minipage}[]{0.4\textwidth}
\caption{The asymptotic normalized size of a maximum matching in $\mathbb{G}\left[n,c\right]$, $\frac{c}{c+\frac{1}{2}}$, is plotted in red, as a function of the parameter $c$. Several simulations (blue crosses) for different values of $c$ and for $n=100$ vertices illustrate that this limit is reached rapidly.}
  \end{minipage}
\label{FIG:OfflineOpti}
\end{figure}

This result is illustrated in Figure \ref{FIG:OfflineOpti} which shows both the theoretical asymptotic value and the normalized size of the maximum matching in $\mathbb{G}\left[n,c\right]$, $\mu_*\left(\mathbb{G}\left[n,c\right]\right)/n$,  in several realizations of random graphs for a variety of parameters $c$. \\

We then study the size of the matching constructed by the online algorithm \closest, which matches any incoming vertex to its closest available neighbor, in graph $\mathbb{G}[n,c]$ after the arrival of $\tau n$ vertices. We show that this quantity normalized by the number of offline vertices, $\mu_\closest\left(\mathbb{G}[n,c]\right)(\tau n)/n$, converges in probability to the solution of an explicit PDE. We do so by exhibiting tractable quantities that can be approximated via the Differential Equation Method \citep{Wormald}. More precisely, we shall prove the following Theorem.\\

\textsc{Theorem \ref{thm:online0/1}} \textbf{(informal)}.
\textit{For any $0\leq \tau\leq1$, the normalized size of the matching constructed by algorithm \closest\ in $\mathbb{G}[n,c]$ after the arrival of $\tau n$\footnote{Throughout the paper we implicitly assume $\tau n$ is an integer.} online vertices, converges in probability:
\[
\frac{1}{n}\mu_\closest\left(\mathbb{G}[n,c]\right)(\tau n)\xrightarrow[n\rightarrow + \infty]{\mathbb{P}}1-\int_0^{+\infty}f(x,\tau)dx
\]
where $f$ is the solution of some explicit PDE, described later in Equation \eqref{PDE}.}\\

\begin{figure}[htb]
\begin{minipage}{0.6\textwidth}
\includegraphics[scale=0.33]{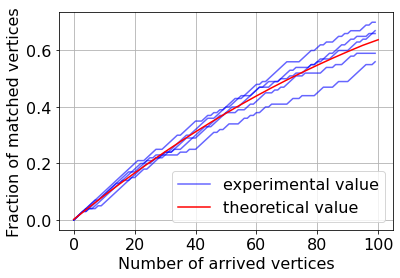}
\hspace{0.14cm}\includegraphics[scale=0.33,trim=0.9cm 0 0 0, clip]{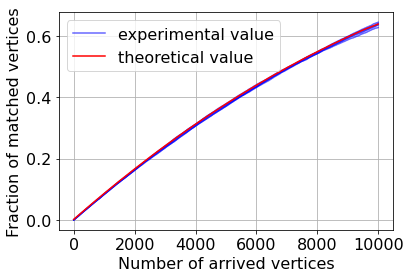}

    \includegraphics[scale=0.33]{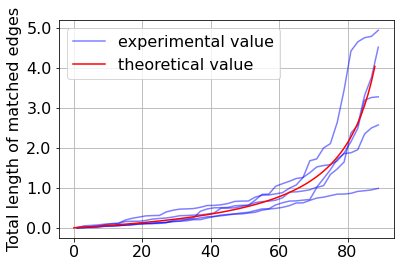}
    \hspace{0.03cm}
        \includegraphics[scale=0.33,trim=0.9cm 0 0 0, clip]{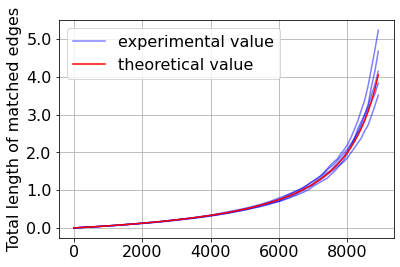}
      
\end{minipage}
\begin{minipage}{0.39\textwidth}
    \caption{First row, from left to right: asymptotic theoretical (red line) vs.\ experimental (blue lines) sizes of the online matching in one-dimensional bipartite geometric random graph $\mathbb{G}[n,c]$ ($c=1$) as a function of the number of arrived vertices, for $n=100$ and $n=10.000$.    
    Second row, from left to right: asymptotic theoretical (red line) vs.\ experimental (blue lines) length of the online matching in the weighted one-dimensional geometric random graph as a function of the number of arrived vertices, for $n=100$ and $n=10.000$.  }
  \label{FIG:Online}
\end{minipage}
\end{figure}

\begin{figure}[htb]
\begin{minipage}[]{0.1\textwidth}
\hfill \break
  \end{minipage}
  \begin{minipage}[]{0.4\textwidth}
    \includegraphics[width=6cm]{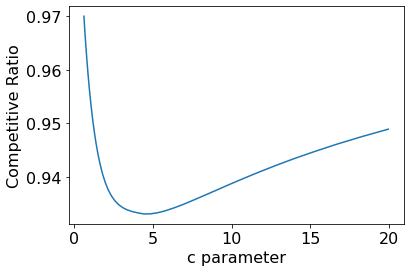}
  \end{minipage}
  \begin{minipage}[]{0.4\textwidth}
\caption{Asymptotic value of the competitive ratio of \closest\ in $\mathbb{G}[n,c]$ for several values of $c$. }
\label{FIG:CR}
  \end{minipage}
\end{figure}

The difference between the theoretical and the actual sizes of the matchings (as a function of the number of online vertices observed) for different values of $n$ is illustrated in Figure \ref{FIG:Online}. Those two results combined give the asymptotic value of the competitive ratio of the \closest\  algorithm in $\mathbb{G}[n,c]$, illustrated in Figure \ref{FIG:CR}.

We also extend the technique developed to the online metric matching problem, and prove the following theorem, illustrated in Figure \ref{FIG:Online}.\\

\textsc{Theorem \ref{thm:onlinelength}} \textbf{(informal)}.
\textit{For any $0\leq \tau<1$, the cost of the matching constructed by algorithm \closest\ in $\mathbb{G}_\rho[n]$ after the arrival of $\tau n$ online vertices, converges in probability:
\[
\kappa_\textsc{\closest}\left(\mathbb{G}_\rho[n]\right)(\tau n)\xrightarrow[n\rightarrow + \infty]{\mathbb{P}}\frac{1}{2}\left[\frac{1}{1-\tau}-1\right].
\]}

\paragraph{Organization of the paper} Section \ref{SE:Off} is dedicated to the offline case. We give the asymptotic formula of the normalized size of a maximum matching in $\mathbb{G}[n,c]$. Our approach is algorithmic: we provide a way to construct a maximum matching, and we analyze the latter by carefully studying some random walk. Section \ref{SE:Online} focuses on the online case. 
We focus on a simple greedy algorithm, \closest, that matches any incoming vertex to the closest available one (as a greedy procedure would do) and we characterize the size/length of the matching it creates by studying its fluid limit. 
We shall prove it satisfies some PDE.

\section{Maximum Matching in one-dimensional Random Geometric Graphs}\label{SE:Off}

\hfill \break

This section is dedicated to the offline case, where the whole underlying bipartite one-dimensional geometric graph is known from the beginning. Consider algorithm \leftfirst\, which iteratively matches the unmatched vertices with the smallest coordinates. The following proposition states this algorithm produces a maximum matching.

\begin{proposition}\label{prop:leftfirtoptimal}
Algorithm \leftfirst\  returns a maximum matching in any bipartite one-dimensional geometric graph $\textsc{G}[U,V,n,c]$.
\end{proposition}

\textit{Proof}: First, there exists a maximum matching in which no two edges cross. Indeed, assume pairs $(x,y')$ and $(x',y)$ belong to the matching with $x<x'$ and $y<y'$. Since the graph is a geometric graph, there are also the edges $(x,y)$ and $(x',y')$, so the matches can be uncrossed without modifying the size of the matching. 

Among those maximum matchings with no crossing edges, there is one in which each vertex is matched to its leftmost neighbor not already matched to another vertex. Again, if this is not the case, the matching can be modified without modifying its size.

This maximum matching with no crossing edges and where every vertex is matched to its leftmost available neighbor is the one returned by the algorithm \leftfirst.
\hfill \(\Box\)\\

Using the optimality of \leftfirst, we can now prove the first main result.
\begin{theorem}\label{thm:sizemaxmath1D}The expected fraction of vertices included in a maximum matching of  $\mathbb{G}[n,c]$ converges:
\[
\lim_{n\rightarrow \infty} \frac{\mathbb{E}\left[ \mu_*\left(\mathbb{G}[n,c]\right)\right]}{n}=\frac{c}{c+\frac{1}{2}},
\]
and, with probability at least $1-O\left(\frac{1}{n}\right)$:
\[
\mu_*\left(\mathbb{G}[n,c]\right)=\frac{c}{c+\frac{1}{2}}n + O\left(\sqrt{n \ln(n)}\right).
\]
\end{theorem}

\textit{Proof}: The proof is decomposed into two steps. In the first step, we show that the size of the maximum matching in $\mathbb{G}[n,c]$ is related to the size of the maximum matching in a geometric graph with sets of vertices generated by Poisson point processes. In the second step, the asymptotic size of the maximum matching in those graphs is derived through the study of a random walk.

\paragraph{Step 1, connection with Poisson point processes.}

Let $\Phi^n$ be an independent homogeneous Poisson point process on the segment $[0,1]$ of intensity $n$. The definition and some standard properties of Poisson Point Processes (PPP) are reported in  \cref{app:poisson} for the sake of completeness. Let $\mathcal{U} \sim \Phi^n$ and $\mathcal{V} \sim \Phi^n$ be two independent PPP. The two vertex sets of $\text{G}[\mathcal{U},\mathcal{V},n,c]$ are points uniformly distributed in $[0,1]$, hence this graph is almost a one-dimensional random geometric graph, except that the cardinality of the vertex sets is not necessarily equal to $n$.

From $\mathcal{U}$, define a new set $\mathcal{U}_n$ of $n$ vertices as follows.
If $|\mathcal{U}|>n$, delete uniformly at random $|\mathcal{U}|-n$ points from $\mathcal{U}$.  
If $n>|\mathcal{U}|$, add $n-|\mathcal{U}|$ points independently and uniformly distributed in $[0,1]$ to $\mathcal{U}$. The set $\mathcal{V}_n$ is constructed from $\mathcal{V}$ similarly. Now, $\text{G}[\mathcal{U}_n,\mathcal{V}_n,n,c]$ and $\mathbb{G}[n,c]$ have the same law. Additionally, the following lemma holds.

\begin{lemma}\label{lem:pppvsunifoff}Under the above notations,  for any $n\geq 10$, the following holds with probability at least $1-\frac{4}{n}$:
\[
|\mu_*\left(\textsc{G}[\mathcal{U}_n,\mathcal{V}_n,n,c]\right)-\mu_*\left(\textsc{G}[\mathcal{U},\mathcal{V},n,c]\right)|\leq 4\sqrt{n\ln n}, 
\]
and $\textsc{G}[\mathcal{U}_n,\mathcal{V}_n,n,c]$ and $\mathbb{G}[n,c]$ have the same law.
\end{lemma}
\textit{Proof:} By \cref{lem:concentratinPPP} (\cref{app:poisson}), for any $n\geq 10$: 

\begin{equation}
\label{eq:numpoints}
  \mathbb{P}\left\{\big||\mathcal{U}|-n\big|\geq 2\sqrt{n\ln n} \right\}\leq \frac{2}{n}.  
\end{equation}
and the same holds for $|\mathcal{V}|$.  Moreover, the transformation from $\mathcal{U}$ to $\mathcal{U}_n$ and from $\mathcal{V}$ to $\mathcal{V}_n$ affects the size of the maximum matching at most by the number of added and removed points. \hfill \(\Box\)
 \endproof

\paragraph{Step 2, estimating $\mu_*\left(\textsc{G}[\mathcal{U},\mathcal{V},n,c]\right)$.}
A possible way to generate 
set  $\mathcal{U}$ from $\Phi^n$ is through a renewal process with exponential holding times. Precisely, let $\left\{F_{k}\right\}_{k \in \mathbb{N}}$ be a sequence of i.i.d. exponential random variables of parameter $n$. Ensemble $\mathcal{U}$ is defined as:
\[
\mathcal{U} = \left\{u_{k}=\sum_{i=1}^{k} F_{i}\ \text{for}\ k \text{ s.t. } \sum_{i=1}^{k} F_{i}<1\right\}.
\]
The same holds from $\mathcal{V}$. 

To compute the size of the maximum matching in a given graph, we introduce a modified version of Algorithm \leftfirst, that generates the graph together with the matching constructed by algorithm \leftfirst. Algorithm \leftfirstgenerative\ (Algorithm \ref{alg:leftfirstgenerative}) proceeds as follows. The positions of the first points in $\mathcal{U}$ and $\mathcal{V}$ are drawn independently from two exponential distributions of parameter $n$. At iteration $t$, denote $u_{i(t)}$ and $v_{j(t)}$ the position of the last generated points in $\mathcal{U}$ and $\mathcal{V}$ respectively. Define the potential function
\[
\psi(t):= n(u_{i(t)}-v_{j(t)}).
\]
Until all the points on one side have been drawn, the following operations are iteratively performed:
\begin{itemize}
    \item if $|\psi(t)|<c$, edge $(u_{i(t)}, v_{j(t)})$ is added to the matching and the next points on both sides of the graph are generated,
    \item if $\psi(t)>c$, the next point in $\mathcal{V}$ is generated,
    \item if $\psi(t)<-c$, the next point in $\mathcal{U}$ is generated.
\end{itemize}

\begin{algorithm}[htb]
\begin{algorithmic}

\State Draw $u_1\sim \text{Exp}(n)$ and $v_1\sim \text{Exp}(n)$
\State Initialize $i(1)\leftarrow1$, $j(1)\leftarrow1$ and $m\gets \emptyset$ 
\State Define for $t=1,\ldots$, $\psi(t)\leftarrow n(u_{i(t)}-v_{j(t)})$
	\While{$u_{i(t)}<1$ and $v_{j(t)}<1$}
	\If{$|\psi(t)|<c$}
	\State $m\gets m \cup (u_{i(t)}, v_{j(t)})$
	\State$i(t+1)\leftarrow i(t)+1$ and $j(t+1)\leftarrow j(t)+1$
	\State Draw $u_{i(t+1)}-u_{i(t)}\sim\text{Exp}(n)$ and $v_{j(t+1)}-v_{j(t)}\sim\text{Exp}(n)$
 \EndIf
	\If{$\psi(t)>c$}
	\State $i(t+1)\leftarrow i(t)$ and $j(t+1)\leftarrow j(t)+1$
	\State Draw $v_{j(t+1)}-v_{j(t)}\sim\text{Exp}(n)$
 \EndIf
	\If{$\psi(t)<-c$}
	\State $i(t+1)\leftarrow i(t)+1$ and $j(t+1)\leftarrow j(t)$
	\State Draw $u_{i(t+1)}-u_{i(t)}\sim\text{Exp}(n)$
  \EndIf
\EndWhile

\caption{\leftfirstgenerative}
\label{alg:leftfirstgenerative}
\end{algorithmic}
\end{algorithm}

The potential function $\psi(t)$ is a Markov chain with the following evolution:
\[
\psi(t+1)-\psi(t) \mathop{\sim}\limits^{\mathcal{L}}\left\{
    \begin{array}{ll}
        \text{Laplace}(0,\frac{1}{n})\mbox{ if } |\psi(t)|\leq c\\
        \text{Exp}(n) \mbox{ if } \psi(t)\leq -c \\
        -\text{Exp}(n) \text{ if } \psi(t)\geq c
    \end{array}
    \right.
\]
with $\text{Laplace}(0,\frac{1}{n})$ the distribution of the difference between two exponential variables of parameter $n$. Let us now study this random walk.

\begin{lemma}\label{lem:stationnarydistrib}
The  Markov chain $\psi(t)$ described above admits the following stationary distribution:
\[
\pi(x)= \left\{
    \begin{array}{ll}
        \frac{1}{2c+2} \mbox{ if } |x|\leq c\\
        \frac{e^{x+c}}{2c+2}\mbox{ if } x\leq -c \\
        \frac{e^{-(x-c)}}{2c+2} \text{ if } x\geq c.
    \end{array}
    \right.
\]
\end{lemma}
\textit{Proof}:
Let us denote by $\Pi$ the transition kernel of random walk $\psi$; then for any
$x \in [-c,c]$:
\begin{align*}
(2c+2)\int_{-\infty}^{+\infty}\Pi(x,y)\pi(y)dy=& \int_{-\infty}^{-c}e^{-(x-y)}e^{y+c}dy+\int_{c}^{+\infty}e^{-(y-x)}e^{-y+c}dy\\
&+\int_{-c}^{x}\frac{1}{2}e^{-(x-y)}dy+\int_{x}^{c}\frac{1}{2}e^{x-y}dy\\
=&1.
\end{align*}

Similarly, for any $x \leq-c$:
\begin{align*}
(2c+2)\int_{-\infty}^{+\infty}\Pi(x,y)\pi(y)dy=& \int_{-\infty}^{x}e^{-(x-y)}e^{y+c}dy+\int_{c}^{+\infty}e^{-(y-x)}e^{-y+c}dy\\
&+\int_{-c}^{c}\frac{1}{2}e^{x-y}dy\\
=&e^{x+c}.
\end{align*}
By symmetry, the computation also holds $\forall x \geq c$. \hfill \(\Box\)

Let  $\tau_n$ be the random number of iterations in Algorithm \leftfirstgenerative\ run with input parameter $n$. Note that \leftfirstgenerative\ does not generate the full graph $\textsc{G}[\mathcal{U},\mathcal{V},n,c]$, as one side of the graph is not fully generated. Still, algorithm \leftfirstgenerative\ constructs exactly the matching that would be constructed on graph $\textsc{G}[\mathcal{U},\mathcal{V},n,c]$ by algorithm \leftfirst, and thus, by   \cref{prop:leftfirtoptimal}, generates  a maximum matching in the partially generated graph $\textsc{G}[\mathcal{U},\mathcal{V},n,c]$. Denote $\mu_{\tau_n}=\frac{1}{\tau_n}\sum_{t=1}^{\tau_n}\mathds{1}_{\{|\psi(t)|\leq c\}}=\frac{1}{\tau_n}\mu_*\left(\textsc{G}[\mathcal{U},\mathcal{V},n,c]\right)$, the size of the constructed matching normalized by $\tau_n$. The next step of the proof consists in showing that this quantity concentrates. In \cref{sec:proofoffline}, we show:
\begin{align}
\mathbb{P}\left(\bigg|\mu_{\tau_n}-\frac{2c}{2c+2}\bigg|\geq \sqrt{\frac{12\ln(n)}{n}}\right)
\leq  \frac{16(c+2)}{n}.\label{eq:mun}
\end{align}
 Now, denote $\gamma_n$ the total number of vertices generated by algorithm \leftfirstgenerative. When $|\psi(t)|\leq c$, $2$ vertices are generated, and when
 $|\psi(t)|> c$, $1$ vertex is generated, thus we have:
\begin{align*}
\gamma_n= \tau_n+\sum_{t=1}^{\tau_n}\mathds{1}_{\{|\psi(t)|\leq c\}}= \tau_n \left(1+\mu_{\tau_n}\right).
\end{align*}
 This gives that the size of the generated matching is:
 \begin{equation}\label{eq:finaleq}
 \mu_*\left(\textsc{G}[\mathcal{U},\mathcal{V},n,c]\right)=\tau_n \mu_{\tau_n}= \gamma_n \frac{\mu_{\tau_n}}{1+\mu_{\tau_n}}.
\end{equation}

An estimate of $\mu_{\tau_n}$ has already been obtained, it remains to estimate $\gamma_n$.  In  \cref{sec:proofoffline}, we show:
\begin{align}
\mathbb{P}\left(\bigg|\gamma_n-2n\bigg|\geq (c+7)\sqrt{n\ln(n)}\right)
\leq  \frac{14}{n}.\label{eq:gamman}
\end{align}  
Combining with \cref{eq:mun,eq:finaleq}, we obtain that with probability $1-O(\frac{1}{n})$,
\begin{align*}
    \mu_*\left(\textsc{G}[\mathcal{U},\mathcal{V},n,c]\right)=&\left(2n+O\left(\sqrt{n \ln(n)}\right)\right)\frac{\frac{2c}{2c+2}+O\left(\sqrt{\frac{\ln(n)}{n}}\right)}{1+\frac{2c}{2c+2}+O\left(\sqrt{\frac{\ln(n)}{n}}\right)}\\
    =&\frac{c}{c+\frac{1}{2}}n+O(\sqrt{n \ln(n)}).
\end{align*}

Combining with \cref{lem:pppvsunifoff} gives the high probability bound in the theorem. For the convergence of the expectation, as $\mu_*(\mathbb{G}[n,c])\leq n$ a.s., we have:
\[
\bigg|\frac{\mathbb{E}[\mu_*(\mathbb{G}[n,c])]}{n}-\frac{c}{c+\frac{1}{2}}\bigg|\leq O\left(\sqrt{\frac{ \ln(n)}{n}}\right)+O\left(\frac{1}{n}\right).
\]
\hfill \(\Box\) 

\section{Algorithm \closest}\label{SE:Online}

\hfill \break

In this section, we study the performances in $\mathbb{G}[n,c]$ of the online matching algorithm \closest, which matches the incoming vertex to its closest available neighbor if there is one. The following theorem states that $\mu_\closest(\mathbb{G}[n,c])(t)$, is closely related to the solution of an explicit PDE.


\begin{theorem}\label{thm:online0/1}
For any $0\leq \tau\leq 1$, the normalized size of the matching constructed by \closest\  in $\mathbb{G}[n,c]$ after the arrival of $\tau n$ vertices converges in probability:
\[
\frac{\mu_\closest(\mathbb{G}[n,c])(\tau n)}{n}\xrightarrow[n\rightarrow + \infty]{\mathbb{P}}1-\int_0^{+\infty}f(x,\tau)dx
\]
with $f(x,\tau)$ the solution of the following differential equation
\begin{align}\label{PDE}
\notag    \frac{\partial f(x,\tau)}{\partial \tau} =&-\min(x,2c)f(x,\tau)-\frac{1}{\int_{0}^{+\infty}f(x',\tau)dx'}\int_{0}^{+\infty}\min(x',2c)f(x',\tau)dx'f(x,\tau)\\
   &+\frac{1}{\int_{0}^{+\infty}f(x',\tau)dx'}\int_{0}^{x}\min(x',2c)f(x',\tau)f(x-x',\tau)dx' 
\end{align}
with the following initial conditions
$$
f(x,0)= e^{-x}.
$$

\end{theorem}

\textbf{Remark}: Function $f$ depends on $c$, we omit that dependency in the notations.

Turning to metric matching, the following theorem gives the asymptotic cost of algorithm \closest. 

\begin{theorem}\label{thm:onlinelength}
For any $0\leq \tau<1$, the cost of the matching constructed by algorithm \closest\ in $\mathbb{G}_\rho[n]$ after the arrival of $\tau n$ online vertices, converges in probability:
\[
\kappa_\textsc{\closest}\left(\mathbb{G}_\rho[n]\right)(\tau n)\xrightarrow[n\rightarrow + \infty]{\mathbb{P}}\frac{1}{2}\left[\frac{1}{1-\tau}-1\right].
\]

\end{theorem}
\textit{Proof structure}: 
The proofs of the two theorems are very similar and we prove them together, providing additional details for each subcase where it is needed. We first show that the score of \closest\ in $\mathbb{G}[n,c]$ (resp. $\mathbb{G}_\rho[n]$) is closely related to its score in a transformed graph $\textsc{G}^{\text{glued}}[\tilde{\mathcal{U}}_k,\mathcal{Y},n,c]$ (resp. $\textsc{G}_\rho^{\text{glued}}[\tilde{\mathcal{U}}_k,\mathcal{Y}]$), where the vertices of the offline side are generated through a Poisson point process and have their coordinates rounded to a discrete grid (Section \ref{sec:graph_rounding}), with $k$ the discretization parameter. We then show that the score of \closest\ on the modified graph is closely related to the solution of a PDE through the differential equation method \citep{Noiry}(Section \ref{sec:PDE}). \hfill \(\Box\)

\subsection{Graph rounding}\label{sec:graph_rounding}

\hfill \break

The \graphrounding\ procedure is introduced purely for the sake of the analysis. Algorithm \closest\ is easier to analyze on the graph transformed by the \graphrounding\ procedure for two key reasons. First, as will be detailed in the description of the procedure, the unit interval is replaced with the unit circle, which removes any potentially hard-to-analyze boundary effect. Second, the position of the offline points is rounded to a discrete grid, which will imply that some key quantities concentrate. 

The procedure involves several steps, and in \cref{prop:roundingproc}, we show that these steps together have a limited effect on the performance of \closest. In other words, the algorithm performs similarly on both the transformed and original graphs. Our main goal is to evaluate the performance of \closest\ in the original graph. Since it is easier to do this analysis on the modified graph, and \cref{prop:roundingproc} shows the performance is roughly the same, we will analyze \closest's performance on the modified graph in the next section.
 \\

\begin{figure}[htb]
\begin{center}

  \includestandalone[width=0.5\textwidth, angle=90]{tikz_folder/test}
  \caption{Graph Rounding}
  \label{fig:graphrounding}
  \end{center}
\end{figure}

Let us now describe the procedure. Take $\mathcal{X}$ and $\mathcal{Y}$ two sets of $n$ points drawn i.i.d.\ uniformly in $[0,1]$. The \graphrounding\ procedure, illustrated in Figure \ref{fig:graphrounding}, transforms graph $\textsc{G}[\mathcal{X}, \mathcal{Y},n,c]$ (resp. $\textsc{G}_\rho[\mathcal{X}, \mathcal{Y}]$) into a graph $\textsc{G}^{\text{glued}}[\tilde{\mathcal{U}}_k,\mathcal{Y},n,c]$ (resp.  $\textsc{G}_\rho^{\text{glued}}[\tilde{\mathcal{U}}_k,\mathcal{Y}]$) through the following steps:
\begin{itemize}
    \item \textbf{Poissonization step:} Take $n_0\sim \text{Poi}(n)$. If $n_0>n$, then let $(u_i)_{i=n+1}^{n_0}$ be $n_0-n$ points drawn uniformly and independently in $[0,1]$ and define $\mathcal{U}= \mathcal{X}\cup \{u_i\ |\ i \in [n+1;n_0] \}$. If $n_0<n$, then $\mathcal{U}$ is obtained be removing $n-n_0$ points selected uniformly at random from $\mathcal{X}$. \footnote{Note that this is the reverse of the procedure used in Section \ref{SE:Off}, where a set of points drawn from a PPP was transformed into a set of $n$ points drawn uniformly and independently in $[0,1]$ . }
    \item  \textbf{Rounding step:} Transform $\mathcal{U}$ in a new ensemble $ \mathcal{U}_k$ by rounding the coordinate of each point $u\in \mathcal{U}$ to $\frac{\lfloor unk\rfloor}{nk}$,
    \vspace{-0.2cm}
    \[
    \mathcal{U}_k:= \left\{\frac{\lfloor unk\rfloor}{nk}\ \bigg|\  u\in \mathcal{U} \right\}. 
   \]
    \item  \textbf{Discarding step:} For any $\ell \in [nk]$, if multiple vertices have their coordinates rounded to $\frac{\ell}{nk}$, a random vertex among those is selected, and all others are removed from the graph. This gives the final ensemble $\tilde{\mathcal{U}}_k$.
\item \textbf{Gluing step:} The interval $[0,1]$ is mapped to the unit circle of circumference one. Formally, the distance is replaced with the following one: $d(x,y)= \min(|x-y|,|x+1-y|)$.
We also add a vertex at coordinate $0$ to $\tilde{\mathcal{U}}_k$ if it is not already in it. 
\end{itemize}

Note that for any $\ell \in [1;nk-1]$, we have: 
\[
\mathbb{P}\left(\exists u \in \tilde{\mathcal{U}}_k \text{ s.t. } u=\frac{\ell}{nk}\right)=\underbrace{1-e^{-1/k}}_{:=p_k},
\]
and the events $\left(\{\exists u \in \tilde{\mathcal{U}}_k \text{ s.t. } u=\frac{\ell}{nk}\}\right)_{\ell \in [1;nk-1]}$ are independent of each other.

The following proposition states that the \graphrounding\ procedure does not impact too much the size of the matchings.
\begin{proposition}\label{prop:roundingproc}
For all $0\leq \tau \leq 1$, $n\geq 10$, with probability at least $1-\frac{8}{n}-\exp(-\frac{\sqrt{n}}{4c})-\exp(-\frac{2n}{k^2})$, we have:
\[
\left|\mu_\closest\left(\textsc{G}[\mathcal{X}, \mathcal{Y},n,c]\right)(\tau n)-\mu_\closest\left(\textsc{G}^{\text{glued}}[\tilde{\mathcal{U}}_k, \mathcal{Y},n,c]\right)(\tau n)\right|\leq 21\frac{n}{k}+6\sqrt{n\ln(n)}+4c.
\]

For any $0\leq \tau < 1$,  $n\geq 10$, $k\geq \frac{2}{\tau(1-\tau)}$ , with probability at least $1-\frac{16\alpha+4}{n^2}-\frac{8}{n}-\exp(-\frac{2n}{k^2})$, we have: 
\[
\left|\kappa_\textsc{\closest}\left(\textsc{G}_\rho[\mathcal{X}, \mathcal{Y}]\right)(\tau n)-\kappa_\textsc{\closest}\left(\textsc{G}^{\text{glued}}_\rho[\tilde{\mathcal{U}}_k , \mathcal{Y}]\right)( \tau n)\right|\leq84\left( \frac{1}{k}+\sqrt{\frac{\ln(n)}{n}}+\frac{\alpha\ln(n)}{n}\right)\left(\alpha\ln(n)\right)^2.
\]
where $\alpha = \frac{128}{(1-\tau)^2\tau^2}$.  
\end{proposition}

The proof of this technical proposition is postponed to Section \ref{sec:proofroundingproc}.

\subsection{Analysis of \closest\ on the modified graph}\label{sec:PDE}
\hfill \break

As announced in the previous section, we now analyze the performance of \closest\ in the transformed graphes, i.e. we analyze: \[\mu_\closest\left(\textsc{G}^{\text{glued}}[\tilde{\mathcal{U}}_k \mathcal{Y},n,c]\right)(\tau n)\text{ and }\kappa_\textsc{\closest}\left(\textsc{G}^{\text{glued}}_\rho[\tilde{\mathcal{U}}_k , \mathcal{Y}]\right)(\tau n).\] This will be done by tracking the number of gaps of a certain length between successive free vertices at every iteration. First we will show that we can compute the expected evolution of those quantities, then that at iteration $0$, those quantities concentrate. We will show in the next section how this implies that we can track them, which in turn implies that we can compute $\mu_\closest\left(\textsc{G}^{\text{glued}}[\tilde{\mathcal{U}}_k \mathcal{Y},n,c]\right)(\tau n)\text{ and }\kappa_\textsc{\closest}\left(\textsc{G}^{\text{glued}}_\rho[\tilde{\mathcal{U}}_k , \mathcal{Y}]\right)(\tau n).$

\subsubsection{Expected evolution and initial condition\\}
Denote $n_t$ be the number of free vertices at iteration $t$. At time $t$, let $u_t(i)$ be the coordinate of the $i^{\text{th}}$ free vertex, with the vertices enumerated according to their coordinates, and the convention $u_t(n_t+1)=u_t(1)$ and $u_t(n_t)=u_t(0)$ . 
 For $\ell \in [kn]$, define 
\[
F_{k,n}(\ell,t) :=\left|\left\{i\in [n_t] \ \text{s.t.}\ n\left(u_t(i+1)-u_t(i)\right)=\frac{\ell}{k} \right\}\right|, 
\]
which is the number of gaps of size $\frac{\ell}{nk}$ between two successive free vertices. First, not that, as the total length of the interval does not change, $\sum_{\ell \in[kn]}\frac{\ell}{kn}F_{k,n}(\ell,t)=1$.
Also note that at iteration $t$, the number of free vertices is equal to $\sum_{\ell \in[kn]}F_{k,n}(\ell,t)$, hence:
\begin{align}\label{eq:linkgapsmatch}
\frac{1}{n}\mu_\closest\left(\textsc{G}^{\text{glued}}[\tilde{\mathcal{U}}_k, \mathcal{Y},n,c]\right)(\tau n)=\frac{|\tilde{\mathcal{U}}_k|}{n}-\frac{1}{n}\sum_{\ell \in[kn]}F_{k,n}(\ell,\tau n),
\end{align}
i.e. from quantities $\left(F_{k,n}\left(\ell,t\right)\right)_{\ell}$ we can compute the size of the constructed matching at iteration $t$.  For maximum cardinality matching, the crux of the proof is therefore to show that quantities $\left(F_{k,n}\left(\ell,t\right)\right)_{\ell}$ can be tracked. For metric matching, there are a few additional technicalities detailed in the next section, but quantities $\left(F_{k,n}\left(\ell,t\right)\right)_{\ell}$  also play a central role.

We now compute the expected evolution of each $\left(F_{k,n}\left(\ell,t\right)\right)_{\ell}$ (\cref{lem:evollaw}) and show concentration of $\left(F_{k,n}\left(\ell,0\right)\right)_{\ell}$ (\cref{lem:initialconditions}). 
Define:
\[
M_{k,n}(\ell_-,\ell_+,t):= \left|\left\{i\in [n_t] \ \text{s.t.}\ \left(u_t(i)-u_t(i-1)\right)=\frac{\ell_-}{nk}\text{ and }\left(u_t(i+1)-u_t(i)\right)=\frac{\ell_+}{nk} \right\}\right|,\]
which is the number of times a gap of size $\frac{\ell_-}{kn}$ is followed by a gap of size $\frac{\ell_+}{kn}$ when enumerating the free vertices according to their coordinates.

Let $\mathcal{F}_t$ be the filtration associated with the values of the sizes of the gaps up to time $t$, $\left(F_{k,n}\left(\ell,t'\right)\right)_{\ell,t'\leq t }$. The following lemma describes how many times two gaps of a certain size follow each other at iteration $t$, in expectation conditionally on $\mathcal{F}_t$.

\begin{lemma}[Gaps repartition]\label{lem:gaps_repartition}
For all $t \in [n]$, for all $\ell_-,\ell_+ \in [kn]^2$,
\[
\mathbb{E}\left[M_{k,n}(\ell_-,\ell_+,t) \bigg | \mathcal{F}_t \right] = F_{k,n}(\ell_-,t)\frac{F_{k,n}(\ell_+,t)-\mathds{1}_{\{l_-=\ell_+\}}}{n_t-1}.
\]
\end{lemma}

\textit{Sketch of Proof}:  
This lemma is implied by a stronger result: conditionally on  $\mathcal{F}_t$, the gaps are ordered uniformly at random. This is proved by induction in Section \ref{SE:gaps_repartition}. In the case of metric matching, this result can be found in \cite{frieze1990greedy}, the proof here focuses on the case of maximum cardinality matching.
\hfill \(\Box \)

Note that the expression on the right-hand side is exactly the expectation obtained by drawing uniformly without replacements two successive gaps among all gaps at time $t$. Now, the following lemma gives the expected evolution of the gaps, still conditionally on $\mathcal{F}_t $. 

\begin{lemma}[Evolution law]\label{lem:evollaw}
For all $t \in [n]$, for all $\ell\in [kn]$, any $w_1>0$,
\begin{align*}
    \bigg|&\underbrace{\sum_{\ell'=1}^\ell\frac{\min(2c,\frac{\ell'}{k})}{n}\frac{F_{k,n}(\ell',t)F_{k,n}(\ell-\ell',t)}{n_t}-\left(\frac{\min(2c,\frac{\ell}{k})}{n}+\sum_{\ell'}\frac{\min(2c,\frac{\ell'}{k})}{n}\frac{F_{k,n}(\ell',t)}{n_t}\right)F_{k,n}(\ell,t)}_{:=A_{\ell,t}}\nonumber \\
       &\hspace{4cm}-\mathbb{E}\left[F_{k,n}(\ell,t+1)-F_{k,n}(\ell,t) \bigg | \mathcal{F}_t,n_t\geq w_1 n \right]\bigg|\leq \frac{6}{w_1n}.
\end{align*}

In the case of metric matching, the expression simplifies to:
\begin{align*}
    \bigg|&\sum_{\ell'=1}^\ell\frac{\ell'}{nk}\frac{F_{k,n}(\ell',t)F_{k,n}(\ell-\ell',t)}{n-t}-\left(\frac{\ell}{kn}+\frac{1}{(1-t)n}\right)F_{k,n}(\ell,t)\nonumber \\
       &\hspace{4cm}-\mathrm{E}\left[F_{k,n}(\ell,t+1)-F_{k,n}(\ell,t) \bigg | \mathcal{F}_t \right]\bigg|\leq \frac{6}{n-t}.
\end{align*}
\end{lemma}
\textbf{Remark}: The following proof holds with $c=+\infty$ in the case of metric matching.

\textit{Proof}:
If the $i^{th}$ free vertex at iteration $t$ is s.t. $u_t(i)-u_t(i-1)=\frac{\ell_-}{k}$ and  $u_t(i+1)-u_t(i)=\frac{\ell_+}{k}$, we use the shorthand 
\[
u_t(i) \in M_{k,n}(\ell_-,\ell_+,t).
\]
At iteration $t$, if no vertex is matched, $F_{k,n}(\ell,t+1)=F_{k,n}(\ell,t)$, for all $\ell \in [nk]$. Denote $i_t$ the index of the matched vertex if there is one. If $u_t(i_t) \in M_{k,n}(\ell_-,\ell_+,t)$, then
\begin{itemize}
    \item   $F_{k,n}(\ell_-,t+1)=F_{k,n}(\ell_-,t)-1$ and $F_{k,n}(\ell_+,t+1)=F_{k,n}(\ell_+,t)-1$,
    \item $F_{k,n}(\ell_+ + \ell_-,t+1)=F_{k,n}(\ell_++ \ell_-,t)+1$.
\end{itemize}

Moreover,
\begin{align*}
  \mathbb{P}\left(v_t \text{ matched to } u_t(i)| u_t(i) \in M_{k,n}\left(\ell_-,\ell_+,t\right), \mathcal{F}_t\right) =&\underbrace{\frac{\min\left(c,\frac{\ell_-}{2k}\right)+\min\left(c,\frac{\ell_+}{2k}\right)}{n}}_{ m(\ell_-,\ell_+)}.  
\end{align*}
Then,
\begin{align*}
  \mathbb{P}\left(v_t \text{ matched in } M_{k,n}\left(\ell_-,\ell_+,t\right)|\mathcal{F}_t\right) =&\sum_{i\in[n_t]}\mathbb{E}\left[\mathds{1}\left\{v_t \text{ matched to } u_t(i), u_t(i) \in M_{k,n}\left(\ell_-,\ell_+,t\right) \right\}|\mathcal{F}_t\right]\\
    =& m(\ell_-,\ell_+)\sum_{i\in[n_t]}\mathbb{E}\left[\mathds{1}\left\{ u_t(i) \in M_{k,n}\left(\ell_-,\ell_+,t\right) \right\}|\mathcal{F}_t\right] \\
  =& m(\ell_-,\ell_+)\mathbb{E}\left[M_{k,n}\left(\ell_-,\ell_+,t\right)|\mathcal{F}_t\right].  
\end{align*}

This implies the following chain of equality:
\begin{align*}
    \mathrm{E}\left[F_{k,n}(\ell,t+1)-F_{k,n}(\ell,t) \bigg | \mathcal{F}_t \right] =& -\underbrace{\left(\sum_{\ell'}m(\ell,\ell')\mathbb{E}\left[M_{k,n}(\ell,\ell',t)|\mathcal{F}_t\right]+m(\ell',\ell)\mathbb{E}\left[M_{k,n}(\ell',\ell,t)|\mathcal{F}_t\right]\right)}_{(a)}\\
    &+\underbrace{\sum_{\ell'\leq \ell} m(\ell',\ell-\ell')\mathbb{E}\left[M_{k,n}(\ell',\ell-\ell',t)|\mathcal{F}_t\right]}_{(b)}.
   \end{align*}
   
   By Lemma \ref{lem:gaps_repartition} $\mathbb{E}\left[M_{k,n}(\ell,\ell',t)|\mathcal{F}_t\right]=\mathbb{E}\left[M_{k,n}(\ell',\ell,t)|\mathcal{F}_t\right]$, so: 
   \begin{align*}
   (a)=& 2\sum_{\ell'} \frac{\min(c,\frac{\ell'}{2k})+\min(c,\frac{\ell}{2k})}{n}\mathbb{E}\left[M_{k,n}(\ell,\ell',t)|\mathcal{F}_t\right]\\
   =&  \frac{\min(2c,\frac{\ell}{k})}{n}\underbrace{\sum_{\ell'}\mathbb{E}\left[M_{k,n}(\ell,\ell',t)|\mathcal{F}_t\right]}_{=F_{k,n}(\ell,t)}+\underbrace{ \sum_{\ell'} \frac{\min(2c,\frac{\ell'}{k})}{n}\mathbb{E}\left[M_{k,n}(\ell,\ell',t)|\mathcal{F}_t\right]}_{=F_{k,n}(\ell,t)\sum_{\ell'}\frac{\min(2c,\ell'/k)}{n}\frac{F_{k,n}(\ell',t)-\mathds{1}_{\ell=\ell'}}{n_t-1}:=(a,ii)},
   \end{align*}
   where the expression for $(a,ii)$ is obtained by injecting the expression of \cref{lem:gaps_repartition}. We can simplify the expression at the cost of a small approximation, using the identity $1/(x-1)-1/x=1/[x(x-1)]$ for all $x>0$:
     \begin{align}\label{eq:approx}
 \bigg| (a,ii)- F_{k,n}(\ell,t)\sum_{\ell'}\frac{\min(2c,\ell'/k)}{n}\frac{F_{k,n}(\ell',t)}{n_t}\bigg|\leq& \frac{1}{n_t}\underbrace{\frac{F_{k,n}(\ell,t)}{n_t-1}}_{\leq 2}\underbrace{\sum_{\ell'}\frac{\min(2c,\ell'/k)}{n}F_{k,n}(\ell',t)}_{\leq1}\\
 &+\underbrace{\frac{\min(2c,\ell/k)F_{k,n}(\ell,t)}{nn_t}}_{\leq 1/n_t}.\notag
  \end{align}
Finally:
     \begin{align}\label{eq:bounda}
 \bigg| (a)- \left(\frac{\min(2c,\frac{\ell}{k})}{n}- \sum_{\ell'}\frac{\min(2c,\ell'/k)}{n}\frac{F_{k,n}(\ell',t)}{n_t}\right)F_{k,n}(\ell,t)\bigg|\leq&\frac{3}{n_t}.
  \end{align}
  
  We now turn to the second term. Using $\mathbb{E}\left[M_{k,n}(\ell',\ell-\ell',t)|\mathcal{F}_t\right]=\mathbb{E}\left[M_{k,n}(\ell-\ell',\ell',t)|\mathcal{F}_t\right]$, we obtain:
 \begin{align*}
 (b)= &\sum_{\ell'\leq \ell}\frac{\min(2c,\ell'/k)}{n}\mathbb{E}\left[M_{k,n}(\ell',\ell-\ell',t)|\mathcal{F}_t\right]\\
 =&\sum_{\ell'=1}^\ell\frac{\min(2c,\ell'/k)}{n}\frac{F_{k,n}(\ell',t)(F _{k,n}(\ell-\ell',t)-\mathds{1}_{\{\ell-\ell'=\ell'\}})}{n_t-1}
 \end{align*}
 
 Repeating the trick of \cref{eq:approx}, we obtain:
 \begin{align*}
 \bigg|(b)-\sum_{\ell'=1}^\ell\frac{\min(2c,\ell'/k)}{n}\frac{F_{k,n}(\ell',t)F _{k,n}(\ell-\ell',t)}{n_t}\bigg|\leq \frac{3}{n_t}.
 \end{align*}
The first equation of the lemma is a combination of the above equation and \cref{eq:bounda}. To get the second, we instantiate $c=+\infty$ and use that in metric matching $n_t=n-t$ always holds.
\hfill \(\Box\)

The following lemma gives a concentration bound on the initial values of the gaps.

\begin{lemma}[Initial Conditions]\label{lem:initialconditions}
 For any $\ell<kn$, with probability at least $1-\frac{4}{n^4}$,
\[
\left|F_{k,n}(\ell,0)-nkp_k^2e^{-\frac{\ell-1}{k}}\right|\leq 2\sqrt{12n\ln(n)}+1.
\]
\end{lemma}

\textit{Proof}:  Recall that by definition,
\[
F_{k,n}(\ell,0)= \sum_{i=1}^{n_0-1}\mathds{1}_{\{u_{i+1}-u_i=\frac{\ell}{nk}\}}+\mathds{1}_{\{u_{1}-u_{n_0}=\frac{\ell}{nk}\}}.
\]
 Let $\text{Geom}(p_k)$ be the geometric distribution with parameter $p_k$. Consider the process of placing a sequence of vertices $(a_i)_{i=1}^{+\infty}$ in $\mathbb{R}$, placing $a_0$  at zero, and having 
\[
nk(a_{i+1}-a_i) \stackrel{\text{i.i.d.}}{\sim}\text{Geom}(p_k), \text{ i.e. }\mathbb{E}\left[\mathds{1}_{\{a_{i+1}-a_i=\frac{\ell}{nk}\}}\right]=p_ke^{-\frac{\ell-1}{k}}.
\]
With that procedure, the law of the vertices placed before $1$ is exactly the law of the vertices in $\tilde{\mathcal{U}}_k$. Denote $\tilde{n}_0= \max\{i \text{ s.t. } a_i<1\}$.
We  have 
\begin{align}\label{eq:equallaw}
    \sum_{i=1}^{n_0-1}\mathds{1}_{\{u_{i+1}-u_i=\frac{\ell}{nk}\}}\stackrel{\mathcal{L}}{=}\sum_{i=1}^{\tilde{n}_0}\mathds{1}_{\{a_{i+1}-a_i=\frac{\ell}{nk}\}}.
\end{align}
By concentration of sums of Bernoulli random variable and since $kp_k\leq 1$:
\begin{align}
&\mathbb{P}\left(\left|\sum_{i=1}^{nkp_k}\mathds{1}_{\{a_{i+1}-a_i=\frac{\ell}{nk}\}}-nkp_k^2e^{-\frac{\ell-1}{k}}\right|>\sqrt{12n\ln(n)}\right)\nonumber\\
&\hspace{3cm}= \mathbb{P}\left(\left|\sum_{i=1}^{nkp_k}\mathds{1}_{\{a_{i+1}-a_i=\frac{\ell}{nk}\}}-nkp_k^2e^{-\frac{\ell-1}{k}}\right|>kp_k^2e^{-\frac{\ell-1}{k}}n\sqrt{\frac{12\ln(n)}{nk^2p_k^4e^{-2\frac{\ell-1}{k}}}}\right), \nonumber\\
&\hspace{3cm}\leq  \frac{2}{n^4} \label{eq:nbvnk}.
\end{align}

We also have:
\begin{align*}
    \mathbb{P}\left(a_{nkp_k+\sqrt{12n\ln(n)}}<1\right)=& \mathbb{P}\left(\big|\tilde{n}_0\big|\geq nkp_k+\sqrt{12n\ln(n)}\right)\\
    \leq& \mathbb{P}\left(\big|\tilde{n}_0\big|\geq nkp_k\left(1+\sqrt{\frac{12\ln(n)}{nkp_k}}\right)\right)\\
    \leq &\frac{1}{n^4}.
\end{align*}

Where the last inequality holds by Chernoff bound for sums of Bernoulli random variables.
Similarly, $\mathbb{P}(a_{nkp_k-\sqrt{12n\ln(n)}}>1)\leq \frac{1}{n^4}$. Thus, with probability at least $1-\frac{2}{n^4}$, there are between $nkp_k-\sqrt{12n\ln(n)}$ and $nkp_k+\sqrt{12n\ln(n)}$ vertices placed before $1$. Hence, combining with \cref{eq:nbvnk}:
\[
\mathbb{P}\left(\left|\sum_{i=1}^{\tilde{n}_0}\mathds{1}_{\{a_{i+1}-a_i=\frac{\ell}{nk}\}}-nke^{-\frac{\ell-1}{k}}p_k^2\right|>2\sqrt{12n\ln(n)}\right)\leq \frac{4}{n^4}.
\]
 Combining with  \cref{eq:equallaw} finishes the proof.
\hfill \(\Box\)

 \subsubsection{Relation to the differential equation}\label{sec:analofclosest}
\hfill \break

The methods employed to relate the discrete process to the differential equation, albeit similar in spirit, now differ for maximal cardinality matching and metric matching. The proofs are separated into two different paragraphs. 
\paragraph{Maximum cardinality matching}\label{sec:01matchingDEM}

The following proposition is obtained by applying the differential equation method (DEM, see Appendix \ref{app:Wormald}) on the discretized graphs.

\begin{proposition}
    [DEM, maximum cardinality matching]\label{lem:resinter}
For any $\tau\in [0,1]$, 
\[
\frac{1}{n}\sum_{\ell=1}^{kn}F_{k,n}\left(\ell, \tau n\right)\xrightarrow[n\rightarrow + \infty]{\mathbb{P}}\sum_{\ell=1}^{+\infty}f_{k}(\ell,\tau).
\]

where the functions $\left(f_{k}(\ell,t)\right)_\ell$ are the solutions of the following system of differential equations:
\begin{align*}
        \frac{\partial f_{k}(\ell,\tau)}{\partial \tau} =&-\min\left(\frac{\ell}{k},2c\right)f_{k}(\ell,\tau)-\frac{1}{\sum_{\ell'=1}^{+\infty}f_{k}(\ell',\tau)}\left[\sum_{\ell'=1}^{+\infty}\min\left(\frac{\ell}{k},2c\right)f_{k}(\ell',\tau)\right]f_{k}(\ell,\tau)\\
   &+\frac{1}{\sum_{\ell'=1}^{+\infty}f_{k}(\ell',\tau)}\sum_{\ell'=1}^{+\infty}\min\left(\frac{\ell'}{k},2c\right)f_{k}(\ell',t)f_{k}(\ell-\ell',\tau),
 \end{align*}
 with the initial conditions:
 $$
 f_{k}(\ell,0)= kp_k^2e^{-\frac{\ell-1}{k}}.
 $$
\end{proposition}
\textbf{Remark}: The functions $f_k$ depend on $c$, we omit that dependency in the notations.

\textit{Sketch of Proof}: The crux of this proof is to check all the conditions necessary to apply \cref{thm:Wormald}. The main conditions follow from \cref{lem:initialconditions,lem:evollaw}, there still however remains some computations to be done, which are deferred to \cref{sec:proof01matchingDEM}. \hfill \(\Box\)

The last step of the proof of \cref{thm:online0/1} is to link the functions $f_k(\ell,t)$ to a single function $f$, that does not depend on $k$.

\begin{lemma}\label{lem:discretetocontinuous}
For any $\tau \in [0,1]$, the following holds:
\[
\bigg|\int_{0}^{+\infty}f(x,\tau)dx-\sum_{\ell=1}^{+\infty}f_{k}(\ell,\tau)\bigg|\leq \frac{\omega}{k}.
\]
with $\omega$ a constant depending only on $c$ and $f(x,\tau)$ the solution of the following PDE 
\begin{align}\label{eq:continuous}
    \frac{\partial f(x,\tau)}{\partial t} =&-\min(x,2c)f(x,\tau)-\frac{1}{\int_{0}^{+\infty}f(x',\tau)dx'}\int_{0}^{+\infty}\min(x',2c)f(x',\tau)dx'f(x,\tau)\\
   &+\frac{1}{\int_{0}^{+\infty}f(x',\tau)dx'}\int_{0}^{x}\min(x',2c)f(x',\tau)f(x-x',\tau)dx'.\nonumber
\end{align}
with initial conditions $f(x,0)= e^{-x}.$

\end{lemma}

\textit{Sketch of proof}: This proof is an application of the finite elements method. The details of the computations are given in  \cref{sec:finiteelements}.  \hfill \(\Box\)

\textbf{Proof of \cref{thm:online0/1}}: It remains to put all the pieces together. Take any $\varepsilon>0$ and $k$ large enough so that $\frac{w}{k}\leq \frac{\varepsilon}{4}$.  This implies
\begin{align*}
\mathbb{P} \left(\left|\frac{1}{n}\mu_\closest\left(\textsc{G}[\mathcal{X}, \mathcal{Y},n,c]\right)(\tau n)-\left(1-\int_0^{+\infty}f(x,\tau)dx\right)\right|\geq \varepsilon\right)&\\
&\hspace{-8cm}=\mathbb{P} \left(\left|\frac{1}{n}\mu_\closest\left(\textsc{G}[\mathcal{X}, \mathcal{Y},n,c]\right)(\tau n)-\left(1-\sum_{\ell=1}^{+\infty}f_{k}(\ell,\tau)\right)\right|\geq \frac{3\varepsilon}{4}\right)\\
&\hspace{-8cm}\leq\mathbb{P} \left(\left|\mu_\closest\left(\textsc{G}[\mathcal{X}, \mathcal{Y},n,c]\right)(\tau n)-\mu_\closest\left(\textsc{G}^{\text{glued}}[\tilde{\mathcal{U}}_k \mathcal{Y},n,c]\right)(\tau n)\right|\geq \frac{n\varepsilon}{4}\right)\\
&\hspace{-7.5cm}+\underbrace{\mathbb{P} \left(\left|\frac{1}{n}\mu_\closest\left(\textsc{G}^{\text{glued}}[\tilde{\mathcal{U}}_k, \mathcal{Y},n,c]\right)(\tau n)-\left(1-\sum_{\ell=1}^{+\infty}f_{k}(\ell,\tau)\right)\right|\geq \frac{\varepsilon}{2}\right)}_{(ii)}.
\end{align*}
If $k$ is also large enough so that $\frac{21}{k}\leq \frac{\varepsilon}{8}$, by \cref{prop:roundingproc}, the first term of the last line vanishes as $n$ gets large. For the second term $(ii)$, recall \cref{eq:linkgapsmatch}:
\begin{align*}
\frac{1}{n}\mu_\closest\left(\textsc{G}^{\text{glued}}[\tilde{\mathcal{U}}_k, \mathcal{Y},n,c]\right)(\tau n)=\frac{|\tilde{\mathcal{U}}_k|}{n}-\frac{1}{n}\sum_{\ell \in[kn]}F_{k,n}(\ell,\tau n).
\end{align*}
It implies:

\begin{align*}
(ii)
   &\leq \mathbb{P} \left(\left|\frac{1}{n}\sum_{\ell=1}^{+\infty}F_{k,n}\left(\ell,\tau n\right)-f_{k}(\ell,\tau)\right|\geq \frac{\varepsilon}{4}\right)+\mathbb{P} \left(\left|\frac{|\tilde{\mathcal{U}}_k|}{n}-1\right|\geq \frac{\varepsilon}{4}\right).
\end{align*}

By \cref{lem:resinter} the first term vanishes as $n$ gets large. By \cref{eq:lowerboundUk}, so does the second. This achieves the proof of the convergence in probability. \hfill \(\Box\)

\paragraph{Metric matching}\label{sec:metricmatching}
In the case of metric matching, more work is still required before the application of the Differential Equation Method. We start by showing that the cost of the matching produced by algorithm \closest\ concentrates around its expectation. Define $c_t[k]$ as the length of the edge added to the matching at time $t$, i.e.:
\[
c_t[k] = \kappa_\textsc{\closest}\left(\textsc{G}^{\text{glued}}_\rho[\tilde{\mathcal{U}}_k , \mathcal{Y}]\right)(t)-\kappa_\textsc{\closest}\left(\textsc{G}^{\text{glued}}_\rho[\tilde{\mathcal{U}}_k , \mathcal{Y}]\right)(t-1).
\]

Note that by this definition \begin{align}
    \kappa_\textsc{\closest}\left(\textsc{G}^{\text{glued}}_\rho[\tilde{\mathcal{U}}_k , \mathcal{Y}]\right)(t)= \sum_{j=1}^tc_j[k].\label{eq:breakdowncostmetric}
    \end{align}
\begin{lemma}\label{lem:costconcentrate}
The cost of the matching produced by algorithm \closest\ concentrates around its expectation:
\[
\mathbb{P}\left(\left|\sum_{j=1}^{t} c_t[k]-\mathbb{E}\left[\sum_{j=1}^{t} c_j[k]\right]\right|\geq \delta+\frac{4\alpha+1}{n}\right)\leq  \frac{8\alpha+4}{n^2},
\]
with $\delta= \left[\sqrt{\frac{8\ln(n)(k+1)}{n}}+\frac{8(4\alpha+1)(k+1)}{n^2}\right]\alpha^2\ln(n)^2$ and $\alpha$ defined in \cref{prop:roundingproc}.
\end{lemma}
\textit{Sketch of Proof}: The proof is an application of McDiarmid's inequality. The technical aspects are deferred to \cref{sec:proofmetricmatchingDEM}. \hfill \(\Box\)

We now show that the total contribution of long edges to the final cost remains small. 

\begin{lemma}\label{lem:boundprobahighcost} [Adaptation of Lemma 12-13 from \cite{balkanski2022power}]
 For any $\eta\in[\frac{16}{3(1-\tau)^2n},1]$, $k\geq \frac{1}{1-\tau}$ and $n\geq \frac{25}{\tau}$, we have:
\begin{align*}
\mathbb{E}\left[\sum_{j\leq \tau n}c_j[k]\mathds{1}\{c_j[k]\geq \frac{\eta}{n}\} \right]
\leq& C_\tau e^{-\eta C'_\tau}.
\end{align*}
with $C_\tau$ and $C'_\tau$ two constants depending only on $\tau$.
\end{lemma}

\textit{Sketch of Proof}: This lemma is an adaptation of similar lemmas in \cite{balkanski2022power}.   As the setting differs slightly and for completeness, we provide a brief proof of the statement in Appendix \ref{app:boundprobahighcost}, which essentially follows the steps of \cite{balkanski2022power}. \hfill \(\Box\)

We now apply the differential equation method to control the contribution of the short edges.

\begin{lemma}\label{lem:wormaldmetric}
    For any $n\geq \frac{1}{(1-\tau)^2}$, $k\geq \frac{2}{\eta}$,  we  have:
\begin{align*}
\left|\mathbb{E}\left[\sum_{t=1}^{n\tau} c_t[k]\mathds{1}\{c_t[k]\leq \frac{\eta}{n}\}\right]-\frac{1}{4n}\sum_{t=0}^{n\tau-1}\sum_{\ell=1}^{k\eta}\left(\frac{\ell}{k}\right)^2g_k(\ell,\frac{t}{n})\right|\leq &kw_3(\eta,\tau)\sqrt{\frac{\ln(n)}{n}},
\end{align*}
with $w_3(\eta,\tau)$ a constant depending on $\eta$ and $\tau$ and $g_k(\ell,t)$ the solution of the following system of differential equations.
\begin{align*}
        \frac{\partial g_{k}(\ell,t)}{\partial t} =&-\left(\frac{\ell}{k}+\frac{1}{1-t}\right)g_{k}(\ell,t)+\frac{1}{1-t}\sum_{\ell'=0}^{+\infty}\frac{\ell'}{k}g_{k}(\ell',t)g_{k}(\ell-\ell',t),
 \end{align*}
 with the initial conditions:
\[
 g_{k}(\ell,0)= kp_k^2e^{-\frac{\ell-1}{k}}.
\]
\end{lemma}
\textbf{Remark}: $ g_{k}$ is the function $f_k$ obtained by setting $c=+\infty$ (the dependency of $f_k$ on $c$ was omitted in the notations).

\textit{Sketch of Proof}: As before,  the key to this proof is to check all the conditions necessary to apply \cref{thm:Wormald}. Again, main conditions follow from \cref{lem:initialconditions,lem:evollaw}, and details can be found in \cref{sec:proofmetricmatchingDEM}. \hfill \(\Box\)

\begin{lemma}[Link with the continuous equation, metric matching version]\label{lem:continuousmetric}
\[
    \left|\int_{t=0}^{\tau}\int_{x=1}^{\eta}x^2g(x,t)dxdt-
    \frac{1}{n}\sum_{t=0}^{n\tau-1}\sum_{\ell=1}^{k\eta}\left(\frac{\ell}{k}\right)^2g_k(\ell,\frac{t}{n})\right| \leq \frac{w_1(\tau,\eta)}{n}+\frac{w_2(\tau,\eta)}{k},
\]
 with $w_1(\tau,\eta)$ and $w_2(\tau,\eta)$ constants depending on $\tau$ and $\eta$ only, and $g(x,t)$ is the solution of the following PDE:
\begin{align*}
        \frac{\partial g(x,t)}{\partial t} =&-\left(x+\frac{1}{1-t}\right)g(x,t)+\frac{1}{1-t}\int_{x'=0}^{x}x'g(x',t)g(x-x',t)dx',
 \end{align*}
 with the initial conditions:
\[
 g(x,0)= e^{-x}.
\]
\end{lemma}

\textit{Sketch of proof}: This proof is an application of the finite elements method. The details of the computations are given in  \cref{sec:finiteelements}.  \hfill \(\Box\)

\begin{lemma}\label{lem:computelength}
We have:
\[
\int_{0}^{\tau}\int_{0}^{+\infty}\frac{x^2}{4}g(x,t)dxdt=\frac{1}{2}\left[\frac{1}{1-\tau}-1\right].
\]
\end{lemma}
The proof of that lemma is again a computation deferred to section \ref{sec:proofmetricmatchingDEM}.

\textbf{Proof of \cref{thm:onlinelength}:} Now that we have all the necessary lemmas, we put the pieces together to prove the main theorem.  By fixing  $k=n^{1/8}$ in \cref{prop:roundingproc}, we get the following convergence:
\begin{align*}
   \kappa_\textsc{\closest}\left(\textsc{G}_\rho[\mathcal{X}, \mathcal{Y}]\right)(\tau n)\xrightarrow[n\rightarrow +\infty]{\mathbb{P}}\kappa_\textsc{\closest}\left(\textsc{G}^{\text{glued}}_\rho[\tilde{\mathcal{U}}_{n^{1/8}} , \mathcal{Y}]\right)( \tau n).
\end{align*}

Again with $k=n^{1/8}$, in \cref{lem:costconcentrate}, we get:
\[
\kappa_\textsc{\closest}\left(\textsc{G}^{\text{glued}}_\rho[\tilde{\mathcal{U}}_{n^{1/8}} , \mathcal{Y}]\right)(\tau n)\xrightarrow[n\rightarrow +\infty]{\mathbb{P}}\mathbb{E}\left[\kappa_\textsc{\closest}\left(\textsc{G}^{\text{glued}}_\rho[\tilde{\mathcal{U}}_{n^{1/8}}, \mathcal{Y}]\right)(\tau n)\right].
\]

Let us now bound:
\begin{align*}
\bigg|\mathbb{E}\left[\kappa_\textsc{\closest}\left(\textsc{G}^{\text{glued}}_\rho[\tilde{\mathcal{U}}_{n^{1/8}}, \mathcal{Y}]\right)(\tau n)\right]-\int_{t=0}^{\tau}\int_{x=0}^{+\infty}x^2g(x,t)dxdt\bigg|:=(a)
\end{align*}

We bound separately the contributions of the small and long edges, using \cref{eq:breakdowncostmetric}:
\begin{align*}
    (a)\leq &\left|\mathbb{E}\left[\sum_{t=1}^{n\tau} c_t[n^{1/8}]\mathds{1}\{c_t[n^{1/8}]\leq \frac{\eta}{n}\}\right]-\frac{1}{4n}\sum_{t=0}^{n\tau-1}\sum_{\ell=1}^{n^{1/8}\eta}\left(\frac{\ell}{n^{1/8}}\right)^2g_{n^{1/8}}(\ell,\frac{t}{n})\right|\\
    &+\left|\int_{t=0}^{\tau}\int_{x=1}^{\eta}x^2g(x,t)dxdt-\frac{1}{4n}\sum_{t=0}^{n\tau-1}\sum_{\ell=1}^{n^{1/8}\eta}\left(\frac{\ell}{n^{1/8}}\right)^2g_{n^{1/8}}(\ell,\frac{t}{n})\right|\\
    &+\int_{t=0}^{\tau}\int_{x=\eta}^{+\infty}x^2g(x,t)dxdt+\mathbb{E}\left[\sum_{j\leq \tau n}c_j[n^{1/8}]\mathds{1}\{c_j[n^{1/8}]\geq \frac{\eta}{n}\} \right],\\
    \leq &w_3(\eta,\tau)\sqrt{\frac{\ln(n)}{n^{3/4}}}+\frac{w_1(\tau,\eta)}{n}+\frac{w_2(\tau,\eta)}{n^{1/8}}+\int_{t=0}^{\tau}\int_{x=\eta}^{+\infty}x^2g(x,t)dxdt+C_\tau e^{-\eta C'_\tau}.
\end{align*}
The second line is a consequence of \cref{lem:boundprobahighcost,lem:wormaldmetric,lem:continuousmetric}. It goes to $0$ as $\eta$ then $n$ get large, and the theorem follows. \hfill \( \Box\)
\section{Technical Steps for the proofs of  \cref{SE:Online}}\label{app:thmonline}
\subsection{Proof of \cref{eq:mun,eq:gamman}}\label{sec:proofoffline}

\paragraph{Proof of \cref{eq:mun}:} Denote $\mu_{k}=\frac{1}{k}\sum_{t=1}^{k}\mathds{1}_{\{|\psi(t)|\leq c\}}$ for any $k\geq 0$, with the convention that if $k\geq \tau_n$, the generating process for $\psi(t)$ remains unchanged ignoring the stopping condition. For any $y\in \mathbb{R}$,
$
\int_{-\infty}^{+\infty}\Pi(x,y)\pi(x)dx \leq \frac{1}{2}
$
and for any $x\in \mathbb{R}$,
$
\int_{-\infty}^{+\infty}\Pi(x,y)\pi(y)dy \leq \frac{1}{2}.
$
Thus by the Schur test lemma, the operator norm of the kernel is bounded as $|| \Pi ||_\pi\leq \frac{1}{2}$.  By a version of Hoeffding's inequality adapted to Markov chains \citep{miasojedow2014hoeffding}, we get for any $\delta>0$:
\[
\mathbb{P}\left(|\mu_{k}-\pi(|x|\leq c)\right|\geq \delta) \leq 4(c+1)e^{-\frac{\delta^2 k}{3}}.
\]

The algorithm generates completely at least one side of the graph, either $\mathcal{U}$ or $\mathcal{V}$, and generates at most one vertex in each side at every iteration. Thus, by \cref{eq:numpoints}, with probability at least $1-\frac{4}{n}$, \[\tau_n\geq n-2\sqrt{n\ln(n)}\geq \frac{n}{2} \text{ and } \tau_n\leq 2n+4\sqrt{n\ln(n)}\leq 4n,\] for any $n\geq 10$. In addition, by \cref{lem:stationnarydistrib}, $\pi(|x|\leq c)=\frac{2c}{2c+2}$. This, and setting $\delta=\sqrt{\frac{12\ln(n)}{n}}$, gives that for any $n\geq 10$:
\begin{align}
\mathbb{P}\left(\bigg|\mu_{\tau_n}-\frac{2c}{2c+2}\bigg|\geq \sqrt{\frac{12\ln(n)}{n}}\right) \leq&\mathbb{P}\left(\tau_n\leq \frac{n}{2}\right)+\mathbb{P}\left(\tau_n\geq 4n\right)+\sum_{k= \frac{n}{2}}^{4n}\mathbb{P}\left(|\mu_{k}-\pi(|x|\leq c)\right|\geq \delta) \notag\\
\leq & \frac{4}{n}+ 4(c+1)\sum_{k=\frac{n}{2}}^{4n} e^{-\frac{4\ln(n)}{n}k}\notag\\
\leq & \frac{4}{n}+ 16n(c+1)e^{-2\ln(n)}
\leq  \frac{16(c+2)}{n}.\notag
\end{align}\hfill \(\Box\)

\paragraph{Proof of \cref{eq:gamman}:}Note that as the algorithm stops when all vertices on one side have been generated, we have $\gamma_n\leq |\mathcal{U}|+|\mathcal{V}|$. Three cases are possible: all vertices on both sides are generated, in which case $\gamma_n= |\mathcal{U}|+|\mathcal{V}|$, or some vertices in $\mathcal{U}$ or $\mathcal{V}$ are not generated. 

By symmetry, we can assume w.l.o.g. that the algorithm stopped because of the condition $u_{i(\tau_n+1)}>1$, i.e. when all vertices in $\mathcal{U}$ have been generated. This implies that a vertex in $\mathcal{U}$ has been generated at iteration $\tau_n$, hence $v_{j(\tau_n)}\geq u_{i(\tau_n)}-\frac{c}{n}$. We thus have:
\begin{align*}
|\gamma_n-|\mathcal{U}|-|\mathcal{V}||\leq & |\mathcal{V}\cap(v_{j(\tau_n)};1]|\\
\leq& |\mathcal{V}\cap(\max_{u\in \mathcal{U}}u-\frac{c}{n};1]|.
\end{align*}

We indicated $\mathcal{U}$ and $\mathcal{V}$ could be generated by a renewal process with exponential holding time starting from $0$, but the same holds starting from $1$. This implies that $1-\max_{u \in \mathcal{U}}u \sim \text{Exp}(n)$ . By the property of exponential random variables:
\[
\mathbb{P}\left(1-\max_{u \in \mathcal{U}}u\geq \frac{\ln(n)}{n}\right)\leq \frac{1}{n}.
\]

Also, by Chernoff bound,

\[
\mathbb{P}\left(\bigg|\mathcal{V}\cap\left(1-\frac{c+\ln(n)}{n};1\right]\bigg|\geq c+\ln(n)+2\sqrt{n \ln(n)} \right)\leq \frac{4}{n}.
\]
Therefore, by union bound, and using $c+\ln(n)+2\sqrt{n \ln(n)}\leq  (c+3)\sqrt{n \ln(n)}$, we get:

\begin{equation}\label{eq:boundmissingvertices}
\mathbb{P}\left(\bigg|\mathcal{V}\cap\left(\max_{u\in \mathcal{U}}u-\frac{c}{n};1\right]\bigg|\geq (c+3)\sqrt{n \ln(n)} \right)\leq \frac{5}{n}.\notag
\end{equation}
By symmetry, the above bounds hold with $\mathcal{U}$ and $\mathcal{V}$ inverted. Combining the above inequality with \cref{eq:numpoints}, with probability at least $1-\frac{14}{n}$, for any $n\geq10$, we have:
\begin{align*}
|\gamma_n-2n|\leq& |\gamma_n-|\mathcal{U}|-|\mathcal{V}||+4\sqrt{n\ln(n)}\\
\leq& (c+7)\sqrt{n \ln(n)}.
\end{align*}
\hfill \(\Box\)
\subsection{Proof of Proposition \ref{prop:roundingproc}}\label{sec:proofroundingproc}

\hfill \break

The proposition is a consequence of four lemmas that bound the impact in matching sizes/costs incurred in each step. We start with two technical lemmas that are used in the proofs of the four following ones. 

The first one, helpful for the maximum cardinality matching setting, shows that slightly modifying the set of offline vertices does not affect too much the size of the matching generated by \closest.

\begin{lemma}\label{lem:addvertex}
Adding or removing a vertex to the offline side of the graph modifies the score of \closest\ by at most one, i.e. for any sets $\mathcal{X}$, $\mathcal{Y}$, any $n$, $c$, any vertex $x_0$ and any $\tau$ s.t. $\tau n\leq |\mathcal{Y}|$ we have:
\[
|\mu_\closest(\textsc{G}[\mathcal{X}\cup x_0, \mathcal{Y},n,c])(\tau n)-\mu_\closest(\textsc{G}[\mathcal{X}, \mathcal{Y},n,c])(\tau n)|\leq 1.
\]
\end{lemma}

\textit{Proof}: Denote $\mathcal{X}_t$ and  $\mathcal{X}^+_t$ the sets of free vertices at iteration $t$, in $\textsc{G}[\mathcal{X}, \mathcal{Y},n,c]$ and $\textsc{G}[\mathcal{X}\cup x_0, \mathcal{Y},n,c]$ respectively. We will show by induction that at every iteration, one of the following properties holds:
\[
\text{(P1): \{$\mu_\closest(\textsc{G}[\mathcal{X}\cup x_0, \mathcal{Y},n,c])(\tau n)=\mu_\closest(\textsc{G}[\mathcal{X}, \mathcal{Y},n,c])(\tau n)$ and $\exists\ x_t^+ \in \mathcal{X}\cup{x_0}$ s.t. $\mathcal{X}^+_t=\mathcal{X}_t\cup{x_t^{+}}$}\}\,
\]
or
\[
 \text{ (P2): \{$\mu_\closest(\textsc{G}[\mathcal{X}\cup x_0, \mathcal{Y},n,c])(t)=\mu_\closest(\textsc{G}[\mathcal{X}, \mathcal{Y},n,c])(t)+1$ and $\mathcal{X}^+_t=\mathcal{X}_t$}\}.
\]

If (P2) is true at some iteration $t$, it remains true until the end of the run and the proof is over. If (P1) is true at iteration $t$, the following cases are possible:
\begin{enumerate}
    \item if $y_t$ has no neighbor in $\mathcal{X}^+_t$, it is unmatched in both graphs,
    \item if $y_t$'s closest neighbor in $\mathcal{X}^+_t$ is not $x_t^{+}$, it is matched to the same vertex in both graphs,
    \item if $y_t$'s  closest neighbor in $\mathcal{X}^+_t$ is $x_t^{+}$ and $x^+_{t+1}$ in $\mathcal{X}_t$, it is matched to $x_t^{+}$ in $\textsc{G}[\mathcal{X}\cup x_0, \mathcal{Y},n,c]$ and to $x^+_{t+1}$ in $\textsc{G}[\mathcal{X}, \mathcal{Y},n,c]$,
    \item $y_t$'s  only neighbor in $\mathcal{X}^+_t$ is $x_t^{+}$, in which case it is matched to $x_t^{+}$ in $\textsc{G}[\mathcal{X}\cup x_0, \mathcal{Y},n,c]$ and unmatched in $\textsc{G}[\mathcal{X}, \mathcal{Y},n,c]$.
\end{enumerate}
Cases 1 to 3 imply that (P1) remains true at iteration $t+1$, case 4 implies that (P2) is true at iteration $t+1$. (P1) is true at iteration 0, thus either (P1) or (P2) holds at every iteration.

\hfill \(\Box\)

The second lemma is helpful for the metric matching setting. Denote $\mathcal{Y}_{\tau n}=(y_i)_{i=1}^{\tau n}$ the sets of vertices that have arrived before iteration $\tau n$. The lemma shows that,  w.h.p., for any large enough interval $I$, for all the intermediate graphs of the \graphrounding\ procedure, the number of offline vertices of the graph in $I$ is larger than $|\mathcal{Y}_{\tau n}\cap I|$, plus a bit of slack.  Moreover, $|\mathcal{Y}_{\tau n}\cap I|$ is not too large w.h.p.. In the whole proof, we use the convention that for any $a,b\in \mathbb{R}^+$, $a\leq b$, $a\leq n$, $I=[\frac{a}{n};\frac{b}{n}]=\left[\frac{a}{n};1\right]\cup\left[0;\frac{b-n}{n}\right]$ if $b\geq n$ and $|I| = \frac{b-a}{n}$. We also assume $n\geq 10$ throughout.
\begin{lemma}\label{lem:metricmatchingnbverticeslargeintervals}
    Let $\alpha = \frac{128}{(1-\tau)^2\tau^2}$, as in \cref{prop:roundingproc}.  For any  $k\geq \frac{2}{\tau(1-\tau)}$ probability at least $1-(4\alpha+1)/n^2$, for any interval  s.t. $|I|\geq \frac{\alpha \ln(n)+2}{n}$, 
    \[|\mathcal{X}\cap I|\geq |\mathcal{Y}_{\tau n}\cap I|+8\ln(n)+4,\]
    
    and the same lower bound holds for $|\mathcal{U}\cap I|$ and $|\tilde{\mathcal{U}}_k\cap I|$. Moreover:
    \[|\mathcal{Y}_{\tau n}\cap I|\leq \tau n\left(1+\frac{1-\tau}{2}\right)|I| + 8\ln(n)+2 \text{ and } \big|\mathcal{Y}_{\tau n}\cap\left[\frac{i}{n};\frac{i+1}{n}\right]\big|\leq 3\ln(n)+1, \forall i \in [n].\] We denote $\mathcal{E}_\mathcal{I}$ that event.
\end{lemma}
\textit{Proof}: We start by showing that the number of online points in any small interval is small. By Chernoff bound, we have:
\begin{align*}
\mathbb{P}\left(\sum_{j=1}^{n\tau}\mathds{1}\left\{y_j \in\left[\frac{i}{n};\frac{i+1}{n}\right]\right\}\geq \left(1+\frac{4\ln(n)}{\tau}\right)\tau\right) &\leq e^{-3\ln(n)}\leq \frac{1}{n^3}.
\end{align*}
By union bound over all $n$ possible values of $i$, this gives:
\begin{align}\label{eq:endintery_t}
\mathbb{P}\left(\exists i \in [n] \text{ s.t. }\big|\mathcal{Y}_{\tau n}\cap\left[\frac{i}{n};\frac{i+1}{n}\right]\big|\geq 1+4\ln(n)\right) &\leq \frac{1}{n^2}.
\end{align}

Now, consider any $i \in [n]$, any $\ell \in [\lceil\alpha\ln(n)\rceil;n]$. Take interval $I=\left[\frac{i}{n};\frac{i+\ell}{n}\right]$. Again by Chernoff bound:
\begin{align}\label{eq:boundy_tI}
\mathbb{P}\left(|\mathcal{Y}_{\tau n}\cap I|\geq \tau\left(1+\frac{1-\tau}{2}\right)\ell\right) \leq e^{-\frac{(1-\tau)^2\tau}{12} \ell}\leq e^{-\frac{(1-\tau)^2\tau^2}{32} \ell}.
\end{align}

We have:
\begin{align}\label{eq:unionboundell}
\sum_{\ell = \lceil\alpha \ln(n)\rceil}^{+\infty} e^{-\frac{\tau^2(1-\tau)^2}{32} \ell}= \frac{e^{-\frac{\tau^2(1-\tau)^2}{32} }}{e^{-\frac{\tau^2(1-\tau)^2}{32} }-1}  e^{-\frac{\tau^2(1-\tau)^2}{32} \lceil\alpha \ln(n)\rceil}\leq \frac{64}{\tau^2(1-\tau)^2}e^{-\frac{\tau^2(1-\tau)^2}{32} \alpha\ln(n)} \leq \frac{\alpha}{n^4}.
\end{align}
The first inequality holds since $\frac{e^{-x}}{e^{-x}-1}\leq \frac{2}{x}$ for any $x\leq1$, and the second one by definition of $\alpha$. With an upper bound over all possible values of $\ell$ and $i$, we obtain:

\begin{align}\label{eq:boundy_tI}
\mathbb{P}\left(\exists i \in [n],  \ell \in [\lceil\alpha\ln(n)\rceil;n],  I=\left[\frac{i}{n};\frac{i+\ell}{n}\right] \text{ s.t. }|\mathcal{Y}_{\tau n}\cap I|\geq \tau\left(1+\frac{1-\tau}{2}\right)\ell\right) \leq \frac{\alpha}{n^2}.
\end{align}

Now, for any interval $I=[\frac{a}{n};\frac{b}{n}]$, we have:
\[
I\subseteq \left[\frac{\lfloor a\rfloor}{n};\frac{\lceil a\rceil}{n}\right] \cup \left[\frac{\lceil a\rceil}{n};\frac{\lfloor b\rfloor}{n}\right] \cup \left[\frac{\lfloor b\rfloor}{n};\frac{\lceil b\rceil}{n}\right].
\]
If $b-a\geq \alpha \ln(n)+2$, then $\lfloor b\rfloor-\lceil a\rceil\geq \alpha \ln(n)$, thus applying an union bound over \cref{eq:boundy_tI,eq:endintery_t}, we obtain:
\begin{align}\label{eq:boundy_tallI}
\mathbb{P}\left(\exists  b-a\geq \alpha \ln(n)+2 \text{ s.t. }|\mathcal{Y}_{\tau n}\cap [\frac{a}{n};\frac{b}{n}]|\geq \tau\left(1+\frac{1-\tau}{2}\right)(\lfloor b\rfloor-\lceil a\rceil)+8\ln(n)+2\right) \leq \frac{\alpha+1}{n^2}.
\end{align}

We now lower bound $|\mathcal{X}\cap I|$, $|\mathcal{U}\cap I|$ and  $|\tilde{\mathcal{U}}\cap I|$ for any interval $I=\left[\frac{i}{n};\frac{i+\ell}{n}\right]$ with $i \in [n]$ and $\ell \in [\lceil\alpha\ln(n)\rceil;n]$. 
Fist note that:
\begin{equation}\label{eq:boundinterUI}
\mathbb{P}\left(|\mathcal{X}\cap I|\leq \tau\left(1+\frac{1-\tau}{2}\right)\ell+ 16 \ln(n)+6\right)
\leq \mathbb{P}\left(|\mathcal{X}\cap I|\leq \left(\tau+\frac{\tau(1-\tau)}{2}+(1-\tau)^2\right)\ell\right),
\end{equation}
as, by definition of $\alpha$, for any  $n\geq 2$, $
16\ln(n)+6\leq (1-\tau)^2\alpha\ln(n)\leq \ell(1-\tau)^2$. The same holds for $\mathcal{U}$ and $\tilde{\mathcal{U}}_k$.
By Chernoff bound:
\begin{align*}
\mathbb{P}\left(|\mathcal{X}\cap I|\leq \left(\tau+\frac{\tau(1-\tau)}{2}+(1-\tau)^2\right)\ell\right)
&= \mathbb{P}\left(|\mathcal{X}\cap I|\leq \left(1-\frac{\tau(1-\tau)}{2}\right)\ell\right)\\
&\leq e^{-\frac{\tau^2(1-\tau)^2}{12}\ell}.
\end{align*}
This gives:
\begin{equation}\label{eq:boundXI}
\mathbb{P}\left(|\mathcal{X}\cap I|\leq \tau\left(1+\frac{1-\tau}{2}\right)\ell+ 16 \ln(n)+6\right)
\leq e^{-\frac{\tau^2(1-\tau)^2}{32}\ell}.
\end{equation}
By tail bounds for Poisson random variables:
\begin{align*}
\mathbb{P}\left(|\mathcal{U}\cap I|\leq \left(1-\frac{\tau(1-\tau)}{2}\right)\ell\right)
&\leq e^{-\frac{\tau^2(1-\tau)^2\ell^2}{4\times2(\ell+\frac{\tau(1-\tau)}{2})\ell}}\leq  e^{-\frac{\tau^2(1-\tau)^2}{12}\ell}.
\end{align*}
So, combining with \cref{eq:boundinterUI}:
\begin{align}\label{eq:boundUI}
\mathbb{P}\left(|\mathcal{U}\cap I|\leq \tau\left(1+\frac{1-\tau}{2}\right)\ell+ 16 \ln(n)+6\right)
\leq e^{-\frac{\tau^2(1-\tau)^2}{32}\ell}.
\end{align}
We have $|\tilde{\mathcal{U}}_k\cap I|\sim\text{Bin}(p_k,k\ell)$. Take $k\geq \frac{2}{\tau(1-\tau)}$.
As $1\geq p_kk \geq 1-\frac{1}{2k}$, this implies $kp_k\geq 1-\frac{\tau(1-\tau)}{4}$, hence:
\begin{align*}
 \mathbb{P}\left(|\tilde{\mathcal{U}}_k\cap I|\leq \left(1-\frac{\tau(1-\tau)}{4}\right)kp_k\ell\right)
&\geq  \mathbb{P}\left(|\tilde{\mathcal{U}}\cap I|\leq \left(1-\frac{\tau(1-\tau)}{4}\right)^2\ell\right),\\
&\geq  \mathbb{P}\left(|\tilde{\mathcal{U}}\cap I|\leq \left(1-\frac{\tau(1-\tau)}{2}\right)\ell\right).
\end{align*}

Hence, by Chernoff bound and combining with \cref{eq:boundinterUI}:
\begin{align}\label{eq:boundtildeUI}
\mathbb{P}\left(|\tilde{\mathcal{U}}_k\cap I|\leq \tau\left(1+\frac{1-\tau}{2}\right)\ell+ 16 \ln(n)+6\right)
\leq e^{-\frac{\tau^2(1-\tau)^2}{32}\ell}.
\end{align}
As any interval $I=[\frac{a}{n};\frac{b}{n}]$ contains $I=[\frac{\lceil a\rceil}{n};\frac{\lfloor b\rfloor}{n}]$, a union bound over all possible values of $i$ and $\ell$ in \cref{eq:boundXI,eq:boundUI,eq:boundtildeUI},  gives:
\[
\mathbb{P}\left(\exists  b-a\geq \alpha \ln(n)+2 \text{ s.t. }|\mathcal{X}\cap [\frac{a}{n};\frac{b}{n}]|\leq \tau\left(1+\frac{1-\tau}{2}\right)(\lfloor b\rfloor-\lceil a\rceil)+16\ln(n)+6\right) \leq \frac{\alpha}{n^2},
\]
and the same holds for $\mathcal{U}$ and $\tilde{\mathcal{U}}_k$.
Combining with \cref{eq:boundy_tallI} gives the result of the lemma.

\hfill\(\Box\)

\begin{lemma}[Poissonization]\label{lem:poissonization}
With probability at least $1-\frac{2}{n}$, it holds that:
\[
\big|\mu_\closest\left(\textsc{G}[\mathcal{X}, \mathcal{Y},n,c]\right)(\tau n)-\mu_\closest\left(\textsc{G}[\mathcal{U}, \mathcal{Y},n,c]\right)(\tau n)\big|\leq 2\sqrt{n\ln(n)}.
\]
With probability at least $1-\frac{2}{n}-\frac{4\alpha+1}{n^2}$ :
\[
|\kappa_\textsc{\closest}\left(\textsc{G}_\rho[\mathcal{X}, \mathcal{Y}]\right)(\tau n)-\kappa_\textsc{\closest}\left(\textsc{G}_\rho[\mathcal{U}, \mathcal{Y}]\right)(\tau n)|\leq 8\sqrt{\frac{\ln(n)}{n}}(\alpha\ln(n))^2. 
\]
\end{lemma}

\textit{Proof:} To obtain the first equation, we combine equation \ref{eq:numpoints}, which implies that the poissonisation step only adds or removes $2\sqrt{n\ln(n)}$ vertices with probability at least $1-\frac{2}{n}$,  with  Lemma \ref{lem:addvertex}.

For the second equation, let us first assume that $|\mathcal{X}|\leq |\mathcal{U}|$ and that event $\mathcal{E}_{\mathcal{I}}$ holds. Consider $x$ and $x'$, two consecutive free vertices remaining at iteration $\tau n$ of \closest\ in $\textsc{G}_\rho[\mathcal{X}, \mathcal{Y}]$. Let us study the impact of adding an offline vertex between $x$ and $x'$. This can only modify the match of vertices lying in the interval $[x,x']$ to another vertex within that interval. An upper bound on the modification to the length of the matching is thus $|[x,x']|\times |\mathcal{Y}_{\tau n}\cap[x,x']|$. All the vertices in $\mathcal{X}\cap[x,x']$ have been matched to vertices in $\mathcal{Y}_{\tau n}\cap[x,x']$, so $|\mathcal{Y}_{\tau n}\cap[x,x']|\geq|\mathcal{X}\cap[x,x']|$. Under event $\mathcal{E}_\mathcal{I}$, this implies 
\[|[x,x']|\leq \frac{\alpha\ln(n)+2}{n} \text{ and }|\mathcal{Y}_{\tau n}\cap[x,x']|\leq \tau\left(1+\frac{1-\tau}{2}\right)(\alpha\ln(n)+2) + 8\ln(n)+2.\] 

So, by upper bounding the multiplication of the two bounds, we obtain:

\begin{align*}
    |[x,x']|\times |\mathcal{Y}_{\tau n}\cap[x,x']|\leq&  \frac{4(\alpha\ln(n))^2}{n}.
\end{align*}

To modify $\mathcal{X}$ into $\mathcal{U}$, we need to add $|\mathcal{U}-\mathcal{X}|$ vertices, which upon each addition modifies the length of the matching by at most $\frac{4(\alpha\ln(n))^2}{n}$. The same reasoning can be applied if $|\mathcal{X}|\geq |\mathcal{U}|$.
\hfill \( \Box \)

\begin{lemma}[Rounding]\label{lem:rounding}
With probability at least $1-\exp(-\frac{2n}{k^2})-\frac{2}{n}$, it holds that:
\[
\big|\mu_\closest\left(\textsc{G}[\mathcal{U}_k, \mathcal{Y},n,c]\right)(\tau n)-\mu_\closest\left(\textsc{G}[\mathcal{U}, \mathcal{Y},n,c]\right)(\tau n)\big|\leq \frac{20n}{k},
\]
and, for any  $k\geq \frac{2}{\tau(1-\tau)}$, with probability at least $1-\exp(-\frac{2n}{k^2})-\frac{2}{n}-\frac{4\alpha+1}{n^2}$:
\[
\big|\kappa_\textsc{\closest}\left(\textsc{G}_\rho[\mathcal{U}_k, \mathcal{Y}]\right)(\tau n)-\kappa_\textsc{\closest}\left(\textsc{G}_\rho[\mathcal{U}, \mathcal{Y}]\right)(\tau n)\big|\leq \frac{80\left(\alpha\ln(n)\right)^2}{k}. 
\]
\end{lemma}

\textit{Proof:}
Consider the \textsc{($i$)}-\closest\ algorithm, which matches the incoming vertex to the closest rounded vertex up to iteration $i$ (included), then to the closest vertex. Ties with the rounded coordinates are broken following the unrounded ones' order. The runs of algorithms \textsc{($i-1$)}-\closest\ and \textsc{($i$)}-\closest\  only differ  if
 $y_i$'s closest neighbor upon arrival is modified by the rounding of the coordinates. We denote this event $\mathcal{C}_i$. This can only happen if  $y_i$ falls within $2/nk$ of the middle of the segment between two consecutive free vertices at iteration $i$. As $y_i\sim \mathcal{U}[0,1]$,
\[
\mathbb{P}\left(\mathcal{C}_i\bigg|\mathcal{U}, y_1,\ldots,y_{i  -1}, |\mathcal{U}|\leq n+2\sqrt{n\ln(n)}\right)\leq \frac{4|\mathcal{U}|}{nk}\leq \frac{8}{k}.
\]
By Azuma-Hoeffding,
\begin{equation*}
\mathbb{P}\left(\sum_{i=1}^{n}\mathds{1}\left\{\mathcal{C}_i\right\}\geq \frac{10n}{k}\bigg||\mathcal{U}|\leq n+2\sqrt{n\ln(n)}\right)\leq e^{-\frac{2n}{k^2}}.
\end{equation*}
Hence, combining with \cref{eq:numpoints}:
\begin{equation}\label{eq:nbmodifiedmatches}
\mathbb{P}\left(\sum_{i=1}^{n}\mathds{1}\left\{\mathcal{C}_i\right\}\geq \frac{10n}{k}\right)\leq e^{-\frac{2n}{k^2}}+\frac{2}{n}.
\end{equation}

In the maximum cardinality matching problem, by \cref{lem:addvertex},  as each modification is equivalent to adding and removing a vertex in the offline set remaining at the step of the modification, each one changes the size of the matching produced at the end by at most two. This gives the first equation of the lemma.

We now turn to bounding the change in cost incurred upon each modification in the metric matching case. Let us assume $\mathcal{C}_i$ holds, and denote $u^{i-1}$ the match of $y_i$  by \textsc{($i-1$)}-\closest\ and $u^i$ the match by \textsc{($i$)}-\closest, i.e. $u^{i-1}$ is $y_i$'s closest free neighbor in the original coordinates, $u^i$ is $y_i$'s closest free neighbor in the rounded ones. In the proof of the lemma, denote $\mathcal{U}_{i-1}\setminus\{u^i\}$ the set of free vertices remaining at iteration $\tau n$ when running \textsc{($i-1$)}-\closest\ with $u^i$ removed. Note that this is the same set as $\mathcal{U}_{i}\setminus\{u^{i-1}\}$, the one obtained when running \textsc{$i$}-\closest\ with $u^{i-1}$ removed. Indeed, up to iteration $i$, both algorithms make the same decisions, and leave both $u^{i-1}$ and $u^i$ free, which means that removing either does not modify the behavior. At iteration $i$, \textsc{($i-1$)}-\closest\ matches $y_i$ with $u^{i-1}$ while \textsc{($i$)}-\closest\ matches $y_i$ with $u^i$. Thus, with $u^{i}$ removed for \textsc{($i-1$)}-\closest\ and $u^{i-1}$ removed for \textsc{($i$)}-\closest\, they both have the same set of free vertices at iteration $i+1$. From this point until the end, they make the same decision, hence $\mathcal{U}_{i-1}\setminus\{u^{i}\}=\mathcal{U}_{i}\setminus\{u^{i-1}\}$. Another implication of this reasoning is:
\[
|\kappa_\textsc{\textsc{($i-1$)}-\closest}\left(\textsc{G}_\rho[\mathcal{U}\setminus\{u^{i}\}, \mathcal{Y}_{\tau n}]\right)-\kappa_\textsc{\textsc{($i$)}-\closest}\left(\textsc{G}_\rho[\mathcal{U}\setminus\{u^{i-1}\}, \mathcal{Y}_{\tau n}]\right)|\leq \big||y_i-u^{i-1}|-|y_i-u^{i}|\big|.
\]

    Let $u_a$ and $u_b$ be the two closest vertices in $\mathcal{U}_{i-1}\setminus\{u^{i}\}$ s.t. $u^{i} \in [u_a;u_b]$. As $u_a$ and $u_b$ also remain unmatched when \textsc{($i$)}-\closest\ is run and matches $y_i$ with $u^{i}$, this also implies  $y^{i}\in [u_a-\frac{1}{n};u_b]$. Hence:
    \[
  |y_i-u^{i}|\leq |u_b-u_a|+\frac{1}{n}.
    \]
    As $u_a$ and $u_b$ remain unmatched without $u^{i}$, adding back $u^{i}$ can only modify the match of the vertices in $\left[u_a-\frac{1}{n};u_b+\frac{1}{n}\right]\cap \mathcal{Y}_{\tau n}$. Moreover, the match of those vertices can only be modified to another vertex within that interval,  i.e.:
   {\small \[
    |\kappa_\textsc{\textsc{($i-1$)}-\closest}\left(\textsc{G}_\rho[\mathcal{U}\setminus\{u^i\}, \mathcal{Y}_{\tau n}]\right)-\kappa_\textsc{\textsc{($i-1$)}-\closest}\left(\textsc{G}_\rho[\mathcal{U}, \mathcal{Y}_{\tau n}]\right)|\leq   \bigg|\left[u_a-\frac{1}{n};u_b+\frac{1}{n}\right]\cap \mathcal{Y}_{\tau n}\bigg|\left(|u_a-u_b|+\frac{2}{n}\right).
    \]}
    By definition of \textsc{($i-1$)}-\closest, all vertices in $]u_a;u_b[\cap \mathcal{U}\setminus\{u^i\}$ are matched to vertices in $\left[\frac{\lfloor n u_a\rfloor}{n};u_b\right]\cap\mathcal{Y}_{\tau n}$, which implies that 
    \begin{align*}
        \bigg|\left]u_a;u_b\right[\cap \mathcal{U}\bigg|-1\leq &\bigg|\left[\frac{\lfloor nu_a\rfloor}{n};\frac{\lfloor nu_a\rfloor+1}{n}\right]\cap\mathcal{Y}_{\tau n}\bigg|+\bigg|\left]u_a;u_b\right[\cap \mathcal{Y}_{\tau n}\bigg|.
    \end{align*}

    Under event $\mathcal{E}_{\mathcal{I}}$ , this implies that $|u_a-u_b|n\leq \alpha \ln(n)+2$, which means:
    \begin{align*}
    \bigg|\left[u_a-\frac{1}{n};u_b+\frac{1}{n}\right]\cap \mathcal{Y}_{\tau n}\bigg|\leq& \tau\left(1+\frac{1-\tau}{2}\right)(\alpha \ln(n)+2) + 12\ln(n)+4\\
        \leq &2\alpha \ln(n).  \end{align*}
Putting the three above inequalities together we obtain:
\begin{align*}
|\kappa_\textsc{\textsc{($i-1$)}-\closest}\left(\textsc{G}_\rho[\mathcal{U}\setminus\{u^i\}, \mathcal{Y}_{\tau n}]\right)-\kappa_\textsc{\textsc{($i-1$)}-\closest}\left(\textsc{G}_\rho[\mathcal{U}, \mathcal{Y}_{\tau n}]\right)|\leq   2\alpha \ln(n)\left(\frac{\alpha \ln(n)+4}{n}\right).
\end{align*}
    
    Repeating the same reasoning with $u^{i-1}$, we get $  |y_i-u^{i-1}|\leq \frac{\alpha \ln(n)+3}{n}$ and:
  \[
|\kappa_\textsc{\textsc{($i$)}-\closest}\left(\textsc{G}_\rho[\mathcal{U}\setminus\{u^{i-1}\}, \mathcal{Y}_{\tau n}]\right)-\kappa_\textsc{\textsc{($i$)}-\closest}\left(\textsc{G}_\rho[\mathcal{U}, \mathcal{Y}_{\tau n}]\right)|\leq   2\alpha \ln(n)\left(\frac{\alpha \ln(n)+4}{n}\right).
\]  
Finally:
\[
|\kappa_\textsc{\textsc{($i-1$)}-\closest}\left(\textsc{G}_\rho[\mathcal{U}, \mathcal{Y}_{\tau n}]\right)-\kappa_\textsc{\textsc{($i$)}-\closest}\left(\textsc{G}_\rho[\mathcal{U}, \mathcal{Y}_{\tau n}]\right)|\leq \left(4\alpha \ln(n)+2\right)\left(\frac{\alpha \ln(n)+4}{n}\right).
\]
Now, in the case where $\mathcal{C}_i$ does not hold, $y_i$ is matched to the same vertex by \textsc{($i-1$)}-\closest\ and \textsc{$i$}-\closest, which means that only the modification of the length of the edge by the rounding has an impact, i.e.:
\[
|\kappa_\textsc{\textsc{($i-1$)}-\closest}\left(\textsc{G}_\rho[\mathcal{U}, \mathcal{Y}_{\tau n}]\right)-\kappa_\textsc{\textsc{($i$)}-\closest}\left(\textsc{G}_\rho[\mathcal{U}, \mathcal{Y}_{\tau n}]\right)|\leq \frac{1}{nk}.
\]
By \cref{eq:nbmodifiedmatches,lem:metricmatchingnbverticeslargeintervals}, for any  $k\geq \frac{2}{\tau(1-\tau)}$,
$\sum_{i=1}^{n}\mathds{1}\left\{\mathcal{C}_i\right\}\leq \frac{10n}{k}$ and $\mathcal{E}_\mathcal{I}$ both hold  with probability at least $1-\exp(-\frac{2n}{k^2})-\frac{2}{n}-\frac{4\alpha+1}{n^2}$, in which case:
\begin{align*}
    &\big|\kappa_\textsc{\closest}\left(\textsc{G}_\rho[\mathcal{U}_k, \mathcal{Y}]\right)(\tau n)-\kappa_\textsc{\closest}\left(\textsc{G}_\rho[\mathcal{U}, \mathcal{Y}]\right)(\tau n)\big|\\&\hspace{4cm}\leq \sum_{i=1}^{\tau n}|\kappa_\textsc{\textsc{($i-1$)}-\closest}\left(\textsc{G}_\rho[\mathcal{U}, \mathcal{Y}_{\tau n}]\right)-\kappa_\textsc{\textsc{($i$)}-\closest}\left(\textsc{G}_\rho[\mathcal{U}, \mathcal{Y}_{\tau n}]\right)|\\
    &\hspace{4cm}\leq \sum_{i=1}^{\tau n}\mathds{1}\left\{\mathcal{C}_i\right\}\left(4\alpha \ln(n)+2\right)\left(\frac{\alpha \ln(n)+4}{n}\right)+\sum_{i=1}^{n}\mathds{1}\left\{\bar{\mathcal{C}_i}\right\}\frac{1}{nk}\\
    &\hspace{4cm}\leq 80\alpha^2\frac{\ln(n)^2}{k}.
\end{align*}

\hfill \(\Box\)

\begin{lemma}[Discard]\label{lem:discard}
With probability at least $1-\frac{4}{n}$, it holds that:
\[
\left|\mu_\closest\left(\textsc{G}[\tilde{\mathcal{U}}_k, \mathcal{Y},n,c]\right)(\tau n)-\mu_\closest\left(\textsc{G}[\mathcal{U}_k, \mathcal{Y},n,c]\right)(\tau n)\right|\leq \frac{n}{k}+ 4\sqrt{n\ln(n)}.
\]
For any  $k\geq \frac{2}{\tau(1-\tau)}$, with probability at least $1-\frac{4}{n}-\frac{4\alpha+1}{n^2}$,
\[
\big|\kappa_\textsc{\closest}\left(\textsc{G}_\rho[\tilde{\mathcal{U}}_k, \mathcal{Y}]\right)(\tau n)-\kappa_\textsc{\closest}\left(\textsc{G}_\rho[\mathcal{U}_k, \mathcal{Y}]\right)(\tau n)\big|\leq \left(\frac{1}{2k}+ 4\sqrt{\frac{\ln(n)}{n}}\right)4(\alpha\ln(n))^2.
\]
\end{lemma}

\textit{Proof:}
For any $\ell \in [kn]$, let 
$$
n_{\ell}:=\left\{\frac{\lfloor unk\rfloor}{nk}=\frac{\ell}{nk}\ \bigg|\ u \in \mathcal{U}\right\}.
$$
The points in $\mathcal{U}$ are generated through a Poisson point process of intensity $n$ in $[0,1]$, thus, for any $\ell \in [nk]$:
$$
\mathbb{P}(n_{\ell}>0) = 1-e^{-1/k},
$$
and the $(n_{\ell})_{\ell \in [nk]}$ are independent of each other. The number of  points in $\tilde{\mathcal{U}}_k$ is
$$
|\tilde{\mathcal{U}}_k|= \sum_{\ell=1}^{nk}\mathds{1}_{\{n_{\ell}>0\}}.
$$
We have
\begin{align*}
  \mathbf{E}\left[|\tilde{\mathcal{U}}_k|\right]=&nk(1-e^{-1/k})\\
  \geq& n-\frac{n}{2k}.
\end{align*}
By Chernoff bound, 
\begin{align}
\mathbb{P}\left(|\tilde{\mathcal{U}}_k|\leq n-\frac{n}{2k}-2\sqrt{n\ln(n)}\right)\leq \frac{2}{n}.\label{eq:lowerboundUk}
\end{align}
The number of removed points is:
\[
n_{\text{removed}}:=|\mathcal{U}|- |\tilde{\mathcal{U}}_k|.
\]

We have already obtained by concentration of Poisson random variables
\begin{align*}
    \mathbb{P}\left(\big|n-|\mathcal{U}|\big|\geq 2\sqrt{n\ln(n)}\right)\leq&\frac{2}{n}.
\end{align*}
Thus,
$$
\mathbb{P}\left(n_{\text{removed}}\geq 4\sqrt{n\ln(n)}+\frac{n}{2k}\right) \leq \frac{4}{n}.
$$
Combining this with Lemma \ref{lem:addvertex} terminates the proof of the first equation.
For the second, we proceed in the same way as in the proof of Lemma \ref{lem:poissonization}: using this time the lower bound on the number of vertices in $\tilde{\mathcal{U}}_k$ per interval, under event $\mathcal{E}_{\mathcal{I}}$, each vertex deletion modifies the length of the matching by  at most $\frac{4(\alpha\ln(n))^2}{n}$.
 \hfill \(\Box\)

\begin{lemma}[Gluing]\label{lem:gluing}
With probability at least $1-\exp(-\frac{\sqrt{n}}{4c})$, it holds that:
$$
\left|\mu_\closest\left(\textsc{G}[\tilde{\mathcal{U}}_k, \mathcal{Y},n,c]\right)(\tau n)-\mu_\closest\left(\textsc{G}^{\text{glued}}[\tilde{\mathcal{U}}_k, \mathcal{Y},n,c]\right)\right|\leq 2 \sqrt{n}+4c.
$$
For any  $k\geq \frac{2}{\tau(1-\tau)}$, with probability at least $1-\frac{4\alpha+1}{n^2}$:
\[
\left|\kappa_\textsc{\closest}\left(\textsc{G}_\rho[\tilde{\mathcal{U}}_k , \mathcal{Y}]\right)(\tau n)-\kappa_\textsc{\closest}\left(\textsc{G}^{\text{glued}}_\rho[\tilde{\mathcal{U}}_k , \mathcal{Y}]\right)(\tau n)\right|\leq \frac{16\alpha^3\ln(n)^3}{n}.
\]
\end{lemma}
\textit{Proof:} We first consider the maximum cardinality matching setting. The match of some vertex $y \in \mathcal{Y}$ may be modified by the gluing step only if $y<\frac{c}{n}$ or $y>1-\frac{c}{n}$. Let
\[
n_{\text{gluing}}:=\sum_{y \in \mathcal{Y}}\mathds{1}_{\{\text{the match of $y$ is modified during the gluing step}\}}.
\]
By Chernoff bound, 
\[
\mathbb{P}(n_{\text{gluing}}\geq 2c+\sqrt{n})\leq e^{-\frac{\sqrt{n}}{4c}}.
\]
Lemma \ref{lem:addvertex} concludes the proof of the first inequality, as each modification is equivalent to adding and removing a vertex in the offline set remaining at the step of the modification.

Let us now prove the second inequality. Consider algorithm \textsc{($i$)}-\closest'\ which matches the incoming vertex to the closest vertex in the glued space up to iteration $i$ (included), then to the closest on the line.  Denote $\mathcal{C}'_i$ the event that $y_i$'s closest neighbor is modified by the gluing step. This can happen only if $y_i$ lands between the remaining vertex of highest and smallest coordinates at iteration $i$. This interval is included in the interval $[u_h,u_\ell]$ between the remaining vertices of highest and smallest coordinates at iteration $n \tau$ when running the \closest\ algorithm with the glued coordinates at every iteration. This implies:
\[
\sum_{i=1}^{n\tau}\mathds{1}\{\mathcal{C}'_i\}\leq \left|\mathcal{Y}_{\tau }\cap(u_h,u_\ell)\right|.
\]

By definition of \closest, all vertices in $\mathcal{U}\cap(u_h,u_\ell)$ are matched to vertices in $\mathcal{Y}_{\tau }\cap[u_h,u_\ell]$, hence:
\[
\mathcal{U}\cap(u_h,u_\ell)\leq \mathcal{Y}_{\tau }\cap(u_h,u_\ell)
\]

By the same reasoning we employed before, under event $\mathcal{E}_{\mathcal{I}}$:
\[
|[u_h,u_\ell]|\leq \frac{\alpha\ln(n)+2}{n}\text{ and } |\mathcal{Y}_{\tau n}\cap[u_h,u_\ell]|\leq \tau\left(1+\frac{1-\tau}{2}\right)(\alpha \ln(n)+2) + 6\ln(n)+2\leq 2\alpha \ln(n).
\]
Hence, under $\mathcal{E}_{\mathcal{I}}$:
\[
\sum_{i=1}^{n\tau}\mathds{1}\{\mathcal{C}'_i\}\leq 2\alpha \ln(n).
\]

Let us now bound the impact of each event $\mathcal{C}'_i$. This step of the proof is very similar to the proof of \cref{lem:rounding}, in which the arguments are fleshed out in more detaims. Assume that $\mathcal{C}'_i$ holds, and denote $u^{i-1}$ the match of $y_i$  by \textsc{($i-1$)}-\closest'\ and $u^i$ the match by \textsc{($i$)}-\closest', i.e. $u^{i-1}$ is $y_i$'s closest free neighbor in the glued coordinates, $u^i$ is $y_i$'s closest free neighbor in the unglued ones. 
We have:
\[
|\kappa_\textsc{\textsc{($i-1$)}-\closest'}\left(\textsc{G}_\rho[\mathcal{U}\setminus\{u^i\}, \mathcal{Y}_{\tau n}]\right)-\kappa_\textsc{\textsc{($i$)}-\closest'}\left(\textsc{G}_\rho[\mathcal{U}\setminus\{u^{i-1}\}, \mathcal{Y}_{\tau n}]\right)|\leq \big||y_i-u^i|-|y_i-u^{i-1}|\big|.
\]

Denote $\mathcal{U}_{i-1}\setminus\{u^{i}\}$ the set of free vertices remaining at iteration $\tau n$ when running \textsc{($i-1$)}-\closest'\ with $u^{i}$ removed, which is the same set as $\mathcal{U}_{i}\setminus\{u^{i-1}\}$, the one obtained when running \textsc{$i$}-\closest'\ with $u^{i-1}$ removed. Let $u_a$ and $u_b$ be the two closest vertices in $\mathcal{U}_{i-1}\setminus\{u^{i}\}$ s.t. $u^i \in [u_a;u_b]$. We have:
   {\small \[
    |\kappa_\textsc{\textsc{($i-1$)}-\closest'}\left(\textsc{G}_\rho[\mathcal{U}\setminus\{u^i\}, \mathcal{Y}_{\tau n}]\right)-\kappa_\textsc{\textsc{($i-1$)}-\closest'}\left(\textsc{G}_\rho[\mathcal{U}, \mathcal{Y}_{\tau n}]\right)|\leq   \bigg|\left[u_a;u_b\right]\cap \mathcal{Y}_{\tau n}\bigg|\times |u_a-u_b|.
    \]}
    By definition of \textsc{($i-1$)}-\closest, all vertices in $]u_a;u_b[\cap \mathcal{U}\setminus\{u^i\}$ are matched to vertices in $\left[u_a;u_b\right]\cap\mathcal{Y}_{\tau n}$, which implies that 
    \begin{align*}
        \bigg|\left]u_a;u_b\right[\cap \mathcal{U}\bigg|-1\leq &\bigg|\left]u_a;u_b\right[\cap \mathcal{Y}_{\tau n}\bigg|.
    \end{align*}

    Under event $\mathcal{E}_{\mathcal{I}}$ , this implies that $|u_a-u_b|n\leq \alpha \ln(n)+2$ and:
    \begin{align*}
    \bigg|\left[u_a-\frac{1}{n};u_b+\frac{1}{n}\right]\cap \mathcal{Y}_{\tau n}\bigg|\leq& \tau\left(1+\frac{1-\tau}{2}\right)(\alpha \ln(n)+2) + 12\ln(n)+4,\\
        \leq &2\alpha \ln(n).  \end{align*}
Putting the three above inequalities together we obtain $|y_i-u^i|\leq \alpha \ln(n)+2$ and:
\begin{align*}
|\kappa_\textsc{\textsc{($i-1$)}-\closest'}\left(\textsc{G}_\rho[\mathcal{U}\setminus\{u^i\}, \mathcal{Y}_{\tau n}]\right)-\kappa_\textsc{\textsc{($i-1$)}-\closest'}\left(\textsc{G}_\rho[\mathcal{U}, \mathcal{Y}_{\tau n}]\right)|\leq   2\alpha \ln(n)\left(\frac{\alpha \ln(n)+2}{n}\right).
\end{align*}
    
    Repeating the same reasoning with $u^{i-1}$, we get $|y_i-u^{i-1}|\leq \alpha \ln(n)+2$ and:
  \[
|\kappa_\textsc{\textsc{($i$)}-\closest'}\left(\textsc{G}_\rho[\mathcal{U}\setminus\{u^{i-1}\}, \mathcal{Y}_{\tau n}]\right)-\kappa_\textsc{\textsc{($i$)}-\closest'}\left(\textsc{G}_\rho[\mathcal{U}, \mathcal{Y}_{\tau n}]\right)|\leq   2\alpha \ln(n)\left(\frac{\alpha \ln(n)+2}{n}\right).
\]  
Finally:
\[
|\kappa_\textsc{\textsc{($i-1$)}-\closest'}\left(\textsc{G}_\rho[\mathcal{U}, \mathcal{Y}_{\tau n}]\right)-\kappa_\textsc{\textsc{($i$)}-\closest'}\left(\textsc{G}_\rho[\mathcal{U}, \mathcal{Y}_{\tau n}]\right)|\leq \left(4\alpha \ln(n)+2\right)\left(\frac{\alpha \ln(n)+2}{n}\right).
\]
Now, in the case where $\mathcal{C}_i$ does not hold, $y_i$ is matched to the same vertex by \textsc{($i-1$)}-\closest'\ and \textsc{$i$}-\closest', which means that the cost is not modified.
By \cref{lem:metricmatchingnbverticeslargeintervals}, for any  $k\geq \frac{2}{\tau(1-\tau)}$,
and $\mathcal{E}_\mathcal{I}$ both hold  with probability at least $1-\frac{4\alpha+1}{n^2}$, in which case:
\begin{align*}
    &\big|\kappa_\textsc{\closest}\left(\textsc{G}_\rho[\mathcal{U}_k, \mathcal{Y}]\right)(\tau n)-\kappa_\textsc{\closest}\left(\textsc{G}^\text{glued}_\rho[\mathcal{U}_K, \mathcal{Y}]\right)(\tau n)\big|\\&\hspace{4cm}\leq \sum_{i=1}^{n}|\kappa_\textsc{\textsc{($i-1$)}-\closest'}\left(\textsc{G}_\rho[\mathcal{U}, \mathcal{Y}_{\tau n}]\right)-\kappa_\textsc{\textsc{($i$)}-\closest'}\left(\textsc{G}_\rho[\mathcal{U}, \mathcal{Y}_{\tau n}]\right)|\\
    &\hspace{4cm}\leq \sum_{i=1}^{n}\mathds{1}\left\{\mathcal{C}_i\right\}\left(4\alpha \ln(n)+2\right)\left(\frac{\alpha \ln(n)+4}{n}\right)\\
    &\hspace{4cm}\leq 2\alpha \ln(n)\left(4\alpha \ln(n)+2\right)\left(\frac{\alpha \ln(n)+4}{n}\right)\leq \frac{16\alpha^3\ln(n)^3}{n}.
\end{align*}

\hfill \(\Box\)

\subsection{Proof of Lemma \ref{lem:gaps_repartition} (Gaps repartition)}\label{SE:gaps_repartition}

\hfill \break

Denote $|H|$ the length of a sequence of integers  $H$ and $\mathfrak{S}([n])$ the ensemble of all permutations over $[n]$. For a sequence $H=(h_i)_{i\in [|H|]}$ let  $\mathcal{A}_t\left(H\right)$ be the event that $H$ is the sequence of the values of the gaps at time $t$:
\[
\left\{n_t=|H|,\exists \sigma \in \mathfrak{S}([n_t]) \text{ s.t.}\ \forall i \in [n_t],u_t(i+1)-u_t(i) = \frac{h_{\sigma(i)}}{nk}\right\},
\]
and $\mathcal{A}_t\left(\sigma,H\right)$  the event that $H$ is the sequence of the values of the gaps, and that the gaps follow a particular order at time $t$:
\[
\left\{ n_t=|H|, \forall i \in [n_t],u_t(i+1)-u_t(i) = \frac{h_{\sigma(i)}}{nk}\right\}.
\]
 Note that $\mathcal{A}_t\left(H\right)= \cup_{\sigma \in \mathfrak{S}([H])}\mathcal{A}_t\left(\sigma,H\right).$  For a list of sequences $H_1, \ldots,H_t$, we denote $\mathcal{A}_{1:t}\left(H_{1:t}\right)$ the event that $A_s(H_s)$ hold for all $s\leq t$:
\[
\mathcal{A}_{1:t}\left(H_{1:t}\right):= \cap_{1\leq s \leq t} A_s(H_s).
\]

Lemma \ref{lem:gaps_repartition} is a consequence of the following stronger Lemma.
\begin{lemma}\label{lem:gaps_repartition_aux}
For any iteration $t$, any list of sequences $H_1, \ldots,H_t$, any two permutations $\sigma,\sigma'\in \mathfrak{S}\left(\left[|H_t|\right]\right)$,
\begin{equation}
\mathbb{P}\left(\mathcal{A}_t\left(\sigma,H_t\right)\ \bigg|\ \mathcal{A}_{1:t}\left(H_{1:t}\right)\right)=\mathbb{P}\left(\mathcal{A}_t\left(\sigma',H_t\right)\ \bigg|\ \mathcal{A}_{1:t}\left(H_{1:t}\right)\right).
\label{eq:equiprobaallt}
\end{equation}
\end{lemma}
\textit{Proof}:  We prove this lemma by induction. 
Let $H=(h_i)_{i\in[|H|]}$ be any sequence of integers s.t. 
$\sum_{i=1}^{|H|} \frac{h_i}{nk} = 1. $ Since $u_0(1)=0$,  the knowledge of the sizes and the ordering of the gaps determines the position of the points. Thus, for any $\sigma \in \mathfrak{S}\left(\left[|H|\right]\right)$:
\begin{align*}
\mathbb{P}\left( \forall i \in [|H|],\ u_0(i+1)-u_0(i) = \frac{h_{\sigma(i)}}{nk}\right)=p_k^{|H|-1}(1-p_k)^{nk-|H|}.
\end{align*}
This does not depend on the choice of permutation $\sigma$. Thus for any sequence $H$ s.t. $\sum_{i=1}^{|H|} \frac{h_i}{nk} = 1 $ and any two permutations $(\sigma, \sigma') \in \mathfrak{S}\left(\left[|H|\right]\right)^2$:
\begin{equation}
\label{eq:equiproba_ini}
    \mathbb{P}\left(\mathcal{A}_0\left(\sigma,H\right)\bigg|\mathcal{A}_0\left(H\right)\right)=\mathbb{P}\left(\mathcal{A}_0\left(\sigma',H\right)\bigg|\mathcal{A}_0\left(H\right)\right).
\end{equation}

Thus,  \cref{eq:equiprobaallt} holds at iteration $0$. Let us assume that  \cref{eq:equiprobaallt} holds at all iterations until the $t$-th one. We will show this implies that it also holds at iteration $t+1$. There are two cases possible, depending on whether or not a vertex is matched. 

We first show the implication in the case where the incoming vertex is not matched.  We have:

\begin{align*}
    \mathbb{P}\left(\mathcal{A}_{t+1}\left(\sigma,H_t\right)\ \bigg|\ \mathcal{A}_{1:t}\left(H_{1:t}\right)\right)= \mathbb{P}\left(y_t \text{ is not matched},\mathcal{A}_{t}\left(\sigma,H_t\right)\ \bigg|\ \mathcal{A}_{1:t}\left(H_{1:t}\right)\right).
\end{align*}

The incoming vertex at iteration $t$, $y_t$, is not matched iff it lays at a distance larger than $c/n$ of any free vertex. This depends only on the lengths between the free vertices, not their ordering. Hence, by the induction hypothesis, the right hand side of the above equation does not depend on $\sigma$, so neither does the left hand side.

We now turn to the case where $y_t$ is matched. Given a sequence $H$ and a couple $j<j'\leq|H|$, we define an admissible sequence for sequence $H$ as a sequence $\tilde{H}(j, j')$ s.t. for any $i \in [|H]-1]$
$$
\tilde{h}^{j,j'}_i = \begin{cases} h_{i} &\text{ if } i<j' \text{ and } i \neq j\\
    h_{j}+ h_{j'}&\text{ if } i=j\\
    h_{i+1}  &\text{ if } i\geq j'.\\
    \end{cases}
$$
Note it is the sequence of gaps obtained  when event $\mathcal{A}_t\left(H\right)$ is true and a vertex $u_t(i)$ with $u_t(i)-u_t(i-1)=h_{j}$ and $u_t(i+1)-u_t(i)=h_{j'}$ is matched. For any $H_{1:t}$, any $j<j'\leq |H_t|$, any $\sigma \in \mathfrak{S}([|H_t|-1])$ define events:
\begin{align*}\mathcal{B}_t(H_t,i,j,j')=&\left\{u_t(i)-u_t(i-1)=\frac{h_{j}}{nk} \text{ and } u_t(i+1)-u_t(i)=\frac{h_{j'}}{nk}\right\}\\
&\cup\left\{u_t(i)-u_t(i-1)=\frac{h_{j'}}{nk} \text{ and } u_t(i+1)-u_t(i)=\frac{h_{j}}{nk}\right\},
\end{align*}
the event that gaps around $u_t(i)$ take values $\frac{h_{j'}}{nk}$ and $\frac{h_{j}}{nk}$. Also define
\begin{align*}\mathcal{C}_t(\sigma,H_t,i,j,j')=&\{\forall k<i-1, u_t(k+1)-u_t(k)=\frac{\tilde{h}^{j,j'}_{\sigma(k)}}{nk}\}\cap\{ \forall k>i, u_t(k+1)-u_t(k)=\frac{\tilde{h}^{j,j'}_{\sigma(k-1)}}{nk}\},
\end{align*}
the event that the rest of the gaps follow a particular order.
Denote $\mathcal{E}_t(\sigma,H_t,i,j,j')=\mathcal{B}_t(H_t,i,j,j')\cap\mathcal{C}_t(\sigma,H_t,i,j,j')$.
For any $H_{1:t}$, any $j<j'\leq |H_t|$, any $\sigma \in \mathfrak{S}([|H_t|-1])$, it holds that:
\begin{align*}
    \mathrm{P}&\left(\mathcal{A}_{t+1}\left(\sigma,\tilde{H}_t(j,j')\right)\ \bigg|\ \mathcal{A}_{1:t}\left(H_{1:t}\right)\right)= \sum_{i \text{ s.t. }\tilde{h}_{\sigma(i-1)}=\tilde{h}_j}\mathrm{P}\left(u_t(i) \text{ is matched},\mathcal{E}_t(\sigma,H_t,i,j,j')\ \bigg|\ \mathcal{A}_{1:t}\left(H_{1:t}\right)\right)\\
    &\hspace{4cm}=\frac{\min(c,\frac{h_{j}}{2k})+\min(c,\frac{h_{j'}}{2k})}{n}\sum_{i \text{ s.t. }\tilde{h}_{\sigma(i-1)}=\tilde{h}_j}\mathrm{P}\left(\mathcal{E}_t(\sigma,H_t,i,j,j')\ \bigg|\ \mathcal{A}_{1:t}\left(H_{1:t}\right)\right).
\end{align*}

By the induction property, the right term does not depend on $\sigma$. This implies that the induction property remains true when a vertex is matched as well, which implies that Equation \eqref{eq:equiprobaallt} holds for all $t\in [n]$.\hfill \(\Box\)

We now show that Lemma \ref{lem:gaps_repartition_aux} implies Lemma \ref{lem:gaps_repartition}. Let $\mathcal{F}_t$ be the event associated with the values $\left(F_{k,n}(\ell,t')\right)_{\ell \in [nk]}$ for all $t'\leq t$, which also determines the value of $n_t$. Let $H$ be any sequence s.t. $\mathcal{A}_t(H)$ holds. Equation \eqref{eq:equiprobaallt} implies:
\begin{align*}
    \mathrm{E}\left[M_{k,n}(\ell_-,\ell_+,t) \bigg | \mathcal{F}_t \right] =&\sum_{i=1}^{n_t}\frac{1}{|\mathfrak{S}([n_t])|}\sum_{\sigma \in \mathfrak{S}([n_t])}\mathds{1}_{\{h_{\sigma(i)}=\ell_-, h_{\sigma(i+1)}=\ell_+\}}\\
     =&\sum_{i=1}^{n_t}\frac{F_{k,n}(\ell_-,t)}{n_t}\left(\mathds{1}_{\{\ell_-\neq \ell_+\}}\frac{F_{k,n}(\ell_+,t)}{n_t-1}+\mathds{1}_{\{\ell_-= \ell_+\}}\frac{F_{k,n}(\ell_+,t)-1}{n_t-1}\right)\\
    =&F_{k,n}(\ell_-,t)\left(\mathds{1}_{\{\ell_-\neq \ell_+\}}\frac{F_{k,n}(\ell_+,t)}{n_t-1}+\mathds{1}_{\{\ell_-= \ell_+\}}\frac{F_{k,n}(\ell_+,t)-1}{n_t-1}\right).
\end{align*}

 \hfill \( \Box \)

 \subsection{Proofs of the technical lemmas for section \ref{sec:01matchingDEM}\\}\label{sec:proof01matchingDEM}

 As mentioned in \ref{sec:01matchingDEM}, some computations are necessary before proving \cref{lem:resinter}. The goal is to check all the conditions listed in \cref{thm:Wormald}. We first establish the trend condition, which will be a consequence of \cref{lem:evollaw} after some computations. Denote
\[
r_{k,n}(\tau)=\sum_{\ell= kn^{1/8}}^{+\infty}f_{k}(\ell,\tau) \text{ and } \tilde{r}_{k,n}(\tau)= \sum_{\ell= kn^{1/8}}^{+\infty}\min\left(\frac{\ell}{k},2c\right)f_{k}(\ell,\tau). 
\]
Define the function $\Phi_{k,n}:\mathbb{R}^{kn^{1/8}+1}\rightarrow \mathbb{R}^{kn^{1/8}}$ as:

\begin{align*}
        \Phi_{k,n}^{\ell} \left(\tau,\left(y_\ell\right)_{\ell=1}^{kn^{1/8}}\right) =&-\frac{\sum_{\ell'=1}^{kn^{1/8}}\min(\frac{\ell'}{k},2c)y_{\ell'}+\tilde{r}_{k,n}(\tau)}{\sum_{\ell'=1}^{kn^{1/8}}y_{\ell'}+r_{k,n}(\tau)}y_\ell-\min\left(\frac{\ell}{k},2c\right)y_\ell\\
   &+\frac{\sum_{\ell'=0}^{\ell}\min(\frac{\ell'}{k},2c)y_{\ell'}y_{\ell-\ell'}}{\sum_{\ell'=1}^{kn^{1/8}}y_{\ell'}+r_{k,n}(\tau)},
 \end{align*}

and consider the domain:
\[
\mathcal{D}:= \left\{(a_\ell)_{\ell=1}^{kn^{1/8}} \text{ s.t. } \frac{1}{2}e^{-4c}\leq \sum_\ell a_\ell\leq 2, a_\ell\in [0,1] \forall \ell \right\}.
\]
 \begin{lemma}[Trend, maximum cardinality matching]\label{lem:evollawfinal}
 It holds that for any $t\leq n$:
     \begin{align*}
   & \sum_{\ell=1}^{kn^{1/8}}\left|\mathbb{E}\left[F_{k,n}(\ell,t+1)-F_{k,n}(\ell,t) \bigg | \mathcal{F}_t,\left(\frac{F_{k,n}(\ell,t)}{n}\right)_{\ell=1}^{kn^{1/8}}\in \mathcal{D}\right]-\Phi_{k,n}^{\ell} \left(\frac{t}{n},\left(\frac{F_{k,n}(\ell,t)}{n}\right)_{\ell=1}^{kn^{1/8}}\right) \right|\nonumber\\
   &\hspace{9cm}\leq \delta_1 n^{-1/8}, 
\end{align*}
with $\delta_1$ a constant independent of $n$ and $k$. 

 \end{lemma}

 \textit{Proof}: This proof is lengthy, but purely computational. In this proof, we use the shorthand $a_\ell= \frac{F_{k,n}(\ell,t)}{n}$ and $\tau =t/n$. Restating the first result of  \cref{lem:evollaw}, we have, for all $t \in [n]$, for all $\ell\in [kn]$,
\begin{align}\label{eq:restateevollaw}
    \bigg|A_{\ell,t}-\mathrm{E}\left[F_{k,n}(\ell,t+1)-F_{k,n}(\ell,t) \bigg | \mathcal{F}_t,n_t\geq w_1 n \right]\bigg|\leq \frac{6}{w_1n}.
\end{align}
Rewriting $A_{\ell,t}$ with the shorthand $a_\ell$, we get:
\[
A_{\ell,t} = \frac{\sum_{\ell'=1}^{kn}\min(2c,\frac{\ell'}{k}) a_{\ell'}}{n_t/n}a_\ell-\min\left(2c,\frac{\ell}{k}\right)a_\ell+\sum_{\ell'=0}^\ell\frac{\min(2c,\frac{\ell'}{k})a_{\ell'}a_{\ell-\ell'}}{n_t/n}.
\]

Hence, for any $\ell \leq kn^{1/8}$, we have:
\begin{align*}
    \left|A_{\ell,t}-\Phi_{k,n}^{\ell} \left(\tau,\left(a_\ell\right)_{\ell=1}^{kn^{1/8}}\right) \right|\leq 
    &\underbrace{\left|\frac{\sum_{\ell'=1}^{kn}\min(2c,\frac{\ell'}{k}) a_{\ell'}}{n_t/n}-\frac{\sum_{\ell'=1}^{kn^{1/8}}\min\left(\frac{\ell'}{k},2c\right)a_{\ell'}+\tilde{r}_{k,n}(\tau)}{\sum_{\ell'=1}^{kn^{1/8}}a_{\ell'}+r_{k,n}(\tau)}\right|}_{(i)}a_\ell\\
    &+2c\underbrace{\left| \frac{1}{n_t/n}-\frac{1}{\sum_{\ell'=1}^{kn^{1/8}}a_{\ell'}+r_{k,n}(\tau)}\right|}_{(ii)}\sum_{\ell'=0}^{\ell}a_{\ell'}a_{\ell-\ell'}
\end{align*}

We start by bounding $(ii)$. Recall that for any $t\leq n$, $\sum_{\ell=1}^{nk}\frac{\ell}{k}a_\ell=1$. Some computations on $f_k$ deferred to the appendix (\cref{lem:total_length}) show that $ \sum_{\ell=1}^{+\infty}\frac{\ell}{k}f_k(\ell,t)=1$. Hence:

\begin{equation}\label{eq:boundendsum}
\sum_{\ell= kn^{1/8}}^{nk}a_\ell\leq n^{-1/8} \quad \text{ and } \quad r_{k,n}(\tau)\leq n^{-1/8}.
\end{equation}
Those inequalities are key to the proof. At the high level, both $\sum_{\ell= kn^{1/8}}^{nk}a_\ell \quad \text{ and } \quad r_{k,n}(\tau)$ are so small that approximating one by the other results in a small overall change. The rest of the proof is dedicated to making that intuition formal.
By \cref{eq:boundendsum},  $|\sum_{\ell= kn^{1/8}}^{nk}a_\ell-r_{k,n}(t)|\leq 2n^{-1/8}$. If $(a_\ell)_{\ell=1}^{kn^{1/8}}\in \mathcal{D}$, $\sum_{\ell=1}^{kn^{1/8}}a_\ell\geq  \frac{1}{2}e^{-4c}$, therefore, 
\begin{equation}\label{eq:boundN_tevollaw}
   (ii)=\left| \frac{1}{\sum_{\ell=1}^{kn^{1/8}}a_\ell+\sum_{\ell= kn^{1/8}}^{nk}a_\ell} -\frac{1}{\sum_{\ell=1}^{kn^{1/8}}a_\ell+r_{k,n}(\tau)}\right|\leq w_2 n^{-1/8},
\end{equation}
with $w_2$ a constant that does not depend on $n$.

We now bound $(i)$. First,  $\sum_{\ell'}\min\left(\frac{\ell'}{k},2c\right)a_{\ell'} \leq 1$. By Equation \ref{eq:boundN_tevollaw}, this entails:
\begin{align*}
(i)\leq& \left|\frac{\sum_{\ell'}\min\left(\frac{\ell'}{k},2c\right)a_{\ell'}}{\sum_{\ell'=1}^{kn^{1/8}}a_{\ell'}+r_{k,n}(\tau)}-\frac{\sum_{\ell'=1}^{kn^{1/8}}\min\left(\frac{\ell'}{k},2c\right)a_{\ell'}+\tilde{r}_{k,n}(\tau)}{\sum_{\ell'=1}^{kn^{1/8}}a_{\ell'}+r_{k,n}(\tau)}\right|+w_2n^{-1/8}\\
=&\left|\frac{\sum_{\ell'=kn^{1/8}}^{nk}\min\left(\frac{\ell'}{k},2c\right)a_{\ell'}}{\sum_{\ell'=1}^{kn^{1/8}}a_{\ell'}+r_{k,n}(\tau)}-\frac{\tilde{r}_{k,n}(\tau)}{\sum_{\ell'=1}^{kn^{1/8}}a_{\ell'}+r_{k,n}(\tau)}\right|+w_2n^{-1/8}\\
\leq &2e^{4c}\left|\sum_{\ell'=kn^{1/8}}^{nk}\min\left(\frac{\ell'}{k},2c\right)a_{\ell'}-\tilde{r}_{k,n}(\tau)\right|+w_2n^{-1/8}\\
\leq& \left(4e^{4c}+w_2\right)n^{-1/8}.
\end{align*}

The third inequality holds as $\sum_{\ell'=1}^{kn^{1/8}}a_{\ell'}\geq 0.5e^{-4c}$ when $(a_\ell)_{\ell=1}^{kn^{1/8}}\in \mathcal{D}$, and the last one by  \cref{eq:boundendsum}. Combining the bounds on $(i)$ and $(ii)$, we obtain:
\begin{align*}
   &  \left|A_{\ell,t}-\Phi_{k,n}^{\ell} \left(\tau,\left(a_\ell\right)_{\ell=1}^{kn^{1/8}}\right) \right|\leq 
   \left(4e^{4c}+w_2\right)n^{-1/8}a_\ell+2c w_2n^{-1/8}\sum_{\ell'=0}^{\ell}a_{\ell'}a_{\ell-\ell'}.
\end{align*}
By Young's convolution inequality, $\sum_{\ell}\sum_{\ell'=0}^{\ell}a_{\ell'}a_{\ell-\ell'}\leq (\sum_{\ell}a_{\ell})^2\leq 1$, so:
\begin{align*}
   & \sum_{\ell=1}^{kn^{1/8}}\left|A_{\ell,t}-\Phi_{k,n}^{\ell} \left(\tau,\left(a_\ell\right)_{\ell=1}^{kn^{1/8}}\right) \right|\leq  \left(4e^{4c}+(2c+1)w_2\right)n^{-1/8}
\end{align*}
We now use \cref{eq:restateevollaw} with $w_1=\frac{1}{2}e^{-4c}$ and obtain:

\begin{align*}
    \sum_{\ell=1}^{kn^{1/8}}\left|\mathrm{E}\left[F_{k,n}(\ell,t+1)-F_{k,n}(\ell,t) \bigg | \mathcal{F}_t,\left(a_\ell\right)\in\mathcal{D}\right]-\Phi_{k,n}^{\ell} \left(\tau,\left(a_\ell\right)_{\ell=1}^{kn^{1/8}}\right) \right|\leq& \frac{4e^{4c}+(2c+1)w_2}{n^{1/8}}+ \frac{24e^{4c}}{n}\\
    \leq& \frac{\delta_1}{n^{1/8}}. 
\end{align*}
Note that $\delta_1$ depends on $c$ but not on $n$ or $k$.

\hfill \(\Box\)

We now prove $\Phi_{k,n}$ is Lipschitz on $\mathcal{D}$.

\begin{lemma}[Lipschitz, maximum cardinality matching]\label{lem:lipshcitz}
 On domain $\mathcal{D}$, the $\Phi_{k,n}$ is $L_c$-lipshcitz with respect to the $L_1$ norm, with $L_c$ a constant independent of $n$. Moreover, for any sequence $(a_\ell)_{\ell=1}^{kn^{1/8}} \in \mathcal{D}$, $\forall \ell' \leq kn^{-1/8}$, $\forall \tau \in [0,1]$,
\[
|\Phi^{\ell'}_{k,n}\left(\tau,(a_\ell)_{\ell=1}^{kn^{1/8}} \right)|\leq 16c.
\]
\end{lemma}
\textit{Proof}:  Again, the proof is purely computational. First, 
the bound on $\Phi^{\ell'}_{k,n}\left(\tau,(a_\ell)_{\ell=1}^{kn^{1/8}} \right)$ is obtained by upper bounding $\min(\frac{\ell}{k},2c)$ with $2c$ in the definition.

Consider two sequences $\mathbf{a}=(a_\ell)_{\ell=1}^{kn^{1/8}}$ and $\mathbf{b}=(b_\ell)_{\ell=1}^{kn^{1/8}}$ in $\mathcal{D}$. 
\begin{align*}
        & ||\Phi_{k,n} \left(\tau,\mathbf{a}\right) -\Phi_{k,n} \left(\tau,\mathbf{b}\right)||_1\leq 2c||\mathbf{a}-\mathbf{b}||_1\\
        &\hspace{1cm}+\underbrace{\sum_{\ell}\left|\frac{\sum_{\ell'=1}^{kn^{1/8}}\min(\frac{\ell'}{k},2c)a_{\ell'}+\tilde{r}_{k,n}(\tau)}{\sum_{\ell'=1}^{kn^{1/8}}a_{\ell'}+r_{k,n}(\tau)}a_\ell-\frac{\sum_{\ell'=1}^{kn^{1/8}}\min(\frac{\ell'}{k},2c)b_{\ell'}+\tilde{r}_{k,n}(\tau)}{\sum_{\ell'=1}^{kn^{1/8}}b_{\ell'}+r_{k,n}(\tau)}b_\ell\right|}_{(a)}\\
   &\hspace{1cm}+\underbrace{\sum_{\ell}\left|\frac{\sum_{\ell'=0}^{\ell}\min(\frac{\ell'}{k},2c)a_{\ell'}a_{\ell-\ell'}}{\sum_{\ell'=1}^{kn^{1/8}}a_{\ell'}+r_{k,n}(\tau)}-\frac{\sum_{\ell'=0}^{\ell}\min(\frac{\ell'}{k},2c)b_{\ell'}b_{\ell-\ell'}}{\sum_{\ell'=1}^{kn^{1/8}}b_{\ell'}+r_{k,n}(\tau)}\right|}_{(b)}.
 \end{align*}

Before bounding $(a)$ and $(b)$, we do some preliminary computations. Assume w.l.o.g. $\sum_{\ell=1}^{kn^{1/8}}b_\ell\geq\sum_{\ell=1}^{kn^{1/8}}a_\ell $. We have:
\begin{align*}
\frac{1}{\sum_{\ell}a_\ell+ r_{k,n}(\tau)}\geq& \frac{1}{\sum_{\ell}b_\ell+r_{k,n}(\tau)}\\
=& \frac{1}{\sum_{\ell}a_\ell+r_{k,n}(\tau)}\left(\frac{1}{1+\frac{\sum_\ell(b_\ell-a_\ell)}{\sum_{\ell}a_\ell+r_{k,n}(\tau)}}\right)\\
\geq&\frac{1}{\sum_{\ell}a_\ell+r_{k,n}(\tau)}\left(1-\frac{\sum_{\ell}b_\ell-a_\ell}{\sum_{\ell}a_\ell+r_{k,n}(\tau)}\right).
\end{align*}
As $\sum_\ell a_\ell\geq \frac{1}{2}e^{-4c}$, we have:
\begin{equation}\label{eq:bounddenominator}
\left|\frac{1}{\sum_{\ell}a_\ell+ r_{k,n}(\tau)}-\frac{1}{\sum_{\ell}b_\ell+ r_{k,n}(\tau)}\right|\leq 4e^{8c}||\mathbf{a}-\mathbf{b}||_1.
\end{equation}

\paragraph{Bound of (a)} It holds that $\sum_{\ell=1}^{kn^{1/8}}\min(\frac{\ell}{k},2c)b_{\ell}+\tilde{r}_{k,n}(t)\leq 6c$, which together with  \cref{eq:bounddenominator} implies:
\begin{align*}
    \left|\frac{\sum_{\ell'=1}^{kn^{1/8}}\min(\frac{\ell'}{k},2c)b_{\ell'}+\tilde{r}_{k,n}(t)}{\sum_{\ell'=1}^{kn^{1/8}}a_{\ell'}+r_{k,n}(\tau)}b_\ell-\frac{\sum_{\ell'=1}^{kn^{1/8}}\min(\frac{\ell'}{k},2c)b_{\ell'}+\tilde{r}_{k,n}(t)}{\sum_{\ell'=1}^{kn^{1/8}}b_{\ell'}+r_{k,n}(\tau)}b_\ell\right| \leq 24ce^{8c}||\mathbf{a}-\mathbf{b}||_1b_\ell
\end{align*}
so that:

\begin{align*}
    (a)\leq& 24ce^{8c}||\mathbf{a}-\mathbf{b}||_1\underbrace{\sum_{\ell}b_\ell}_{\leq 2}+ \underbrace{\frac{\tilde{r}_{k,n}(\tau)}{\sum_{\ell'=1}^{kn^{1/8}}a_{\ell'}+r_{k,n}(\tau)}}_{\leq 2c}||\mathbf{a}-\mathbf{b}||_1\\
    &+\sum_{\ell}\frac{1}{\sum_{\ell'=1}^{kn^{1/8}}a_{\ell'}+r_{k,n}(\tau)}\underbrace{\left|\sum_{\ell'=1}^{kn^{1/8}}\min(\frac{\ell'}{k},2c)a_{\ell'}a_\ell-\sum_{\ell'=1}^{kn^{1/8}}\min(\frac{\ell'}{k},2c)b_{\ell'}b_\ell\right|}_{(ii)}.
\end{align*}
We now bound $(ii)$.
\begin{align*}
    (ii) \leq& \left|\sum_{\ell'=1}^{kn^{1/8}}\min(\frac{\ell'}{k},2c)a_{\ell'}a_\ell-\sum_{\ell'=1}^{kn^{1/8}}\min(\frac{\ell'}{k},2c)a_{\ell'}b_\ell\right|+\left|\sum_{\ell'=1}^{kn^{1/8}}\min(\frac{\ell'}{k},2c)a_{\ell'}b_\ell-\sum_{\ell'=1}^{kn^{1/8}}\min(\frac{\ell'}{k},2c)b_{\ell'}b_\ell\right|\\
    \leq & 2c \sum_{\ell'=1}^{kn^{1/8}}a_{\ell'}|a_\ell-b_\ell|+2c b_\ell||\mathbf{a}-\mathbf{b}||_1,
\end{align*}
so
\begin{align*}
    (a)\leq& 2c(24e^{8c}+1)||\mathbf{a}-\mathbf{b}||_1+2c\sum_\ell\frac{ \sum_{\ell'=1}^{kn^{1/8}}a_{\ell'}|a_\ell-b_\ell|+ b_\ell||\mathbf{a}-\mathbf{b}||_1}{\sum_{\ell'=1}^{kn^{1/8}}a_{\ell'}+r_{k,n}(\tau)}\\
    \leq &2c(24e^{8c}+4e^{4c}+2)||\mathbf{a}-\mathbf{b}||_1.
\end{align*}

\paragraph{Bound of (b)} By Young's inequality, it holds that $\sum_{\ell}|\sum_{\ell'=0}^{\ell}\min(\frac{\ell'}{k},2c)b_{\ell'}b_{\ell-\ell'}|\leq 8c$, so we have by equation \ref{eq:bounddenominator}:
\begin{align*}
   \sum_\ell\left|\frac{\sum_{\ell'=0}^{\ell}\min(\frac{\ell'}{k},2c)b_{\ell'}b_{\ell-\ell'}}{\sum_{\ell'=1}^{kn^{1/8}}a_{\ell'}+r_{k,n}(\tau)}-\frac{\sum_{\ell'=0}^{\ell}\min(\frac{\ell'}{k},2c)b_{\ell'}b_{\ell-\ell'}}{\sum_{\ell'=1}^{kn^{1/8}}b_{\ell'}+r_{k,n}(\tau)}\right| \leq 32ce^{8c}||\mathbf{a}-\mathbf{b}||_1
 \end{align*}
This implies:
\begin{align*}
    (b)\leq& 32ce^{8c}||\mathbf{a}-\mathbf{b}||_1+\sum_{\ell}\left|\frac{\sum_{\ell'=0}^{\ell}\min(\frac{\ell'}{k},2c)a_{\ell'}a_{\ell-\ell'}}{\sum_{\ell'=1}^{kn^{1/8}}a_{\ell'}+r_{k,n}(\tau)}-\frac{\sum_{\ell'=0}^{\ell}\min(\frac{\ell'}{k},2c)b_{\ell'}b_{\ell-\ell'}}{\sum_{\ell'=1}^{kn^{1/8}}a_{\ell'}+r_{k,n}(\tau)}\right|\\
    \leq& 32ce^{8c}||\mathbf{a}-\mathbf{b}||_1+2e^{4c}\underbrace{\sum_\ell\left|\sum_{\ell'=0}^{\ell}\min(\frac{\ell'}{k},2c)a_{\ell'}a_{\ell-\ell'}-\sum_{\ell'=0}^{\ell}\min(\frac{\ell'}{k},2c)b_{\ell'}b_{\ell-\ell'}\right|}_{(iii)}.
\end{align*}
We have:
\begin{align*}
    (iii)\leq& \sum_\ell\left|\sum_{\ell'=0}^{\ell}\min(\frac{\ell'}{k},2c)a_{\ell'}(a_{\ell-\ell'}-b_{\ell-\ell'})\right|+\sum_\ell\left|\sum_{\ell'=0}^{\ell}\min(\frac{\ell'}{k},2c)b_{\ell-\ell'}(a_{\ell'}-b_{\ell'})\right|\\
    \leq &2c \left(||\mathbf{a}||_1+||\mathbf{b}||_1\right)||\mathbf{a}-\mathbf{b}||_1.
\end{align*}
So:
\begin{align*}
    (b) \leq 48ce^{8c}||\mathbf{a}-\mathbf{b}||_1.
\end{align*}
\hfill \(\Box\)

With those two technical lemmas, we can prove \cref{lem:resinter}.

\paragraph{Proof of \cref{lem:resinter}:}  We are now ready to check all of the technical conditions and apply the Differential Equation Method, \cref{thm:Wormald}.

The trend condition is \cref{lem:evollawfinal}. The bounded jumps conditions holds with \[\bigg|\sum_{\ell=1}^{kn^{1/8}}F_{k,n}(\ell,t)-F_{k,n}(\ell,t+1)\bigg|\leq 3:=\beta\]
as at most one vertex is matched at every iteration. By  \cref{lem:lipshcitz}, $\Phi_{k,n}$ is $L_c$-Lipschitz with respect to the $L_1$ norm on $\mathcal{D}$, and upper bounded by $16c:=R_1$.

We now turn to the initial condition. Define event 
\[\mathcal{A}_1:=\left\{\sum_{\ell<kn^{1/8}}\left|F_{k,n}(\ell,0)-nkp_k^2e^{-\frac{\ell-1}{k}}\right|\leq 3kn^{-3/8} \sqrt{12\ln(n)}\times n\right \}.\]
By union bound on the result of  \cref{lem:initialconditions}, it holds with probability at least $1-\frac{4kn^{1/8}}{n^4}$. Also, under event $\mathcal{A}_1$, the initial condition hypothesis holds with \[\lambda_1(n)=\max\left(3kn^{-1/4}\sqrt{12\ln(n)},\left(\delta_1+R_1L_ckn^{-3/4}\right)\min(1,L_c^{-1}) \right)n^{-1/8}.\]

It only remains to prove that   $(f_k(\ell,\tau))_{\ell=1}^{kn^{1/8}}$ remains far enough  from the boundaries of $\mathcal{D}$ for $\tau\in[0,1]$. Note that we have:
\[
0\geq\frac{\partial \sum_\ell f_k(\ell,\tau)}{\partial t}\geq -4c \sum_\ell f_k(\ell,\tau).
\]
For any $k\geq 4$,  $\sum_\ell f_k(\ell,0)\geq1- \frac{1}{2k}\geq \frac{7}{8}e^{-4c}$.
By  \cref{eq:boundendsum}, $r_{k,n}(\tau)\leq n^{-1/8}$ for any $\tau\in[0,1]$. Thus, for any $n$ s.t. $n^{-1/8}\leq \frac{1}{8}e^{-4c}$ and $3 e^{L_c} \lambda(n)\leq \frac{1}{4}e^{-4c}$, for any $\tau\in[0,1]$,  $(f_k(\ell,\tau))_{\ell=1}^{kn^{1/8}}$ has $L_1$-distance at least $3 e^{L_c} \lambda(n)$  from the boundaries of $\mathcal{D}$.

Theorem \ref{thm:Wormald} implies that, for any large enough $n$ s.t. $n^{-1/8}\leq \frac{1}{8}e^{-4c}$ and $3 e^{L_c} \lambda(n)\leq \frac{1}{4}e^{-4c}$, with probability at least $\underbrace{1-2 ne^2e^{-\frac{n \lambda(n)^2}{2\beta^2} }-\frac{3kn^{1/8}}{n^4}}_{\xrightarrow[n\rightarrow + \infty]{} 1}$ we have
\[
\max _{0 \leqslant \tau \leqslant 1} \sum _{1 \leqslant \ell \leqslant kn^{1/8}}\left|\frac{1}{n}F_{k,n}(\ell,\tau n)-f_k(\ell,\tau)\right|<\underbrace{3 e^{L_c} \lambda(n) }_{\xrightarrow[n\rightarrow + \infty]{} 0}.
\]

The proposition follows. \hfill \(\Box\)

\subsection{Proofs of the technical lemmas for section \ref{sec:metricmatching}\\}\label{sec:proofmetricmatchingDEM}

\paragraph{Proof of \cref{lem:costconcentrate}:} Define $a_i= \mathds{1}\{\frac{i}{nk}\in \tilde{U}_k\}$. The cost of the matching obtained by \closest\ at time $t$ is a function of $\left(a_i\right)_{\ell=1}^{nk}$ and $(y_j)_{j=1}^{t}$, which we denote $h\left(\left(a_i\right)_{\ell=1}^{nk},(y_j)_{j=1}^{t}\right)$.

Consider two sequence $(y_j)_{j=1}^{t}$ and $(y'_j)_{j=1}^{t}$ and two sequences $\left(a_i\right)_{\ell=1}^{nk}$ and $\left(a'_i\right)_{\ell=1}^{nk}$ s.t. each pairs differ by at most one element and $\mathcal{E}_\mathcal{I}$ holds with all sequences. Then, repeating the reasoning employed in the proof of \cref{lem:poissonization}, we get:

\[
|h\left(\left(a_i\right)_{\ell},(y_j)_{j}
\right)-h\left(\left(a'_i\right)_{\ell},(y_j)_{j}\right)|\leq \frac{4\alpha^2\ln(n)^2}{n},
\]
and:
\[
|h\left(\left(a_i\right)_{\ell},(y'_j)_{j}
\right)-h\left(\left(a_i\right)_{\ell},(y'_j)_{j}\right)|\leq \frac{8\alpha^2\ln(n)^2}{n}.
\]
 Thus, by McDiarmid's inequality:
\begin{align*}
\mathbb{P}\left(\left|h\left(\left(a_i\right)_{\ell},(y_j)_{j=1}^t\right)-\mathbb{E}\left[\sum_{j=1}^t c_j[k]\bigg|\mathcal{E}_{\mathcal{I}}\right]\right|\geq \delta\right)\leq& 2\mathbb{P}(\overline{\mathcal{E}_{\mathcal{I}}})+2\exp{-\frac{n\left(\delta-\mathbb{P}(\overline{\mathcal{E}_{\mathcal{I}}})8n(k+1)\alpha^2\ln(n)^2\right)^2}{(k+1)32\alpha^4\ln(n)^4}} \\
\leq & \frac{8\alpha+4}{n^2}.
\end{align*}
The second inequality is obtained as $\mathbb{P}(\overline{\mathcal{E}_{\mathcal{I}}})\leq \frac{4\alpha+1}{n^2}$ by \cref{lem:metricmatchingnbverticeslargeintervals}, and by setting $\delta= \left[\sqrt{\frac{8\ln(n)(k+1)}{n}}+\frac{8(4\alpha+1)(k+1)}{n^2}\right]\alpha^2\ln(n)^2$. On the other hand:

\[
\left|\mathbb{E}\left[\sum_{j=1}^t c_j[k]\bigg|\mathcal{E}_{\mathcal{I}}\right]-\mathbb{E}\left[\sum_{j=1}^t c_j[k]\right]\right|\leq n\mathbb{P}(\overline{\mathcal{E}_{\mathcal{I}}})\leq \frac{4\alpha+1}{n}.
\]
\hfill\(\Box\)

\paragraph{Proof of \cref{lem:wormaldmetric}:} Here, as in the proof of \cref{lem:resinter}, the goal is to check all the technical conditions to apply \cref{thm:Wormald}. At a given iteration, we have:
\[
\mathbb{E}\left[c_t\mathds{1}\{c_t\leq \frac{\eta}{n}\}\big|\mathcal{F}_t\right]= \frac{1}{4}\sum_{\ell=1}^{k\eta}\left(\frac{\ell}{kn}\right)^2 F_{k,n}\left(\ell,t\right).
\]
Define the function $\Gamma_{k,n}:\mathbb{R}^{k\eta}\rightarrow \mathbb{R}^{k\eta}$ as:
\begin{align*}
        \Gamma_{k,n}^{\ell} \left(\tau,\left(y_\ell\right)_{\ell=1}^{k\eta}\right) =&-\left(\frac{\ell}{k}+\frac{1}{1-\tau}\right)y_\ell+\frac{\sum_{\ell'=0}^{\ell}\frac{\ell'}{k}y_{\ell'}y_{\ell-\ell'}}{1-\tau}.
 \end{align*}
By Lemma \ref{lem:evollaw}, we have the trend condition:
\begin{align}
    \bigg|\Gamma_{k,n}^{\ell} \left(\tau,\left(\frac{F_{k,n}(1,\tau n)}
    {n}\right)_{\ell=1}^{k\eta}\right)-\mathbb{E}\left[F_{k,n}(\ell,\tau n+1)-F_{k,n}(\ell,t) \bigg | \mathcal{F}_t \right]\bigg|\leq \underbrace{\frac{6}{(1-\tau)}}_{\delta_2}\frac{1}{n}.
  \label{eq:evollawinter}
\end{align}

By Lemma \ref{lem:initialconditions}, it holds that  with probability at least $1-4k\eta/n^4$,
\begin{equation}\label{eq:initmetric}
\sum_{\ell=1}^{k\eta}\left|F_{k,n}(\ell,0)-nke^{-\frac{\ell-1}{k}}p_k^2\right|\leq 3k\eta \sqrt{12n\ln(n)}\leq 12k\eta\sqrt{\frac{\ln(n)}{n}}n:=n\lambda_2(n),
\end{equation}
Denote $\mathcal{A}_2$ that event, which is the initial condition. As before, the bounded jumps conditions holds with \[\bigg|\sum_{\ell=1}^{kn^{1/8}}F_{k,n}(\ell,t)-F_{k,n}(\ell,t+1)\bigg|\leq 3\]
as at most one vertex is matched at every iteration.
Define domain $\mathcal{D}_2$ as:
\[
\mathcal{D}_2=\left\{\left(y_\ell\right)_{\ell=1}^{k\eta}, \sum_{\ell} y_\ell\leq 2, \sum_{\ell}\frac{\ell}{k}y_\ell \leq 2\right\}.
\]
By some computations defered to the appendix (\cref{lem:boundL_1norm,lem:total_length}), $(g_k(\ell,t))_\ell\in \mathcal{D}_2, \forall t\in[0,1]$. Let us study the regularity of the function $\Gamma_{k,n}$ on $\mathcal{D}_2$. For any two sequences $(y'_1,\ldots, y'_{k\eta})$ and  $(y_1,\ldots, y_{k\eta})$ in $\mathcal{D}_2$, we have:

\begin{align*}
    \sum_{\ell}\left|\Gamma_{k,n}^{\ell} \left(\tau,y_1, \ldots,y_{k\eta}\right)-\Gamma_{k,n}^{\ell} \left(\tau,y'_1, \ldots,y'_{k\eta}\right)\right|\leq& \left(\eta+\frac{1}{1-\tau}\right)\sum_{\ell}|y_\ell-y'_\ell|\\
    &+\frac{1}{1-\tau}\underbrace{\sum_\ell\left|\sum_{\ell'=1}^\ell\frac{\ell'}{k}(y_{\ell'} y_{\ell-\ell'}-y'_{\ell'} y'_{\ell-\ell'})\right|}_{(iv)}.
\end{align*}
We work on bounding $(iv)$.
\begin{align*}
    (iv)\leq &\sum_\ell\left|\sum_{\ell'=1}^\ell\frac{\ell'}{k}y_{\ell'} (y_{\ell-\ell'}-y'_{\ell-\ell'})\right|+\left|\sum_{\ell'=1}^\ell\frac{\ell'}{k} y'_{\ell-\ell'}(y_{\ell'}-y'_{\ell})\right|,\\
    \leq &\eta\left(||\mathbf{y}||_1+||\mathbf{y'}||_1\right)||\mathbf{y}-\mathbf{y'}||_1\leq 4\eta ||\mathbf{y}-\mathbf{y'}||_1.
\end{align*}

$\Gamma_{k,n}(t,.)$ is thus $\underbrace{(\eta+\frac{4\eta+1}{1-\tau})}_{L_\eta}$-lipschitz with respect to the $L_1$ norm on $\mathcal{D}_2$ for all $t\leq \tau$. We also have:
\begin{align}
\sum_{\ell=1}^{k\eta}\left|\Gamma_{k,n}^{\ell} \left(t,y_1, \ldots,y_{k\eta}\right)\right|\leq&\left|\sum_{\ell=1}^{k\eta}\left(\frac{\ell}{k}+\frac{1}{1-t}\right)y_\ell\right|+\left|\sum_{\ell=1}^{k\eta}\frac{\sum_{\ell'=0}^{\ell}\frac{\ell'}{k}y_{\ell'}y_{\ell-\ell'}}{1-t}\right|\nonumber
\\
\leq & 2\left(1+\frac{1}{1-t}\right)+ \frac{4}{1-t}\leq \underbrace{2+\frac{6}{1-\tau}}_{R_\tau}.\label{eq:boundphimetric}
\end{align}

For any $n\geq \frac{1}{(1-\tau)^2}$, $\left(\delta_2/L_\eta +R_\tau k\eta\right)/ n\leq \lambda_2(n) $.  
By Theorem \ref{thm:Wormald}, w.p. at least $1-2 ne^2e^{-8\ln(n) k^2\eta^2 }-4k\eta/n^4$, for all $t\leq n \tau$:
\[
\sum_{\ell=1}^{k\eta}\left|F_{k,n}(\ell,t)-ng_k(\ell,\frac{t}{n})\right|\leq 3\lambda_2(n) e^{L_\eta} n.
\]
We denote that event $\mathcal{C}$. Under event $\mathcal{C}$, 
\begin{align*}
&\sum_{t=1}^{n\tau}   \left|\frac{1}{4}\sum_{\ell=1}^{k\eta}\left(\frac{\ell}{kn}\right)^2 ng_{k}\left(\ell,\frac{t}{n}\right)-\frac{1}{4}\sum_{\ell=1}^{k\eta}\left(\frac{\ell}{kn}\right)^2 F_{k,n}\left(\ell,t\right)\right|\\
&\hspace{3cm}\leq \sum_{t=1}^{n\tau}   \frac{1}{4}\left(\frac{\eta}{n}\right)^2\sum_{\ell=1}^{k\eta}\left| ng_{k}\left(\ell,\frac{t}{n}\right)-\sum_{\ell=1}^{k\eta}F_{k,n}\left(\ell,t\right)\right|\\
    &\hspace{3cm}\leq \frac{3}{4}\eta^2\lambda_2(n) e^{L_\eta}.
\end{align*}
For $k\geq\frac{1}{\eta}$, for any $n\geq 2$, we have:
\[
\mathbb{P}(\overline{\mathcal{C}}) \leq 2 ne^2e^{-8\ln(n) k^2\eta^2 }-4k\eta/n^4\leq \left(k\eta+1\right)\sqrt{\frac{\ln(n)}{n}}.
\]
Thus:
\begin{align*}
\left|\mathbb{E}\left[\sum_{t=1}^{n\tau} c_t\mathds{1}\{c_t\leq \frac{\eta}{n}\}\right]-\frac{1}{4n}\sum_{t=0}^{n\tau-1}\sum_{\ell=1}^{k\eta}\left(\frac{\ell}{k}\right)^2g_k(\ell,\frac{t}{n})\right|\leq &12k\eta^3e^{L_\eta}\sqrt{\frac{\ln(n)}{n}} + 2\eta \mathbb{P}(\overline{\mathcal{C}})\\
\leq& k \underbrace{\left(12\eta^3e^{L_\eta}+\eta+1\right)}_{w_3(\eta,\tau)}\sqrt{\frac{\ln(n)}{n}}.
\end{align*}
\hfill \(\Box\)

\bibliographystyle{informs2014} 
\bibliography{biblio} 

\begin{thebibliography}{25}
\providecommand{\natexlab}[1]{#1}
\providecommand{\url}[1]{\texttt{#1}}
\providecommand{\urlprefix}{URL }

\bibitem[{Aamand et~al.(2022)Aamand, Chen, \protect\BIBand{}
  Indyk}]{aamand2022optimal}
Aamand A, Chen J, Indyk P (2022) (optimal) online bipartite matching with
  degree information. \emph{Advances in Neural Information Processing Systems}
  35:5724--5737.

\bibitem[{Akbarpour et~al.(2021)Akbarpour, Alimohammadi, Li, \protect\BIBand{}
  Saberi}]{https://doi.org/10.48550/arxiv.2104.03219}
Akbarpour M, Alimohammadi Y, Li S, Saberi A (2021) The value of excess supply
  in spatial matching markets.
  \urlprefix\url{http://dx.doi.org/10.48550/ARXIV.2104.03219}.

\bibitem[{Balkanski et~al.(2022)Balkanski, Faenza, \protect\BIBand{}
  Perivier}]{balkanski2022power}
Balkanski E, Faenza Y, Perivier N (2022) The power of greedy for online minimum
  cost matching on the line.

\bibitem[{Birnbaum \protect\BIBand{} Mathieu(2008)}]{rankingmadesimple}
Birnbaum B, Mathieu C (2008) On-line bipartite matching made simple.
  \emph{SIGACT News} 39(1):80–87, ISSN 0163-5700,
  \urlprefix\url{http://dx.doi.org/10.1145/1360443.1360462}.

\bibitem[{Blaszczyszyn(2017)}]{baszczyszyn:cel-01654766}
Blaszczyszyn B (2017) {Lecture Notes on Random Geometric Models --- Random
  Graphs, Point Processes and Stochastic Geometry},
  \urlprefix\url{https://hal.inria.fr/cel-01654766}, lecture.

\bibitem[{Borodin et~al.(2018)Borodin, Karavasilis, \protect\BIBand{}
  Pankratov}]{borodin2018experimental}
Borodin A, Karavasilis C, Pankratov D (2018) An experimental study of
  algorithms for online bipartite matching.

\bibitem[{Brubach et~al.(2019)Brubach, Sankararaman, Srinivasan,
  \protect\BIBand{} Xu}]{brubach2019online}
Brubach B, Sankararaman KA, Srinivasan A, Xu P (2019) Online stochastic
  matching: New algorithms and bounds.

\bibitem[{Devanur et~al.(2013)Devanur, Jain, \protect\BIBand{}
  Kleinberg}]{rankingprimaldualanalysis}
Devanur N, Jain K, Kleinberg R (2013) Randomized primal-dual analysis of
  ranking for online bipartite matching. \emph{Proc. SODA '13},
  \urlprefix\url{http://dx.doi.org/10.1137/1.9781611973105.7}.

\bibitem[{Enriquez et~al.(2019)Enriquez, Faraud, M{\'e}nard, \protect\BIBand{}
  Noiry}]{Noiry}
Enriquez N, Faraud G, M{\'e}nard L, Noiry N (2019) Depth first exploration of a
  configuration model. \emph{arXiv preprint arXiv:1911.10083} .

\bibitem[{Frieze et~al.(1990)Frieze, McDiarmid, \protect\BIBand{}
  Reed}]{frieze1990greedy}
Frieze A, McDiarmid C, Reed B (1990) Greedy matching on the line. \emph{SIAM
  Journal on Computing} 19(4):666--672.

\bibitem[{Goel \protect\BIBand{} Mehta(2008)}]{GoelMehta}
Goel G, Mehta A (2008) Online budgeted matching in random input models with
  applications to adwords. \emph{Proceedings of the Annual ACM-SIAM Symposium
  on Discrete Algorithms}, 982--991,
  \urlprefix\url{http://dx.doi.org/10.1145/1347082.1347189}.

\bibitem[{Gupta et~al.(2019)Gupta, Guruganesh, Peng, \protect\BIBand{}
  Wajc}]{https://doi.org/10.48550/arxiv.1904.09284}
Gupta A, Guruganesh G, Peng B, Wajc D (2019) Stochastic online metric matching.
  \urlprefix\url{http://dx.doi.org/10.48550/ARXIV.1904.09284}.

\bibitem[{Hayes(2005)}]{hayes2005large}
Hayes TP (2005) A large-deviation inequality for vector-valued martingales.
  \emph{Combinatorics, Probability and Computing} .

\bibitem[{Huang et~al.(2022)Huang, Shu, \protect\BIBand{}
  Yan}]{10.1145/3519935.3520046}
Huang Z, Shu X, Yan S (2022) The power of multiple choices in online stochastic
  matching. \emph{Proceedings of the 54th Annual ACM SIGACT Symposium on Theory
  of Computing}, 91–103, STOC 2022 (New York, NY, USA: Association for
  Computing Machinery), ISBN 9781450392648,
  \urlprefix\url{http://dx.doi.org/10.1145/3519935.3520046}.

\bibitem[{Jaillet \protect\BIBand{} Lu(2014)}]{JailletLu}
Jaillet P, Lu X (2014) Online stochastic matching: New algorithms with better
  bounds. \emph{Mathematics of Operations Research} 39(3):624--646.

\bibitem[{Karp et~al.(1990)Karp, Vazirani, \protect\BIBand{}
  Vazirani}]{ranking}
Karp RM, Vazirani UV, Vazirani VV (1990) An optimal algorithm for on-line
  bipartite matching. \emph{Proceedings of the Twenty-Second Annual ACM
  Symposium on Theory of Computing}, 352–358, STOC '90 (New York, NY, USA:
  Association for Computing Machinery), ISBN 0897913612,
  \urlprefix\url{http://dx.doi.org/10.1145/100216.100262}.

\bibitem[{Mahdian \protect\BIBand{} Yan(2011)}]{rankingrandomorder}
Mahdian M, Yan Q (2011) Online bipartite matching with random arrivals: An
  approach based on strongly factor-revealing lps. \emph{Proceedings of the
  Annual ACM Symposium on Theory of Computing}, 597--606,
  \urlprefix\url{http://dx.doi.org/10.1145/1993636.1993716}.

\bibitem[{Manshadi et~al.(2012)Manshadi, Gharan, \protect\BIBand{}
  Saberi}]{Manshadi}
Manshadi VH, Gharan SO, Saberi A (2012) Online stochastic matching: Online
  actions based on offline statistics. \emph{Mathematics of Operations
  Research} 37(4):559--573.

\bibitem[{Mastin \protect\BIBand{} Jaillet(2013)}]{MastinJaillet}
Mastin A, Jaillet P (2013) Greedy online bipartite matching on random graphs.
  \emph{arXiv preprint arXiv:1307.2536} .

\bibitem[{Mehta(2012)}]{Mehta}
Mehta A (2012) Online matching and ad allocation. \emph{Theoretical Computer
  Science} 8(4):265--368.

\bibitem[{Miasojedow(2014)}]{miasojedow2014hoeffding}
Miasojedow B (2014) Hoeffding’s inequalities for geometrically ergodic markov
  chains on general state space. \emph{Statistics \& Probability Letters}
  87:115--120.

\bibitem[{Noiry et~al.(2021)Noiry, Sentenac, \protect\BIBand{}
  Perchet}]{Noiry2021OnlineMI}
Noiry N, Sentenac F, Perchet V (2021) Online matching in sparse random graphs:
  Non-asymptotic performances of greedy algorithm. \emph{NeurIPS}.

\bibitem[{Warnke(2019)}]{warnke2019wormalds}
Warnke L (2019) On wormald's differential equation method.

\bibitem[{Wormald(1995)}]{Wormald}
Wormald NC (1995) Differential equations for random processes and random
  graphs. \emph{The annals of applied probability} 5(4):1217--1235.

\bibitem[{Wormald et~al.(1999)}]{wormald1999differential}
Wormald NC, et~al. (1999) The differential equation method for random graph
  processes and greedy algorithms. \emph{Lectures on approximation and
  randomized algorithms} 73(155):0943--05073.

\end{thebibliography}
\begin{APPENDICES}
\crefalias{section}{appendix}
\section{Poisson Point Processes}\label{app:poisson}

The following definitions and properties of point processes come from lecture notes \cite{baszczyszyn:cel-01654766}, and are reported here for clarity.
\begin{definition}(Homogeneous Poisson point process). A point process $\Phi$ on $[0,1]$ is an homogeneous Poisson point process of intensity $\lambda$ if the following two conditions are satisfied:
\begin{enumerate}
\item For any $(a,b] \in [0,1], \Phi(a,b]$, the number of points in interval $(a,b]$, is a Poisson random variable of intensity $\lambda(b-a)$, i.e.;
$$
P\{\Phi(a,b]=k\}=\frac{[\lambda(b-a)]^{k}}{k !} e^{-\lambda(b-a)}.
$$

\item The number of points in any two disjoint intervals are independent of each other, and this extends to any finite number of disjoint intervals, i.e.;
$$
P\left\{\Phi\left(a_{i}, b_{i}\right]=k_{i}, i=1, \ldots, \ell\right\}=\prod_{i=1}^{\ell} \frac{\left[\lambda\left(b_{i}-a_{i}\right)\right]^{k_{i}}}{k_{i} !} e^{-\lambda\left(b_{i}-a_{i}\right)},$$
for any integer $k\geq 2$, any $a_1<b_1\leq a_2 \ldots <b_k$.
\end{enumerate}

\end{definition}

Note that the second point of the definition implies the following property: if there are $k$ points of the homogeneous Poisson
process in the window B, these points are independently and uniformly distributed in
B.

Let $\Phi^n$ be a Poisson point process of intensity $n$ on $[0,1]$, and $\mathcal{U} \sim \Phi^n$.
Let us enumerate the points of the point process $\mathcal{U}$ according to their coordinates. The sequence $\{u_k\}$ can be constructed as a renewal process with exponential holding times, i.e., $u_{k}=\sum_{i=1}^{k} F_{i}$ for $k \geq 1$, where $\left\{F_{k}: k=1, \ldots\right\}$
is a sequence of independent, identically distributed exponential random variables of parameter $n$. Indeed,
$$
\mathbb{P}\left\{F_{1}>t\right\}=\mathbb{P}\left\{u_{1}>t\right\}=\mathbb{P}\{\Phi((0, t])=0\}=e^{- n t},
$$
and, for $k\geq 2$ by independence (second point of the definition),
\begin{align*}
   \qquad \mathbb{P}\left\{F_{k}>t \mid F_{1}, \ldots, F_{k-1}\right\}&=\mathbb{P}\left\{u_{k}-u_{k-1}>t \mid u_{1}, \ldots, u_{k-1}\right\} \\
   &=\mathbb{P}\left\{\Phi\left(u_{k-1}, u_{k-1}+t\right]=0 \mid u_{k-1}\right\}\\
   &= e^{- n t}.
\end{align*}

\begin{lemma}[Concentration of PPP]\label{lem:concentratinPPP}
    Consider $n\geq 10$, $\mathcal{U}\sim \Phi^n$. For any interval $[a;b]\subseteq [0,1]$, 
    \[
    \mathbb{P}\left(\big||\mathcal{U}\cap [a;b]|-n|b-a|\big|\geq 2\sqrt{n\ln(n)}\right)\leq \frac{2}{n}.
    \]
\end{lemma}
\textit{Proof}: By Chernoff bound, we have:
\begin{align*}
    \mathbb{P}\left(\big||\mathcal{U}\cap [a;b]|-n|b-a|\big|\geq 2\sqrt{n\ln(n)}\right)\leq 2e^{\frac{4n\ln(n)}{2(n|b-a|+2\sqrt{n\ln(n)})}}.
\end{align*}
For any $n\geq 10$, we have $2(n|b-a|+2\sqrt{n\ln(n)})\leq 4n$. \(\hfill \Box\)
\section{A generalized version of Wormald's theorem}\label{app:Wormald}
The following theorem is a generalized version of Wormald's theorem \citep{wormald1999differential}. The main difference with the original theorem is that we apply the concentration inequality to the whole vector rather than coordinate by coordinate, that is  \cref{lem:vectorAzuma} rather than the classical Azuma-Hoeffding inequality. We repeat here the whole proof for completeness. It is largely based on the one in \cite{warnke2019wormalds}.

\begin{lemma}[\cite{hayes2005large}]\label{lem:vectorAzuma}

Let $\mathbf{X}$ be a real-value martingale taking values in $\mathbb{R}^d$ s.t. $X_0=0$ and $||X_i-X_{i-1}||\leq 1$. Then for every $m>0$
$$
\mathbb{P}(||X_m||\geq a)< 2e^2e^{-a^2/2m}.
$$
\end{lemma}

\begin{theorem}\label{thm:Wormald}
Take $a(n)$, $n\geq 1$, a collection of variables $\left(Y_k(i)\right)_{1\leq k\leq a(n)}$, a bounded domain $\mathcal{D}\subseteq \mathbb{R}^{a(n)+1}$, and a function $F: \mathcal{D} \rightarrow \mathbb{R}^{a(n)}$ that is $L$-Lip with respect to the $L_1$ norm on $\mathcal{D}$ and each coordinate is bounded by $R$ on $\mathcal{D}$. Assume the random variables $\left(Y_k(i)\right)_{1\leq k\leq a(n)}$ are $\mathcal{F}_i$-measurable and the following holds when $\left(\frac{i}{n}, \frac{Y_1(i)}{n},\ldots,\frac{Y_{a(n)}(i)}{n}\right)\in \mathcal{D}$:

\begin{itemize}
\item (Trend,i) $\sum_{k\leq a(n)}\left|\mathbb{E}\left[Y_k(i+1)-Y_k(i)|\mathcal{F}_i\right]-F_{k}\left(\frac{i}{n}, \frac{Y_1(i)}{n},\ldots,\frac{Y_{a(n)}(i)}{n}\right)\right|\leq \delta(n)$,
\item (Bounded jumps, ii) $\sum_{k\leq a(n)}|Y_{k}(i+1)-Y_k(i)|\leq \beta$,
\item (Initial condition, iii) $\sum_{k \leq a(n) }|Y_k(0)-ny_k(0)|\leq \lambda(n) n$.
\end{itemize}

Then there is $T=T(\mathcal{D}) \in(0, \infty)$ such that, whenever $\lambda(n) \geqslant$ $(\delta(n) +RL a(n)/ n)\min \left\{T, L^{-1}\right\}$, with probability at least $1-2 T ne^2e^{-n \lambda(n)^2 /\left(2 T \beta^2\right)}$ we have
\[
\max _{0 \leqslant i \leqslant \sigma n} \sum _{1 \leqslant k \leqslant a(n)}\left|Y_k(i)-y_k\left(\frac{i}{n}\right) n\right|<3 e^{L T} \lambda(n) n
\]
where $\left(y_k(t)\right)_{1 \leqslant k \leqslant a(n)}$ is the unique solution to the system of differential equations
$
y_k^{\prime}(t)=F_k\left(t, y_1(t), \ldots, y_a(t)\right) \text { with } y_k(0) \text { defined in (iii) } \forall 1 \leqslant k \leqslant a(n),
$
and $\sigma=\sigma\left(y_1(0), \ldots y(0)_{a(n)}\right) \in[0, T]$ is any choice of $\sigma \geqslant 0$ with the property that $\left(t, y_1(t), \ldots y_{a(n)}(t)\right)$ has $L_1$-distance at least $3 e^{L T} \lambda(n)$ from the boundary of $\mathcal{D}$ for all $t \in[0, \sigma)$.
\end{theorem}
\textit{Proof}: Define $I_{\mathcal{D},n}$ as the minimum of $\lfloor T n\rfloor$ and the smallest integer $i \geqslant 0$ where $\left(i / n, Y_1(i) / n, \ldots, Y_{a(n)}(i) / n\right)\notin \mathcal{D}$ holds. Set $\Delta Y_k(i):=\mathds{1}_{\left\{i<I_{\mathcal{D}}\right\}}\left[Y_k(i+1)-Y_k(i)\right]$ and 

\[M_k(j):=\sum_{0 \leqslant i<j}\left[\Delta Y_k(i)-\mathbb{E}\left(\Delta Y_k(i) \mid \mathcal{F}_i\right)\right].
\]

 Since the event $\left\{i<I_{\mathcal{D}}\right\}$ is $\mathcal{F}_i$-measurable (determined by all information of the first $i$ steps), we have
\[
Y_k(j)=M_k(j)+Y_k(0)+\sum_{0 \leqslant i<j} \mathbb{E}\left(Y_k(i+1)-Y_k(i) \mid \mathcal{F}_i\right) \quad \text { for all } 0 \leqslant j \leqslant I_{\mathcal{D}}.
\]

Furthermore, for all $i \geqslant 0$, the 'tower property' of conditional expectations implies 

\[\mathbb{E}\left(M_k(i+1)-M_k(i) \mid \mathcal{F}_i\right)=\mathds{1}_{\left\{i<I_{\mathcal{D}}\right\}} \mathbb{E}\left(Y_k(i+1)-Y_k(i)-\mathbb{E}\left(Y_k(i+1)-Y_k(i) \mid \mathcal{F}_i\right) \mid \mathcal{F}_i\right)=0.\]
The 'Boundedness hypothesis' (ii) implies 
\begin{align*}
\sum_{k\leq n(a)}\left|M_k(i+1)-M_k(i)\right| &= \sum_{k\leq n(a)}\left|\Delta Y_k(i+1)-\mathbb{E}\left(\Delta Y_k(i+1) \mid \mathcal{F}_i\right)\right|
&\leqslant 2 \beta.
\end{align*}
Defining $\mathcal{M}$ as the event
\[
\max _{0 \leqslant j \leqslant I_{\mathcal{D}}}\sum_{k\leq n(a)}\left|M_k(j)\right|<\lambda(n) n. 
\]
 The vector Azuma-Hoeffding inequality (Lemma \ref{lem:vectorAzuma} with $m:=\lfloor T n\rfloor)$ and a union bound thus yields $\mathbb{P}(\neg \mathcal{M}) \leqslant  2 T ne^2e^{-n \lambda(n)^2 /\left(2 T \beta^2\right)}$.

The final deterministic part of the argument is based on a discrete variant of Gronwall's inequality (and induction). 

Assuming that the event $\mathcal{M}$ holds, for all $\left(0, {y}_1(0), \ldots, {y}_{a(n)}(0)\right) \in \mathcal{D}$ satisfying $\sum _{1 \leqslant k \leqslant a(n)}\left|Y_k(0)-y_k(0) n\right| \leqslant$ $\lambda(n) n$,  it remains to prove by induction that, for all integers $0 \leqslant m \leqslant \sigma n$, we have
\[
\sum _{1 \leqslant k \leqslant a(n)}\left|Y_k(m)-y_k\left(\frac{m}{n}\right) n\right|<3 \lambda(n) n e^{L T} \text {. }
\]
The base case $m=0$ holds since $\sum _{1 \leqslant k \leqslant a(n)}\left|Y_k(0)-y_k(0) n\right|\leqslant$ $\lambda(n) n$ by assumption. 

Turning to the induction step $1 \leqslant m \leqslant \sigma n$, note that $m-1<\lfloor\sigma n\rfloor \leqslant\lfloor T n\rfloor$ by definition. So, by choice of $\sigma$, the induction hypothesis implies $m-1<I_{\mathcal{D}}$ and thus $m \leqslant I_{\mathcal{D}}$.

Fix $0 \leqslant j \leqslant m$. We have:
\begin{align*}
\left|Y_k(j)-y_k\left(\frac{j}{n}\right) n\right| \leqslant&\left|M_k(j)\right|+\left|Y_k(0)-y_k(0) n\right|\\
&+\sum_{0 \leqslant i<j}\left|\mathbb{E}\left(Y_k(i+1)-Y_k(i) \mid \mathcal{F}_i\right)-\left[y_k\left(\frac{i+1}{n}\right)-y_k\left(\frac{i}{n}\right)\right] n\right| 
\end{align*}

Under event $\mathcal{M}$ and (iii), we have:
\[
\sum_{k\leq n(a)}\left|M_k(j)\right|+\left|Y_k(0)-y_k(0) n\right|\leq 2\lambda(n) n.
\]
On the other hand:
\begin{align*}
\sum_{k\leq a(n)}&\sum_{0 \leqslant i<j}\left|\mathbb{E}\left(Y_k(i+1)-Y_k(i) \mid \mathcal{F}_i\right)-\left[y_k\left(\frac{i+1}{n}\right)-y_k\left(\frac{i}{n}\right)\right] n\right| \\
\leq&\sum_{0 \leqslant i<j}\delta(n) +\sum_{k\leq a(n)}\sup_{\varepsilon \in [\frac{i}{n};\frac{i+1}{n}]}\left|F_k(i/n,\tilde{Y}_i/n)-y'_k(\varepsilon)\right|,
\end{align*}

where $\tilde{Y}_i/n$ is a shorthand for $Y_1(i),\ldots Y_{a(n)}(i)$.For any $\varepsilon \in [\frac{i}{n};\frac{i+1}{n}]$, we have:
\begin{align*}
    |y'_k(\varepsilon)-y'_k(\frac{i}{n})|= |F_k(\varepsilon,\tilde{y}(\varepsilon))-F_k\left(\frac{i}{n},\tilde{y}(\frac{i}{n})\right)|\leq \frac{LR}{n}.
\end{align*}

It follows that:
\begin{align*}
\sum_{k\leq a(n)}\left|Y_k(j)-y_k\left(\frac{j}{n}\right) n\right| \leqslant&\sum_{0 \leqslant i<j}\left(LR \frac{a(n)}{n}+\delta(n) +\sum_{k\leq a(n)}\left|F_k(i/n,\tilde{Y}_i/n)-F_k(i/n,y(i/n))\right|\right)\\
&+2\lambda(n) n\\
\leq& \sum_{0 \leqslant i<j}\left(LR \frac{a(n)}{n}+\delta(n)+\frac{L}{n}\sum_{k\leq a(n)}\left|{Y}_k(i/n)-y_k(i/n)n\right|\right)\\
&+2\lambda(n) n.
\end{align*}

We now use the following discrete version of Gronwall's Lemma:

\begin{lemma}
    Assume that there are $b, c \geqslant 0$ and $a>0$ such that $x_j<$ $c+\sum_{0 \leqslant i<j}\left(a x_i+b\right)$ for all $0 \leqslant j \leqslant m$. Then $x_m<\left(c+b \min \left\{m, a^{-1}\right\}\right) e^{a m}$.
\end{lemma}

It yields:
\begin{align*}
\sum_{k\leq a(n)}\left|Y_k(m)-y_k\left(\frac{m}{n}\right) n\right|\leq& \left(2 \lambda(n)n+\left(LR \frac{a(n)}{n}+\delta(n) \right)\min(m,\frac{n}{L})\right)e^{\frac{Lm}{n}}\\
&\leq 3\lambda(n) ne^{LT}.
\end{align*}

\hfill \(\Box\)

\section{Computational analysis of the solutions of the PDEs}
\hfill \break

In this section, we prove some properties of the functions $f_k,f,g_k$ and $g$.

\begin{lemma}[Total length invariant]\label{lem:total_length}
 For any $\tau\in [0,1]$,
 \[
 \sum_{\ell=1}^{+\infty}\frac{\ell}{k}f_k(\ell,\tau)=1 \text{ and } \int_{x=0}^{+\infty}f(x,\tau)=1.
 \]
The same holds for $g_k$ and $g$.
\end{lemma}
\textit{Proof}: This is true for $\tau=0$ by definition of the initial condition. We now show that this quantity is an invariant of the system of ODEs.
We have: 
\begin{align*}
    \frac{\partial \sum_{\ell=1}^{+\infty}\frac{\ell}{k}f_k(\ell,\tau)}{\partial \tau} =&-\sum_{\ell=1}^{+\infty} \frac{\ell}{k}\min(\frac{\ell}{k},2c)f_k(\frac{\ell}{k},\tau)-\frac{1}{\sum_{\ell=1}^{+\infty}f_k(\ell,\tau)}\sum_{\ell=1}^{+\infty} \min(\frac{\ell}{k},2c)f_k(\frac{\ell}{k},\tau)\sum_{\ell=1}^{+\infty} \frac{\ell}{k}f_k(\frac{\ell}{k},\tau)\\\
   &+\frac{1}{\sum_{\ell=1}^{+\infty}f_k(\ell,\tau)}\sum_{\ell=1}^{+\infty}\sum_{\ell'=0}^{+\infty}\min(\frac{\ell'}{k},2c)f_{k}(\ell',\tau)(\ell-\ell')f_{k}(\ell-\ell',\tau)\\
   &+\frac{1}{\sum_{\ell=1}^{+\infty}f_k(\ell,\tau)}\sum_{\ell=1}^{+\infty}\sum_{\ell'=0}^{+\infty}\ell'\min(\frac{\ell'}{k},2c)f_{k}(\ell',\tau)f_{k}(\ell-\ell',\tau),\\
   =&-\sum_{\ell=1}^{+\infty} \frac{\ell}{k}\min(\frac{\ell}{k},2c)f_k(\frac{\ell}{k},\tau)-\frac{1}{\sum_{\ell=1}^{+\infty}f_k(\ell,\tau)}\sum_{\ell=1}^{+\infty} \min(\frac{\ell}{k},2c)f_k(\frac{\ell}{k},\tau)\sum_{\ell=1}^{+\infty} \frac{\ell}{k}f_k(\frac{\ell}{k},\tau)\\
   &+\frac{1}{\sum_{\ell=1}^{+\infty}f_k(\ell,\tau)}\sum_{\ell=1}^{+\infty} \min(\frac{\ell}{k},2c)f_k(\frac{\ell}{k},\tau)\sum_{\ell=1}^{+\infty}\frac{\ell}{k}f_k(\frac{\ell}{k},\tau)\\
   &+\frac{1}{\sum_{\ell=1}^{+\infty}f_k(\ell,\tau)}\sum_{\ell=1}^{+\infty} \frac{\ell}{k}\min(\frac{\ell}{k},2c)f_k(\frac{\ell}{k},\tau)\sum_{\ell=1}^{+\infty}f_k(\ell,\tau),\\
   =&0.\\
\end{align*}

The computation is similar for $f$, $g$ and $g_k$.
\hfill \( \Box \)

\begin{lemma}[Bound $L_1$ norm]\label{lem:boundL_1norm}
    For any $\tau\in [0,1]$,
    \[
    e ^{-4c} \leq  \sum_{\ell=1}^{+\infty}f_k(\ell,\tau)\leq 1 \text{ and }
e ^{-4c} \leq  \int_{0}^{+\infty}f(x,\tau)dx\leq 1.
    \]
    Also, for any $\tau\in [0,1]$,
    \[
      \sum_{\ell=1}^{+\infty}g_k(\ell,\tau)=1-\tau=  \int_{0}^{+\infty}g(x,\tau)dx.
    \]
\end{lemma}
\textit{Proof}:
\begin{align*}
    \frac{\partial \sum_{\ell=1}^{+\infty}f_k(\ell,\tau)}{\partial \tau} =&-\sum_{\ell=1}^{+\infty} \min(\frac{\ell}{k},2c)f_k(\frac{\ell}{k},\tau)-\frac{1}{\sum_{\ell=1}^{+\infty}f_k(\ell,\tau)}\sum_{\ell=1}^{+\infty} \min(\frac{\ell}{k},2c)f_k(\frac{\ell}{k},\tau)\sum_{\ell=1}^{+\infty} f_k(\frac{\ell}{k},\tau)\\\
   &+\frac{1}{\sum_{\ell=1}^{+\infty}f_k(\ell,\tau)}\sum_{\ell=1}^{+\infty}\sum_{\ell'=0}^{+\infty}\min(\frac{\ell'}{k},2c)f_{k}(\ell',\tau)f_{k}(\ell-\ell',\tau),\\
   =&-\sum_{\ell=1}^{+\infty} \min(\frac{\ell}{k},2c)f_k(\frac{\ell}{k},\tau).
\end{align*}
This implies that for any $\tau \in [0,1]$, $
e ^{-4c} \leq e ^{-4ct}\leq \sum_{\ell=1}^{+\infty}f_k(\ell,\tau)\leq \sum_{\ell=1}^{+\infty}f_k(\ell,0) =1.$

We have:

\begin{align*}
    \frac{\partial \sum_{\ell=1}^{+\infty}g_k(\ell,\tau)}{\partial \tau} 
   =&-\sum_{\ell=1}^{+\infty} \frac{\ell}{k}g_k(\frac{\ell}{k},\tau)=-1.
\end{align*}
So, $\sum_{\ell=1}^{+\infty}g_k(\ell,\tau)=1-\tau$, and the same holds for $g$.

\hfill \(\Box\)

\paragraph{Proof of lemma \ref{lem:computelength} }: We have:
\begin{align*}
\frac{d}{d\tau}\int_{0}^{+\infty}\frac{x^2}{4}g(x,t)dx= &-\int_{0}^{+\infty}\frac{x^2}{4}\left(x+\frac{1}{1-\tau}\right)g(x,\tau)dx+\frac{1}{1-\tau}\underbrace{\int_{0}^{+\infty}\frac{x^2}{4}\int_{0}^{x}x'g(x',\tau)g(x-x',\tau)dx' }_{(a)}.
\end{align*}

Let us simplify $(a)$:
\begin{align*}
    (a)=& \int_{0}^{+\infty}\int_{0}^{x}x'\frac{(x-x')^2}{4}g(x',\tau)g(x-x',\tau)dx' -\int_{0}^{+\infty}\int_{0}^{x}\frac{(x')^3}{4}g(x',\tau)g(x-x',\tau)dx',\\
    &+\int_{0}^{+\infty}\int_{0}^{x}\frac{(x')^2}{2}xg(x',\tau)g(x-x',\tau)dx',\\
    =&\int_{0}^{+\infty}\frac{x^2}{4}g(x,t)dx'\underbrace{\int_{0}^{+\infty}xg(x,\tau)dx'}_{=1}-\int_{0}^{+\infty}\int_{0}^{x}\frac{(x')^3}{4}g(x',\tau)g(x-x',\tau)dx'\\
    &+\int_{0}^{+\infty}\int_{0}^{x}\frac{(x')^2}{2}(x-x')g(x',\tau)g(x-x',\tau)dx'+\int_{0}^{+\infty}\int_{0}^{x}\frac{(x')^3}{2}g(x',\tau)g(x-x',\tau)dx',\\
    =&3\int_{0}^{+\infty}\frac{x^2}{4}g(x,\tau)dx+\int_{0}^{+\infty}\frac{x^3}{4}g(x,\tau)dx\underbrace{\int_{0}^{+\infty}g(x,\tau)dx}_{=1-\tau}.
\end{align*}

Reinjecting in the previous equation we obtain:
\begin{align*}
    \frac{d}{d\tau}\int_{0}^{+\infty}\frac{x^2}{4}g(x,\tau)dx= \frac{2}{1-\tau}\int_{0}^{+\infty}\frac{x^2}{4}g(x,\tau)dx.
\end{align*}

Define $z(\tau)=\int_{0}^{+\infty}\frac{x^2}{4}g(x,\tau)dx$. Note that $z(0)=0.5$. We get:
\begin{align*}
    \frac{z'(\tau)}{z(\tau)}= \frac{2}{1-\tau},
\end{align*}
which integrates to 
    $\ln\left(z(\tau)\right)= -2\ln(1-\tau) +\ln\left(z(0)\right)$. Thus $
z(\tau)=\frac{z(0)}{(1-\tau)^2}$, and the total length of the matching created is:
\[
\int_{0}^\tau z(t)dt= z(0)\left[\frac{1}{1-\tau}-1\right].
\]

\hfill \(\Box\)

\section{Finite elements error bounds}\label{sec:finiteelements}

\paragraph{Proof of Lemma \ref{lem:discretetocontinuous}:}
Define function $\tilde{f}_k$ s.t. for any $x\geq 0$, any $\tau \in [0;1]$:
\[
\tilde{f}_k(x,\tau)= kf_k(\lceil kx\rceil,\tau).
\]
Note that for any $\tau \in [0;1]$,
\[
\int_{0}^{+\infty}\tilde{f}_k(x,\tau)dx=\sum_{\ell=1}^{+\infty}f_k(\ell,\tau),
\]
hence:
\[
\bigg|\int_{0}^{+\infty}f(x,\tau)dx-\sum_{\ell=1}^{+\infty}f_{k}(\ell,\tau)\bigg|\leq \mid\mid \tilde{f}_k(.,\tau)-f(.,\tau) \mid\mid_1.
\]
We have:
\begin{align}
    \mid\mid \tilde{f}_k(.,0)-f(.,0) \mid\mid_1=&\int_{0}^{+\infty}|k^2(1-e^{-1/k})^2e^{\frac{\lceil kx \rceil-1}{k}}-e^{-x}|dx\notag\\
    \leq & \int_{0}^{+\infty}|k^2(1-e^{-1/k})^2-1|e^{-x}dx+\int_{0}^{+\infty}|e^{\frac{\lceil kx \rceil-1}{k}}-e^{-x}|dx\notag\\
    \leq&\frac{3}{k}.\label{eq:boundinitftofk}
\end{align}

Define
\[
\mathcal{L}_c:= \left\{ f \in L_1 \text{ s.t. } e^{-4c}\leq\mid\mid f\mid\mid_1\leq 1 \text{ and } f\geq 0\right\}.
\]
By Lemmas \ref{lem:total_length} and \ref{lem:boundL_1norm}, $\forall \tau \in[0,1]$, $\tilde{f}_k(.,\tau),f(.,\tau) \in \mathcal{L}_c$. Define application $A:\mathcal{L}_c\rightarrow\mathcal{L}_c$ as
\begin{align*}
    A(f)(x)=&-\min(x,2c)f(x,\tau)-\frac{1}{\int_{0}^{+\infty}f(x',\tau)dx'}\int_{0}^{+\infty}\min(x',2c)f(x',\tau)dx'f(x,\tau)\\
   &+\frac{1}{\int_{0}^{+\infty}f(x',\tau)dx'}\int_{0}^{x}\min(x',2c)f(x',\tau)f(x-x',\tau)dx'.
\end{align*}
Let us show it is Lipschitz with respect to the $L_1$ norm, and derive the Lipschitz constant. First, for any two functions $f_1,f_2\in \mathcal{L}_c^2$,
\begin{align}
\left|\frac{1}{\int_{0}^{+\infty}f_1(x,\tau)dx}-\frac{1}{\int_{0}^{+\infty}f_2(x,\tau)dx}\right|\leq e^{4c}\mid\mid f_1-f_2 \mid\mid_1.\label{eq:boundf1f2denom}
\end{align}
Also, denoting $\tilde{f}$ the function $\min(x,2c)f(x)$, we have $\mid\mid \tilde{f}_1-\tilde{f}_2 \mid\mid_1\leq 2c \mid\mid f_1-f_2 \mid\mid_1$, and
\begin{align}
    \mid\mid \tilde{f}_1*f_1-\tilde{f}_2*f_2 \mid\mid_1\leq& \mid\mid (\tilde{f}_1-\tilde{f}_2)*f_1\mid\mid_1
    +\mid\mid\tilde{f}_2*(f_2-f_1) \mid\mid_1\notag\\
    \leq& \mid\mid \tilde{f}_1-\tilde{f}_2\mid\mid_1\mid\mid f_1\mid\mid_1+\mid\mid\tilde{f}_2\mid\mid_1\mid\mid f_2-f_1\mid\mid_1\notag\\
    \leq& 4c \mid\mid f_2-f_1\mid\mid_1,\label{eq:boundconvprod}
\end{align}
where the second line comes from Young's inequality. Also:
\begin{align}
&\int_{0}^{+\infty}\left|\int_{0}^{+\infty}2cf_1(x',\tau)dx'f_1(x,\tau)-\int_{0}^{+\infty}2cf_2(x',\tau)dx'f_2(x,\tau)\right|dx\notag\\
&\leq 2c\int_{0}^{+\infty}\int_{0}^{+\infty}|f_1(x',\tau)-f_2(x',\tau)|dx'f_1(x,\tau)dx+\int_{0}^{+\infty}\int_{0}^{+\infty}2cf_2(x',\tau)dx'|f_1(x,\tau)-f_2(x,\tau)|dx\notag\\
&\leq 4c \mid\mid f_1-f_2 \mid\mid_1.\label{eq:f1f2boundsecondterm}
\end{align}

Putting everything together, we get:
{\small
\begin{align}
   \mid\mid A(f_1)-A(f_2) \mid\mid_1 \leq& \mid\mid \tilde{f}_2-\tilde{f}_1\mid\mid_1+\underbrace{\left|\frac{1}{\int_{0}^{+\infty}f_1(x,\tau)dx}-\frac{1}{\int_{0}^{+\infty}f_2(x,\tau)dx}\right|}_{\leq e^{4c}\mid\mid f_1-f_2 \mid\mid_1, \ (\cref{eq:boundf1f2denom})}\underbrace{\int_{0}^{+\infty}\int_{0}^{+\infty}x'f_1(x',\tau)dx'f_1(x,\tau)dx}_{\leq 1}\notag\\
   &+e^{4c}\underbrace{\int_{0}^{+\infty}\left|\int_{0}^{+\infty}2cf_1(x',\tau)dx'f_1(x,\tau)-\int_{0}^{+\infty}2cf_2(x',\tau)dx'f_2(x,\tau)\right|dx}_{\leq 4c \mid\mid f_1-f_2 \mid\mid_1,\ \cref{eq:f1f2boundsecondterm}}\notag\\
   &+\underbrace{\left|\frac{1}{\int_{0}^{+\infty}f_1(x,\tau)dx}-\frac{1}{\int_{0}^{+\infty}f_2(x,\tau)dx}\right|}_{\leq e^{4c}\mid\mid f_1-f_2 \mid\mid_1, \ (\cref{eq:boundf1f2denom})}\underbrace{\int_{0}^{+\infty}\int_{0}^{x}x'f_1(x',\tau)f_1(x-x',\tau)dx'dx}_{\leq 1}\notag\\
&+e^{4c}\underbrace{\mid\mid \tilde{f}_1*f_1-\tilde{f}_2*f_2\mid\mid_1}_{\leq 4c\mid\mid f_1-f_2 \mid\mid_1, \ (\cref{eq:boundconvprod})}\notag\\
   \leq& 2\left(c+(4c+1)e^{4c}\right)\mid\mid f_2-f_1 \mid\mid_1. \label{eq:boundALip}
\end{align}}
We denote this Lipschitz constant $\text{lip}_c$.
Define application $B_k:\mathcal{L}_c\rightarrow\mathcal{L}_c$
\begin{align*}
    B_k(f)(x)=&-\min\left(\frac{\lceil kx\rceil}{k},2c\right)f(x,\tau)-\frac{1}{\int_{0}^{+\infty}f(x',\tau)dx'}\int_{0}^{+\infty}\min\left(\frac{\lceil kx'\rceil}{k},2c\right)f(x',\tau)dx'f(x,\tau)\\
   &+\frac{1}{\int_{0}^{+\infty}f(x',\tau)dx'}\int_{0}^{\frac{\lceil kx\rceil}{k}}\min\left(\frac{\lceil kx'\rceil}{k},2c\right)f(x',\tau)f\left(\frac{\lceil kx\rceil-\lceil kx'\rceil}{k},\tau\right)dx'.
\end{align*}
Which is defined s.t. for any $x\in \mathbb{R}+$,
$
\frac{\partial \tilde{f}_k(x,\tau)}{\partial t}= B_k\left(\tilde{f}_k\left(.,\tau\right)\right)(x).
$
For any $\tau\in[0,1]$,  we have:

\begin{align*}
    &||B_{k}\left(\tilde{f}_k\left(.,\tau\right)\right)(x)-A\left(\tilde{f}_k\left(.,\tau\right)\right)(x)||_1\\
    & \hspace{1cm}\leq\underbrace{\int_{0}^{2c}\left|\frac{\lceil kx\rceil}{k}-x\right| \tilde{f}_k(x,\tau)dx}_{\leq \frac{1}{k}||\tilde{f}_k||_1\leq \frac{1}{k}}+\underbrace{\frac{1}{\int_{0}^{+\infty}f(x',\tau)dx'}\int_{0}^{+\infty}\int_{0}^{2c}\left|\frac{\lceil kx'\rceil}{k}-x'\right| \tilde{f}_k(x',\tau)dx'\tilde{f}_k(x,\tau)dx}_{\leq \frac{1}{k}}\\
    &\hspace{1.2cm}+\underbrace{\frac{1}{\int_{0}^{+\infty}\tilde{f}_k(x',\tau)dx'}\int_{0}^{+\infty}\int_{x}^{\frac{\lceil kx\rceil}{k}}\min\left(\frac{\lceil kx'\rceil}{k},2c\right)\tilde{f}_k(x',\tau)\tilde{f}_k\left(\frac{\lceil kx\rceil-\lceil kx'\rceil}{k},\tau\right)dx'dx}_{=0, \text{ as }\lceil kx\rceil=\lceil kx'\rceil \text{ for }x'\in [x;\frac{\lceil kx\rceil}{k}]}\\
    &\hspace{1.2cm}+\frac{1}{\int_{0}^{+\infty}\tilde{f}_k(x',\tau)dx'}\underbrace{\int_{0}^{+\infty}\left|\int_{0}^x\left(\frac{\lceil kx'\rceil}{k}-x'\right)\tilde{f}_k(x',\tau)\tilde{f}_k(x-x',\tau)dx'\right|dx}_{\leq \frac{1}{k}\int_{0}^{+\infty}\int_{0}^x\tilde{f}_k(x',\tau)\tilde{f}_k(x-x',\tau)dx'dx\leq \frac{1}{k}||\tilde{f}_k||_1^2\leq \frac{1}{k}} \\
    &\hspace{1.2cm}+\underbrace{\frac{1}{\int_{0}^{+\infty}\tilde{f}_k(x',\tau)dx'}\int_{0}^{+\infty}\int_{0}^x2c\tilde{f}_k(x',\tau)\left|\tilde{f}_k(x-x',\tau)-\tilde{f}_k\left(\frac{\lceil kx\rceil-\lceil kx'\rceil}{k},\tau\right)\right|dx'dx}_{\leq 2c\int_{1/k}^{+\infty}\left|\tilde{f}\left(x,\tau\right)-\tilde{f}\left(x-\frac{1}{k},\tau\right)\right|dx} \\
     &\hspace{1cm}\leq \frac{2c+3}{k}+2c\int_{1/k}^{+\infty}\left|\tilde{f}_k\left(x,\tau\right)-\tilde{f}_k\left(x-\frac{1}{k},\tau\right)\right|dx.
\end{align*}

We now bound $\int_{1/k}^{+\infty}\left|\tilde{f}_k\left(x,\tau\right)-\tilde{f}_k\left(x-\frac{1}{k},\tau\right)\right|dx$. First, we have:
\[
\int_{1/k}^{+\infty}\left|\tilde{f}_k\left(x,0\right)-\tilde{f}_k\left(x-\frac{1}{k},0\right)\right|dx\leq \int_{1/k}^{+\infty} k^2(1-e^{-1/k})^2\left(e^{\frac{\lceil kx \rceil-1}{k}}-e^{\frac{\lceil kx \rceil-2}{k}}\right)\leq \frac{2}{k}.
\]
Also, for any $\tau \in [0;1]$
\begin{align*}
    \int_{1/k}^{+\infty}\left|\frac{\partial \tilde{f}_k\left(x,\tau\right)}{\partial \tau}-\frac{\partial \tilde{f}_k\left(x-\frac{1}{k},\tau\right)}{\partial \tau}\right|dx \leq &6c\int_{1/k}^{+\infty}\left|\tilde{f}_k\left(x,\tau\right)-\tilde{f}_k\left(x-\frac{1}{k},\tau\right)\right|dx
\end{align*}
Hence, by Gronwall's lemma:
\[
\int_{1/k}^{+\infty}\left|\tilde{f}_k\left(x,\tau\right)-\tilde{f}_k\left(x-\frac{1}{k},\tau\right)\right|dx\leq \frac{2}{k}e^{6c}.
\]
Reinjecting in the previous equation, we obtain:
\begin{align}
||B_{k}\left(\tilde{f}_k\left(.,\tau\right)\right)(x)-A\left(\tilde{f}_k\left(.,\tau\right)\right)(x)||_1\leq \frac{2c+3+4ce^{6c}}{k}.\label{eq:bounddiffAB}
\end{align}
Thus, with \cref{eq:boundinitftofk,eq:boundALip,eq:bounddiffAB}, by application of Gronwall's Lemma, for any $\tau \in [0,1]$
\[
\mid\mid f(.,\tau)- \tilde{f}_{k}(.,\tau)\mid\mid_1\leq \frac{3}{k}e^{\text{lip}_c}+\frac{2c+3+4ce^{6c}}{k}.
\]
\hfill \(\Box \)

\paragraph{Proof of Lemma \ref{lem:continuousmetric}}: We first deal with the approximation due to the time discretization.

By equation \ref{eq:boundphimetric}, it holds that:
\begin{align*}
    \sum_{\ell=1}^{k\eta}\left|\frac{\partial g_{k}(\ell,\tau)}{\partial \tau} \right|\leq \sum_{\ell=1}^{k\eta}\left|\Gamma_{k,n}^{\ell} \left(\tau,\left(g_{k}(\ell,\tau)\right)_{\ell=1}^{k\eta}\right)\right|\leq \left(2+\frac{6}{1-\tau} \right).
\end{align*}
This implies:
\begin{align}
    \left|\int_{t=0}^{\tau}\sum_{\ell=1}^{k\eta}\left(\frac{\ell}{k}\right)^2g_k(\ell,t)dt-\frac{1}{n}\sum_{t=0}^{n\tau-1}\sum_{\ell=1}^{k\eta}\left(\frac{\ell}{k}\right)^2g_k(\ell,\frac{t}{n})\right|dt\leq& \eta^2\int_{t=0}^{\tau}\sum_{\ell=1}^{k\eta}\left|g_k(\ell,t)dt-g_k\left(\ell,\frac{\lfloor t n\rfloor}{n}\right)\right|,\nonumber
    \\ 
   \leq & \frac{\eta^2}{n}\left(2+\frac{6}{1-\tau} \right)=\frac{w_1(\tau,\eta)}{n}.\label{eq:boundriemmanmetric}
\end{align}

We now deal with the space discretization. 
For any $\tau \in [0;1)$, we have:
\begin{align}
    \left|\int_{x=0}^{\eta}x^2g(x,\tau)dx-
\sum_{\ell=1}^{k\eta}\left(\frac{\ell}{k}\right)^2g_k(\ell,\tau)\right|\leq&\int_{x=0}^{\eta}\left|\left(\frac{\lceil kx \rceil}{k}\right)^2-x^2\right|g(x,\tau)dx\notag\\
&+\eta^2\int_{x=0}^{\eta}|g(x,\tau)-kg_k(\lceil kx\rceil,\tau)|dx\notag\\
&\leq \frac{\eta\left(2\eta+1\right)}{k}+\eta^2\int_{x=0}^{\eta}|g(x,\tau)-kg_k(\lceil kx\rceil,\tau)|dx.\label{eq:boundspaceintergk}
\end{align}

It remains to bound $\int_{x=0}^{\eta}|g(x,\tau)-kg_k(\lceil kx\rceil,\tau)|dx$. First, by \cref{eq:boundinitftofk}, as $g(x,0)=f(x;0)$ and $g_k(\lceil kx\rceil,\tau)=f_k(x,0)$:
\begin{align}
\int_{x=0}^{\eta}|g(x,\tau)-kg_k(\lceil kx\rceil,\tau)|dx\leq \frac{3}{k}.\label{eq:boundinitggk}
\end{align}
Define domain:

\[
\mathcal{L}_\eta:= \left\{ g:[0,\eta]\rightarrow [0,1]\text{ s.t. } \mid\mid g \mid\mid_1\leq 1,  \mid\mid xg \mid\mid_1\leq 1\text{ and }  g\geq 0\right\}.
\]
By Lemmas \ref{lem:total_length} and \ref{lem:boundL_1norm}, $\forall \tau\in[0,1]$, $g_k(.,\tau),g(.,\tau) \in \mathcal{L}_\eta$. Define application $A_2:\mathcal{L}_\eta\rightarrow\mathcal{L}_\eta$ by
\begin{align*}
    A_2(g)(x)=&-\left(x+\frac{1}{1-\tau}\right)g(x,t)+\frac{1}{1-\tau}\int_{0}^{x}x'g(x',\tau)g(x-x',\tau)dx'.
\end{align*}
Let us show that $A_2$ is Lipschitz with respect to the $L_1$ norm on $\mathcal{L}_\eta$, and derive the Lipschitz constant. Take any $(g_1,g_2) \in \mathcal{L}_\eta^2$. We have: 
\begin{align*}
    \mid\mid (xg_1)*g_1-(xg_2)*g_2 \mid\mid_1\leq& \mid\mid (xg_1-xg_2)*g_1\mid\mid_1+\mid\mid(xg_2)*(g_1-g_2) \mid\mid_1,\\
    \leq& \mid\mid xg_1-xg_2\mid\mid_1\mid\mid g_1\mid\mid_1+\mid\mid xg_2\mid\mid_1\mid\mid g_1-g_2 \mid\mid_1,\\
    \leq& 2\eta \mid\mid g_1-g_2 \mid\mid_1,
\end{align*}
where the second line comes from Young's convolution inequality. We thus get:
\begin{align}
   \mid\mid A_2(g_1)-A_2(g_2) \mid\mid_1 \leq& \left(\eta+\frac{1+2\eta}{1-\tau}\right)\mid\mid g_1-g_2 \mid\mid_1 \label{eq:boundA2Lip}
\end{align}

Define function $\tilde{g}_k(\ell,\tau)$ as:
\[
\tilde{g}_{k}(x,\tau)=g_k\left(\lceil kx\rceil,\tau\right),
\]

Define application $B_{k,2}:\mathcal{L}_\eta\rightarrow\mathcal{L}_\eta$
\begin{align*}
    B_{k,2}(g)(x)=&-\left(\frac{\lceil kx\rceil}{k}+\frac{1}{1-\tau}\right)g(x,\tau)+\frac{1}{1-\tau}\int_{0}^{\frac{\lceil kx\rceil}{k}}\frac{\lceil kx'\rceil}{k}g(x',\tau)g\left(\frac{\lceil kx\rceil-\lceil kx'\rceil}{k},\tau\right)dx'.
\end{align*}
Which is defined s.t. for any $x\in \mathbb{R}+$,
$
\frac{\partial g_k(x,\tau)}{\partial \tau}= B_{k,2}\left(g_k\left(.,\tau\right)\right)(x).
$
For any $\tau\in[0,1]$,  we have:
\begin{align*}
    &||B_{k,2}\left(\tilde{g}_k\left(.,\tau\right)\right)(x)-A_2\left(g_k\left(.,\tau\right)\right)(x)||_1 \\
    & \hspace{1cm}\leq\underbrace{\int_{0}^{\eta}\left|\frac{\lceil kx\rceil}{k}-x\right| \tilde{g}_k(x,\tau)dx}_{\leq \frac{1}{k}||\tilde{g}_k||_1\leq \frac{1}{k}}+\frac{1}{1-\tau}\underbrace{\int_{0}^{\eta}\left|\int_{0}^x\left(\frac{\lceil kx'\rceil}{k}-x'\right)\tilde{g}_k(x',\tau)\tilde{g}_k(x-x',\tau)dx'\right|dx}_{\leq \frac{1}{k}\int_{0}^{\eta}\int_{0}^x\tilde{g}_k(x',\tau)g_k(x-x',\tau)dx'dx\leq \frac{1}{k}||\tilde{g}_k||_1^2\leq  \frac{1}{k}},\\
    & \hspace{1cm}+\frac{1}{1-\tau}\underbrace{\int_{0}^{\eta}\int_{0}^{\frac{\lceil kx\rceil}{k}}\frac{\lceil kx'\rceil}{k}g(x',\tau)\left|\tilde{g}_k\left(\frac{\lceil kx\rceil-\lceil kx'\rceil}{k},\tau\right)-\tilde{g}_k\left(x-x',\tau\right)\right|dx'dx}_{\leq \int_{1/k}^{\eta}\left|\tilde{g}_k\left(x-\frac{1}{k},\tau\right)-\tilde{g}_k\left(x,\tau\right)\right|dx}\\
     &\hspace{1cm}\leq  \frac{2}{k(1-\tau)}+\frac{1}{1-\tau}\int_{1/k}^{\eta}\left|\tilde{g}_k\left(x-\frac{1}{k},\tau\right)-\tilde{g}_k\left(x,\tau\right)\right|dx.
\end{align*}
It remains to bound $\int_{1/k}^{\eta}\left|\tilde{g}_k\left(x-\frac{1}{k},\tau\right)-\tilde{g}_k\left(x,\tau\right)\right|dx$. First, we have:
\[
\int_{1/k}^{\eta}\left|\tilde{g}_k\left(x,0\right)-\tilde{g}_k\left(x-\frac{1}{k},0\right)\right|dx\leq \int_{1/k}^{+\infty} k^2(1-e^{-1/k}\left(e^{\frac{\lceil kx \rceil-1}{k}}-e^{\frac{\lceil kx \rceil-2}{k}}\right)\leq \frac{2}{k}.
\]
Also, for any $\tau \in [0;1]$
\begin{align*}
    \int_{1/k}^{+\infty}\left|\tilde{g}_k\left(x,\tau\right)-\tilde{g}_k\left(x-\frac{1}{k},\tau\right)\right|dx \leq \left(2\eta+\frac{1}{1-\tau}\right)\int_{1/k}^{+\infty}\left|\tilde{g}_k\left(x,\tau\right)-\tilde{g}_k\left(x-\frac{1}{k},\tau\right)\right|dx.
\end{align*}
Hence, by Gronwall's lemma:
\[
\int_{1/k}^{+\infty}\left|\tilde{g}_k\left(x,\tau\right)-\tilde{g}_k\left(x-\frac{1}{k},\tau\right)\right|dx\leq \frac{2}{k}e^{\left(2\eta+\frac{1}{1-\tau}\right)}.
\]
Reinjecting in the previous equation, we obtain:
\begin{align}
||B_{k,2}\left(\tilde{g}_k\left(.,\tau\right)\right)(x)-A_2\left(\tilde{g}_k\left(.,\tau\right)\right)(x)||_1\leq \frac{2}{k(1-\tau)}\left(1+e^{\left(2\eta+\frac{1}{1-\tau}\right)}\right).\label{eq:boundA2B2}
\end{align}
Thus, with \cref{eq:boundA2Lip,eq:boundinitggk,eq:boundA2B2}, by application of Gronwall's Lemma, for any $t \in [0,1]$
\[
\int_{0}^\eta |g(x,\tau)- g_{k}(x,\tau)|dx\leq \frac{3}{k}e^{\eta+\frac{2\eta+1}{1-\tau}}+\frac{2}{k(1-\tau)}\left(1+e^{\left(2\eta+\frac{1}{1-\tau}\right)}\right).
\]

Re-injecting in \cref{eq:boundspaceintergk}, we obtain:

\begin{align}
    \left|\int_{x=0}^{\eta}x^2g(x,\tau)dx-
\sum_{\ell=1}^{k\eta}\left(\frac{\ell}{k}\right)^2g_k(\ell,\tau)\right|\leq&\frac{w_2(\eta,\tau)}{k},\label{eq:boundspacegk}
\end{align}
with $w_2(\eta,\tau)$ a constant that depends on $\tau$ and $\eta$. Finally, combining with \cref{eq:boundriemmanmetric}:
\begin{align*}
    \left|\int_{t=0}^{\tau}\int_{x=1}^{\eta}x^2g(x,t)dxdt-
    \frac{1}{n}\sum_{t=0}^{n\tau-1}\sum_{\ell=1}^{k\eta}\left(\frac{\ell}{k}\right)^2g_k(\ell,\frac{t}{n})\right| \leq& \left|\int_{t=0}^{\tau}\sum_{\ell=1}^{k\eta}\left(\frac{\ell}{k}\right)^2g_k(\ell,t)dt-\frac{1}{n}\sum_{t=0}^{n\tau-1}\sum_{\ell=1}^{k\eta}\left(\frac{\ell}{k}\right)^2g_k(\ell,\frac{t}{n})\right|dt\\ 
    &+\int_{t=0}^{\tau}\left|\int_{x=0}^{\eta}x^2g(x,t)dx-
\sum_{\ell=1}^{k\eta}\left(\frac{\ell}{k}\right)^2g_k(\ell,t)\right|dt\\
    \leq& \frac{w_1(\tau,\eta)}{n}+\frac{w_2(\tau,\eta)}{k}.
\end{align*}

\hfill \(\Box \)

\section{Proof of Lemma \ref{lem:boundprobahighcost}}\label{app:boundprobahighcost}

Consider any integers $a,b\in [n]^2$ and the interval $[\frac{a}{n};\frac{b}{n}]$. Denote $z$ the length of that interval. We have  $z= \frac{b}{n}-\frac{a}{n}$ if $b>a$, else $z =\frac{b}{n}+1-\frac{a}{n} $. Assume
\[
z \geq \frac{16}{3(1-\tau)^2n}.
\]
The number of vertices from the online side that have fallen in this interval before any time $t\leq \tau n$ follows a binomial distribution $\mathcal{B}\left(t-1,z\right)$. By Chernoff bound: 
\begin{align*}
\mathbb{P}\left(\left|\left\{y_i \in \left[\frac{a}{n};\frac{b}{n}\right]\bigg|i<t\right\}\right|\geq \left(1+\frac{1-\tau}{4}\right)(t-1)z\right)&\\
&\hspace{-3cm}\leq \mathbb{P}\left(\left|\left\{y_i \in \left[\frac{a}{n};\frac{b}{n}\right]\bigg|i\neq t, i \neq \tau n \right\}\right|\geq \left(1+\frac{1-\tau}{4}\right)(\tau n-1)z\right)\\
   &\hspace{-3cm} \leq e^{-\frac{(1-\tau)^2}{48}(\tau n-1)z}\leq e^{-\frac{(1-\tau)^2}{64}\tau n z}.
\end{align*}
The last inequality holds as $n\geq \frac{25}{\tau}$.
The number of vertices from the offline side, $\left|\left\{u \in [\frac{a+1}{n};\frac{b-1}{n}]|u\in \tilde{\mathcal{U}}\right\}\right|$ follows a binomial distribution $\mathcal{B}\left(nkz-2k,p_k\right)$.
\begin{align*}
\mathbb{P}\left(\left|\left\{u \in [\frac{a+1}{n};\frac{b-1}{n}]\bigg|u\in \tilde{\mathcal{U}}\right\}\right|\leq \left(1-\frac{1-\tau}{4}\right)(nkz-2)p_k\right)
\leq&e^{-\frac{(1-\tau)^2}{32}(nkz-2k)p_k}\\
\leq& e^{-\frac{(1-\tau)^2}{64}(nz-2)}\\
\leq &e^{-\frac{(1-\tau)^2}{64}nz\tau}.
\end{align*}

The second inequality holds as $kp_k\geq 1-\frac{1}{2k}\geq 0.5$ and the last by the constraint on the value of $z$. We have:
\begin{align*}
   \left(1-\frac{1-\tau}{4}\right)(nkz-2k)p_k- \left(1+\frac{1-\tau}{4}\right)(n\tau-1)z&\geq zn\left[\left(1-\frac{1-\tau}{4}\right)\underbrace{kp_k}_{\geq1-\frac{1}{2k} }-\left(1+\frac{1-\tau}{4}\right)\tau\right]-2,\\
   &\hspace{-1cm}\geq zn\left[\left(1-\frac{1-\tau}{4}\right)\left(1-\frac{1-\tau}{2}\right)-\left(1+\frac{1-\tau}{4}\right)\tau\right]-2,\\
   &\hspace{-1cm}\geq zn\frac{3}{8}(1-\tau)^2-2\geq 0.
\end{align*}

This implies that if $\left|\left\{u \in ]\frac{a+1}{n};\frac{b-1}{n}[\bigg|u\in \tilde{\mathcal{U}}\right\}\right|\geq \left(1-\frac{1-\tau}{4}\right)(nkz-2)p_k$ and $\left|\left\{y_i \in \left[\frac{a}{n};\frac{b}{n}\right]\bigg|i<t\right\}\right|\leq \left(1+\frac{1-\tau}{4}\right)(t-1)z$, then:
\[
\left|\left\{y_i \in \left[\frac{a}{n};\frac{b}{kn}\right]\bigg|i<t\right\}\right|\leq\left|\left\{u \in \left[\frac{a+1}{kn};\frac{b-1}{n}\right]\bigg|u\in \tilde{\mathcal{U}}\right\}\right|.
\]
Let us now bound $\mathbb{P}\left(c_t\geq z|y_t\right)$. Denote $\frac{a^L}{kn}$ and $\frac{b^H}{kn}$ the positions of the two free vertices closest to $y_t$ upon arrival. Note that $c_t\geq \eta$ can only hold if  $\min(|\frac{a^L}{kn}-y_t|; |\frac{b^H}{kn}-y_t||]\geq \eta$ and all online vertices that arrived before $y_t$ in the interval $[\frac{a^L}{kn};\frac{b^H}{kn}]$ have been matched strictly inside that interval. 

\begin{align*}
  \mathbb{P}\left(c_t\geq \eta|y_t\right)\leq& \mathbb{P}\left(\bigcup\limits_{\substack{y_t \in [\frac{a^L}{kn}; \frac{b^H}{kn}] \\ \min(\left|[\frac{a^L}{kn};y_t];[y_t, \frac{b^H}{kn}]\right|)\geq \eta}} \left\{\left|\left\{u \in ]\frac{a^L}{kn};\frac{b^H}{kn}[\bigg|u\in \tilde{\mathcal{U}}\right\}\right|\leq   \left|\left\{y_i \in \left[\frac{a^L}{kn};\frac{b^H}{kn}\right]\bigg|i<t\right\}\right|\right\}\bigg |y_t\right)\\
  \leq& \mathbb{P}\left(\bigcup\limits_{\substack{(a,b)\in [n]^2\\y_t \in [\frac{a}{n}; \frac{b}{n}] \\ \min(\left|[\frac{a}{kn};y_t];[y_t, \frac{b}{kn}]\right|)\geq \eta}} \left\{\left|\left\{u \in [\frac{a+1}{kn};\frac{b-1}{kn}]\bigg|u\in \tilde{\mathcal{U}}\right\}\right|\leq   \left|\left\{y_i \in \left[\frac{a}{kn};\frac{b}{kn}\right]\bigg|i<t\right\}\right|\right\}\bigg |y_t\right)\\
  \leq& \sum_{\ell= \eta n}^{n}\sum_{\substack{ y_t \in [\frac{a}{n}; \frac{b}{n}] \\\left|[\frac{a}{kn}; \frac{b}{kn}]\right|= \frac{\ell}{n}}} \mathbb{P}\left(\left|\left\{u \in [\frac{a+1}{kn};\frac{b-1}{kn}]\bigg|u\in \tilde{\mathcal{U}}\right\}\right|\leq   \left|\left\{y_i \in \left[\frac{a}{kn};\frac{b}{kn}\right]\bigg|i<t\right\}\right|\right)\\
  \leq & \sum_{\ell= \eta n}^{n}\sum_{\substack{y_t \in [\frac{a}{n}; \frac{b}{n}] \\ \left|[\frac{a}{kn}; \frac{b}{n}]\right|= \frac{\ell}{n}}} 2e^{-\frac{(1-\tau)^2}{64}n\tau\frac{\ell}{n}}\\
  \leq&  e^{-C'_\tau n \eta}\sum_{\ell= 1}^{n}\sum_{\substack{y_t \in [\frac{a}{n}; \frac{b}{n}] \\  \left|[\frac{a}{n}; \frac{b}{n}]\right|= \frac{\ell}{n}+\eta}} 2e^{-\frac{(1-\tau)^2}{64}\tau\ell }\\
  \leq&  e^{-C'_\tau n \eta}\underbrace{\sum_{\ell= 1}^{n}2\ell e^{-\frac{(1-\tau)^2\tau}{64} \ell}}_{C^{''}_\tau}.\\
\end{align*}

We have:
\begin{align*}
\mathbb{E}\left[\sum_{t\leq n\tau}c_t[k]\mathds{1}\{c_t[k]\geq \frac{\eta}{n}\} \right]\leq &\sum_{t=1}^{n\tau} \int_{z=\eta}^1 \mathbb{P}(c_t[k]\geq z) dz,\\
\leq &\sum_{t=1}^n\int_{z=\frac{\eta}{n}}^1 C^{''}_\tau e^{-z n C'_\tau }dz,\\
\leq& \frac{C^{''}_\tau}{C'_\tau} e^{-\eta C'_\tau}.
\end{align*}
\hfill \(\Box\)

\end{APPENDICES}
\end{document}